\documentclass[a4paper]{article}

\usepackage{float}
\usepackage[
  a4paper,top=3cm,bottom=2cm,left=3cm,right=3cm,marginparwidth=1.75cm]{geometry}
\usepackage{amsmath}
\usepackage{calrsfs}
\usepackage{amsthm,bm,mathtools,etoolbox,mathrsfs,amsfonts,bbm}
\usepackage{hyperref}
\hypersetup{colorlinks,linkcolor={blue},citecolor={blue},urlcolor={blue}}    
\usepackage[nameinlink,capitalise]{cleveref}
\usepackage[utf8]{inputenc} 
\usepackage[T1]{fontenc}    
\usepackage{url}            
\usepackage{booktabs}       
\usepackage{natbib}
\usepackage{placeins}
\usepackage{adjustbox}
\usepackage{booktabs}
\usepackage{algorithm}
\usepackage{algpseudocode}
\usepackage{multicol} 
\usepackage{makecell, multirow, tabularx} 
\usepackage{rotating}
\usepackage{float}
\usepackage{ulem}
\usepackage{graphicx}
\usepackage{tabularx}
\usepackage{authblk}

\newtheorem{Theorem}{Theorem}[section]

\newtheorem{Lemma}{Lemma}[section]
\newtheorem{Remark}{Remark}[section]
\newtheorem{Corollary}{Corollary}[section]

\title{Bootstrap Prediction Inference of Non-linear Autoregressive Models}

\author[1]{Kejin Wu}
\author[2]{Dimitris N. Politis}

\affil[1]{Department of Mathematics, University of California, San Diego }
\affil[2]{Department of Mathematics and Halicio\u{g}lu Data Science Institute, University of California, San Diego}

\date{}
\begin{document}

\maketitle

\begin{abstract}
The non-linear autoregressive (NLAR) model plays an important role in modeling and predicting time series. One-step ahead prediction is straightforward using the NLAR model, but the multi-step ahead prediction is cumbersome. For instance, iterating the one-step ahead predictor is a convenient strategy for linear autoregressive (LAR) models, but it is suboptimal under NLAR. In this paper,  we first propose a simulation and/or bootstrap algorithm to construct optimal point predictors under an $L_1$ or $L_2$ loss criterion. In addition, we construct bootstrap prediction intervals in the multi-step ahead prediction problem; in particular, we develop an asymptotically valid quantile prediction interval as well as a pertinent prediction interval for future values. In order to correct the undercoverage of   prediction intervals with finite samples, we further employ predictive---as opposed to fitted---residuals in  the bootstrap process. Simulation studies are also given to substantiate the finite sample performance of our methods.
\end{abstract}

\section{Introduction}\label{Sec:1}
In the domain of time series analysis, accurate forecasting based on observed data is an important topic. Such single- or multi-step ahead predictions play an important role in forecasting crop yields, stock prices, traffic volume, etc. For Linear Autoregressive (LAR) models with finite order and independent,
identically distributed (i.i.d.) innovations, it is easy to construct the optimal
(with respect to $L_2$ risk) multi-step ahead predictor by iterating the
one-step ahead predictor. 

Actually, the $L_2$ optimal prediction of linear time series models does not dependent on the distribution of innovations beyond its first two moments \citep{guo1999multi}. The only things that matter are the historical observations and the parameters of the model. When the parameters are unknown, practitioners can apply the Box-Jenkins method of  identifying, fitting, checking and predicting LAR models systematically \citep{box1976time}. However, the LAR model may not be enough to analyze complicated data in the real world. As pointed out by the work of \cite{de1992some} and \cite{tjostheim1994non}, there are various occasions when prior knowledge indicates the data-generating process is in a non-linear form; see the review of \cite{politis2009financial} for example. Furthermore, there are several ways to test the hypothesis of linearity of the data at hand; see the work of \cite{berg2012testing} for traditional and bootstrap/subsampling approaches. 

The analysis of Non-linear Autoregressive (NLAR) models can be traced back to the work of \cite{jones1978nonlinear}. Putting aside the issue of identifying and estimating an NLAR model,  a major application of NLAR models is   forecasting   future values. 
 Although the one-step ahead   optimal (with respect to $L_2$ risk)
 prediction of (causal) NLAR models is usually easy to obtain, the optimal (no matter in $L_2$ or $L_1$ loss) multi-step ahead prediction can not be obtained by the iterative procedure   we employ for LAR models, even
when the NLAR model parameters are known.
 To resolve  this issue, \cite{pemberton1987exact} proposed a numerical integration approach to get the exact solution. However, his approach assumes that the distribution of innovation is known, which is usually not realistic in the real world. Besides, this numerical approach can be very computationally heavy for long-horizon predictions. Instead, some suboptimal ideas were proposed, such as estimating a model by minimizing multi-step ahead mean squared errors and then making predictions directly; see the work of \cite{zhang1998forecasting} and \cite{lee2003new} for a discussion.

The work of \cite{guo1999multi} further shed some light on the multi-step ahead prediction of NLAR models. Taking advantage of the true innovation distribution or the empirical residual distribution, they propose an analytic predictor which asymptotically converges to the optimal predictor. Nevertheless, their analyses are limited to the $L_2$ optimal point prediction and are lacking details when the model is unknown. In several applied areas, e.g. econometrics, 
climate modeling, water resources management, etc., data might not possess a finite 2nd moment in which case optimizing $L_2$ loss is vacuous. For example, financial returns  typically do not possess a finite 4th moment; hence, to predict their volatility (which is a 2nd moment), we can not rely on $L_2$ optimal  prediction.
For all such cases---but also of independent interest---prediction that is optimal with respect to $L_1$ loss should receive more attention in practice; see detailed discussions from Ch. 10 of \cite{politis2015model}.

Unfortunately, the aforementioned numerical integration and analytic methods for NLAR prediction can not be extended to $L_1$ optimal prediction directly. In addition, even for linear autoregressions, the multi-step ahead $L_1$ optimal predictor is elusive since iterating the one-step ahead predictor does not work in the $L_1$ loss setup. Furthermore, we should also be concerned about the accuracy of our point predictions. In analogy to the construction of Confidence Intervals (CI) in estimation problems, we may attempt to measure the accuracy of point predictions by constructing Prediction Intervals (PI); see the formal definition of such measures in \cref{Sec:2}.

In the paper at hand, we provide an algorithm to make prediction inferences for a general class of NLAR models. Then, we focus on analyzing a popular type of NLAR model with a specific structural form that contains the separate parametric mean and volatility/variance functions. When the model and innovation distribution are known, we can deploy Monte Carlo (MC) {\it simulation} to achieve consistent forecasting. When the model is unknown---which is typically the case---we need to fit the model to get estimated parameters and innovation distribution\footnote{Throughout, we assume that the order of the parametric non-linear time series model $p$ is known; when we say the model is unknown, we mean that the corresponding parameter values of this model are unknown.}. Performing MC simulation using the fitted model and estimated innovation distribution effectively becomes a {\it bootstrap} method.

Although the one-step ahead bootstrap optimal predictors and PI for NLAR models were constructed in \cite{politis2015model,pan2016bootstrap}, the methodology for multi-step ahead prediction inference was left open. The challenge for this task is that we must handle the effects of future innovations on predictions appropriately since they play a crucial role due to the non-linearity. In short, our idea is to derive the expression of the multi-step ahead future value by the true/estimated model first and then attempt to approximate this cumbersome expression through simulation/bootstrap so that we can capture the (conditional) distribution of future values.
 
The bootstrap idea was introduced by \cite{efron1979bootstrap} to carry out statistical inference for independent data. After that, many variants of bootstrap were developed to handle time series data which possess an inherent dependence structure; see the book \cite{KreissPaparoditisbook2023} for discussions. For estimation purposes, the core idea of the bootstrap is to mimic the  underlying stochastic structure in order to generate artificial samples in the bootstrap world; on each of those {\it re-samples}, the statistic of interest can be recomputed, thus manifesting the variability of the statistic across samples.

However, for the task of time series  prediction, we should notice that all future predictions are conditional on the latest $p$ observed data where $p$ is the order of the NLAR model. Thus, to construct a reasonable predictor in the bootstrap world, we need to make sure predictors of the bootstrap series are also conditional on the
exact same $p$ data; this is the idea of the forward bootstrap proposed by \cite{politis2015model} and \cite{pan2016bootstrap}. Using the forward bootstrap approach, with consistent estimations on the NLAR model and innovation distribution, we can simulate many future values in the bootstrap world. Then, the empirical distribution of a future value in the bootstrap world can be used to approximate the distribution of the future value in the real world. A
so-called Quantile PI (QPI) can also be built by directly using the
relevant quantile values of this empirical distribution; this method is similar to the density forecast of future value---see \cite{chen2004nonparametric} and \cite{manzan2008bootstrap} who applied   non-parametric approaches to do density forecast. 
Although asymptotically valid, the QPI is typically characterized by finite-sample undercoverage because it does not take the variability of the model estimation into account; 
see \cite{wang2021model} made a related discussion.

To account for the variability in
 model estimation, \cite{politis2015model} introduced the notion
of a so-called Pertinent PI (PPI) that 
has a better empirical Coverage Rate (CVR) in finite-sample cases; see more explanations in \cref{Sec:2}. To implement the PPI, we need to impose more requirements on the bootstrap series, i.e., we require that the estimated model in the bootstrap is also consistent to the true one. 
To check the consistency, 
 the bootstrap series should possess some mixing or weak dependence properties. 
As developed in the work of \cite{franke2002properties}, it is possible to get a self-ergodic bootstrap series that also approximates the true series by a non-parametric autoregressive residual bootstrap (AR bootstrap) approach. Nevertheless, as far as we know, there is no literature about performing AR bootstrap in a non-linear parametric approach to do predictions. For the non-parametric approach within the forward bootstrap frame, only \cite{pan2016bootstrap} took the local constant technique to estimate the model and then make one-step ahead predictions. After that, although some studies devoted to improving the bootstrap prediction performance or combining the bootstrap idea with the prediction of the state-of-the-art technique---neural network, see the work of \cite{trucios2017robust} and \cite{eugriouglu2020new} for example, there is a lack of research in direct non-parametric forward-bootstrap prediction for both theoretical and practical aspects. Beyond applying the local constant technique, we can consider other non-parametric estimation methods, such as local liner/polynomial methods. Since the bootstrap can not capture the bias-type term of the non-parametric estimator exactly, these considerations may be gainful in the prediction process because the inherent estimation bias term of local liner/polynomial estimation is simplified and tends to be smaller than applying local constant estimation. We leave the discussion and development of non-parametric forward-bootstrap prediction in future work. For this paper, we focus on the forward-bootstrap prediction of parametric non-linear models.

We should also notice that the earliest method which is similar to ours is the work of \cite{paparoditis1998backward} where the maximum likelihood estimator was applied to approximate a parametric AR model with ARCH residuals, and then they adopted the local bootstrap proposed by \cite{paparoditis2002local} to generate the bootstrap series. After re-estimating parameters based on the bootstrap series, the prediction in the real world is approximated by its analog in the bootstrap world. \cite{paparoditis1998backward} argued that such bootstrap predictions can capture the variability due to parameter estimations, thus resulting in a pertinent prediction interval.
However, they do not  consider multi-step ahead prediction. 
Finally, \cite{chen2019optimal} applied the bootstrap technique to find approximations of the optimal predictors and build corresponding PIs for the so-called NoVaS transformation and compare to ARCH and GARCH models; they also consider multi-step ahead prediction albeit without formal proof of consistency. The last two works mentioned are both  limited to a specific type of model. 
Instead, the paper at hand rigorously explores one-step ahead 
and multi-step ahead prediction inference under general    
parametric non-linear NLAR models. 

The variability of parameter estimation in finite-sample cases  can be captured by relying on the PPI with the AR bootstrap. Nevertheless, we may still find 
 bootstrap-based PPIs undercovering for small sample sizes. 
To further boost the empirical CVR of our
 bootstrap-based PPIs, we may use predictive (instead of fitted) residuals 
analogously to the successful construction of PI for regression and autoregression with predictive residuals in work of \cite{politis2013model} and \cite{pan2016bootstrap}. In brief, the $t$-th time predictive residual is computed using the delete-$x_t$ dataset; the formal definition of predictive residuals is presented in \cref{Sec:4}. Although the performance of PIs with fitted and predictive residuals are asymptotically the same, the CVR of PIs can typically be improved by the use of predictive residuals when the sample size is small.

The paper is organized as follows. In \cref{Sec:2}, an algorithm to offer prediction inference of general NLAR models will be provided. In \cref{Sec:3}, based on the aforementioned algorithms, we show how to handle a specific form of NLAR model when we have knowledge of  its innovation distribution, mean and variance functions. In \cref{Sec:4}, under standard assumptions, we show the consistency of  optimal point prediction; 
furthermore, we show that  asymptotically valid PI and PPI can be constructed when the parametric model and innovation distribution are unknown. In \cref{Sec:5}, some simulation results will be presented. Conclusions are given in \cref{Sec:6}. The proofs of theorems from \cref{Sec:3,Sec:4} are given in \hyperref[Appendix:Proof]{Appendix}.

\section{Prediction of general NLAR models}\label{Sec:2}
In this paper, we assume that we observe $T+p$ number of real-valued data $\{X_{-p + 1}, X_{-p + 2},\ldots, X_T\}$ from
a general ergodic NLAR model which satisfies the recursion shown below:
\begin{equation}\label{Eq1}
    X_t = G(\bm{X_{t-1}}, \epsilon_t) ;
\end{equation}
 here $\{\epsilon_t\}$ is assumed to be i.i.d. with mean zero, and $\bm{X_{t-1}}$ represents vector $\{X_{t-1},\ldots,X_{t-p}\}$. Throughout this paper, we will assume $X_i$ represents a random variable and $x_i$ represents the observation corresponding to $X_i$. Thus, $\bm{x_{t-1}}$ represents the observation of $\bm{X_{t-1}}$ which is $\{x_{t-1},\ldots,x_{t-p}\}$. $G(\cdot,\cdot)$ can be any continuous (possibly non-linear) function that makes the variance and mean of $\{X_t\}$ finite. 
\begin{Remark}
In section 4 of \cite{tong1990non}, \cref{Eq1} is analyzed as a stochastic difference equation. Assuming that \cref{Eq1} can be decomposable to $\phi(\bm{X_{t-1}}) + \sigma(\bm{X_{t-1}},\epsilon_t)$, they give the conditions under which \cref{Eq1} is ergodic. We will also discuss similar conditions in \cref{Sec:3}. Here, we are interested in providing a methodological algorithm to predict the general NLAR model \cref{Eq1}. 
\end{Remark}
The problem is how can we make multi-step ahead prediction inferences with such a complicated model. As known to us, for a stochastic process $\{X_t\}_{t=-\infty}^{T}$, the $L_2$ optimal predictor of $X_{T+h}$, $h\geq 1$, given the (infinite) past is:
\begin{equation}\label{Eq2}
    \mathbb{E}[X_{T+h}|X_s, s\leq T],
\end{equation}
when it exists. As pointed out by \cite{pemberton1987exact}, this result does not require the stochastic process to be stationary. In our case, since we assume the order of the NLAR model $p$ is finite. \cref{Eq2} can be simplified to:
\begin{equation}\label{Eq3}
     \mathbb{E}[X_{T+h}|X_T,\ldots,X_{T-p+1}].
\end{equation}
Similarly, the $L_1$ optimal predictor of $X_{T+h}$ given past history is the conditional median:
\begin{equation}\label{Eq4}
    Q_{X_{T+h}|X_{T},\ldots,X_{T-p+1}}(1/2),
\end{equation}
where $Q_{X_{T+h}|X_{T},\ldots,X_{T-p+1}}()$ is the conditional quantile function of $X_{T+h}$. 

We will call \cref{Eq3} and \cref{Eq4} the exactly optimal point predictors based on $L_1$ or $L_2$ loss. However, it is hard to compute them directly. Subsequently, we will propose the simulation  or bootstrap-based method to find an approximation of the exactly optimal prediction. Moreover, we also consider the PI of future values;  an asymptotically valid PI of $X_{T+h}$ with $(1-\alpha)100\%$ CVR given past history can be defined as:
\begin{equation}\label{Eq5}
    \mathbb{P}(L \leq X_{T+h} \leq U) \overset{p}{\to} 1-\alpha,~\text{as}~T\overset{}{\to}\infty,
\end{equation}
where $L$ and $U$ are lower and higher PI bounds, respectively. Implicitly, the probability $\mathbb{P}$ should be understood as the conditional probability given the latest $p$ observations. We typically construct a PI that is centered at some meaningful point predictor $
\widehat{X}_{T+h}$. An asymptotically valid centered PI with $(1-\alpha)100\%$ CVR given past history can then be defined as:
\begin{equation}
    \mathbb{P}(\widehat{X}_{T+h} + R_{\alpha/2} \leq X_{T+h} \leq \widehat{X}_{T+h} + R_{1-\alpha/2}  ) \overset{p}{\to} 1-\alpha,~\text{as}~n\overset{}{\to}\infty,
\end{equation}
where $R_{\alpha/2}$ and $R_{1 - \alpha/2}$ denote the lower $\alpha/2$ and $1 - \alpha/2$ quantiles with respect to the conditional distribution of the so-called {\it predictive root}: $X_{T+h} - \widehat{X}_{T+h}$.

\subsection{Simulation-based prediction}
Consider an idealized situation where  the model $G(\bm{X_{t-1}}, \epsilon_t)$ and the distribution of innovations $F_{\epsilon}$ are   known.
In this case, we propose to
approximate the $h$-step  ahead prediction by simulating innovations from $F_{\epsilon}$ and plugging them into the NLAR model. 
To describe the idea, focus on the   2-step  ahead prediction, and
note that the distribution  of the future value $X_{T+2}$ 
(conditional on the observed $\bm{X_T}$)
is identical to the distribution of 
\begin{equation*}
     G(G(\bm{X_T},\epsilon^{*}_{T+1}),\epsilon^{*}_{T+2}),
\end{equation*}
where $\epsilon^{*}_{T+1}$ and $\epsilon^{*}_{T+2}$ are  i.i.d.$\sim F_{\epsilon}$.

Going to the  $h$-step  ahead prediction,  the distribution  of the future value $X_{T+h}$ 
(conditional on the observed $\bm{X_T}$)
is identical to the distribution of the quantity 
\begin{equation}\label{Eq7}
 G(\cdots G(G(G(\bm{X_T},\epsilon^{*}_{T+1}),\epsilon^{*}_{T+2}),\epsilon^{*}_{T+3}),\ldots,\epsilon^{*}_{T+h}),
\end{equation}
where $\epsilon^{*}_{T+1},\ldots, \epsilon^{*}_{T+h}$ are i.i.d.$\sim F_{\epsilon}$.

Of course, in order to obtain the $L_2$ or $L_1$ optimal point predictor, we would need to approximate the mean or median of the quantity 
(\ref{Eq7}). We can do this by Monte Carlo (MC) simulation; the simulation will be based on $M$ replicates of the quantity 
(\ref{Eq7}); these are denoted $\{X^{(m)}_{T+h}\}_{m=1}^{M}$.
 To generate $X^{(m)}_{T+h}$, we need to generate $h$ values 
 i.i.d. from $F_{\epsilon}$; the latter are denoted $\{\epsilon^{(m)}_{T+1},\ldots,\epsilon^{(m)}_{T+h} \} $.
Since we need to generate $X^{(m)}_{T+h}$ for $m=1,\ldots, M$,
all in all, we need to generate $hM$ values 
 i.i.d. from $F_{\epsilon}$; this sums up the computational cost of
the MC simulation.

The $L_2$ or $L_1$ optimal   predictor of $X_{T+h}$ can be approximated by the mean or median, respectively, of $\{X^{(1)}_{T+h},\ldots,X^{(M)}_{T+h}\}$; the approximation will become
more and more accurate as $M$ tends to infinity.
Moreover, we can also find the approximation of the optimal predictor of $f(X_{T+h})$ by taking the mean or median of $\{f(X^{(1)}_{T+h}),\ldots,f(X^{(M)}_{T+h})\}$ where $f$ is  a continuous function
of interest.  Finally, 
the empirical distribution of the
 values $\{X^{(1)}_{T+h},\ldots,X^{(M)}_{T+h}\}$ can also be used to 
approximate the distribution of the future value $X_{T+h}$ 
(conditional on the observed $\bm{X_T}$), leading to the construction
of asymptotically valid PIs; the details will be given shortly.

\subsection{Bootstrap-based prediction}

In the more realistic scenario, both 
  $G(\cdot , \cdot)$ and $F_{\epsilon}$ are unknown but can be estimated 
from the data at hand; denote their estimators by $\widehat{G}(\cdot , \cdot)$ and $\widehat{F}_{\epsilon}$ respectively, and assume they are consistent. 
Typically, in order to construct the estimator $\widehat{F}_{\epsilon}$, 
the practitioner must be able to approximate the unobserved errors
 $\epsilon_t$ by  $\hat \epsilon_t$; in standard situations, 
$\hat \epsilon_t$ have the interpretation of {\it residuals} from fitting
the model \cref{Eq1} to the data. 

\begin{Remark} In order to decompose the estimation from the prediction problem, the  estimated quantities   $\widehat{G}(\cdot, \cdot)$, $\widehat{F}_{\epsilon}$
and  $\hat \epsilon_t$ could be based just on the data $X_{-p+1},\ldots, X_{T-p}$
instead of the full sample $X_{-p+1},\ldots, X_{T}$. This trick may help  with theoretical analysis 
but is not important in  practical applications.
\end{Remark} 

Conducting a simulation as described in the previous subsection
using $\widehat{G}(\cdot , \cdot)$ and $\widehat{F}_{\epsilon}$  in place of the unknown $G(\cdot , \cdot)$ and $F_{\epsilon}$ turns the 
MC simulation into a {\it bootstrap} procedure.
The bootstrap version of the quantity 
(\ref{Eq7}) is now given by 

\begin{equation}\label{Eq8}
 \widehat{G}(\cdots \widehat{G}(\widehat{G}(\widehat{G}(\bm{X_T},\hat{\epsilon}^{*}_{T+1}),\hat{\epsilon}^{*}_{T+2}),\hat{\epsilon}^{*}_{T+3}),\ldots,\hat{\epsilon}^{*}_{T+h}),
\end{equation}
where $\{\hat{\epsilon}^{*}_t\}_{t=T+1}^{T+h}$ are i.i.d.$\sim\widehat{F}_{\epsilon}$; $\widehat{G}(\cdot,\cdot)$ is an estimator to the true model. Next, we can take a similar approach as the simulation-based method previously described   to approximate the optimal predictor of $X_{T+h}$. We summarize this procedure in \cref{algori1}.
\begin{algorithm}[H]
\caption{$h$-step  ahead bootstrap prediction of $X_{T+h}$ for a general NLAR model}
\centering
\label{algori1}
  \begin{tabularx}{\textwidth}{lX}   
    Step 1 & Fit the data $\{X_{-p+1},$ $\ldots,X_{T}\}$ 
with the general autoregressive model $G(\bm{X_{t-1}}, \epsilon_t)$
of \cref{Eq1}, and construct the estimator $\widehat{G}(\cdot,\cdot)$. Furthermore, compute all fitted residuals $\{\hat{\epsilon}_t\}_{t=1}^{T}$. Then, center all residuals by subtracting their sample mean 
$ \hat{\epsilon}_t - T^{-1}\sum_{i=1}^{T}\hat{\epsilon}_i$ for $t = 1, \ldots,T$.  
Denote $\widehat{F}_{\epsilon}$ the empirical distribution of centered residuals.
 \\
    Step 2 & Generate $\{\hat{\epsilon}^{(m)}_{T+1},\ldots,\hat{\epsilon}^{(m)}_{T+h} \}_{m=1}^{M}$ as i.i.d. from $\widehat{F}_{\epsilon}$,  and plug them into \cref{Eq8} to obtain $M$ pseudo-values  $\{{X}^{(1)}_{T+h},\ldots,{X}^{(M)}_{T+h}\}$. \\
    Step 3 & The  $L_2$ or $L_1$ optimal predictor of $X_{T+h}$ can be approximated by $\widehat{X}_{T+h}$ which is the mean or median of $\{X^{(1)}_{T+h},\ldots,X^{(M)}_{T+h}\}$, respectively.\\
  \end{tabularx}
\end{algorithm}

If $\widehat{G}(\cdot,\cdot)$ and $\widehat{F}_{\epsilon}$
are consistent to $ {G}(\cdot,\cdot)$ and $ {F}_{\epsilon}$, respectively. Then, $\widehat{X}_{T+h}$ can be made to be 
arbitrarily close to the true optimal  predictor by taking $M$ large enough. Furthermore, based on \cref{algori1}, the conditional distribution of $X_{T+h}$ is approximated by the empirical distribution of $\{X^{(1)}_{T+h},\ldots,X^{(M)}_{T+h}\}$. It is then straightforward to construct a bootstrap 
QPI that is asymptotically valid.  However, such a QPI will suffer  from finite-sample undercoverage  since the variability of estimating the model is not taken into account. We will prefer constructing  a PPI instead that possesses a stronger property than the asymptotic validity. In other words, the PPI can asymptotically capture the variability stemming from the error in the estimation of the model---see \cite{politis2015model} and \cite{pan2016bootstrap} for an in-depth discussion.  


We present the method to build the PPI for $X_{T+h}$ in \cref{algori2}. In short, we employ the distribution of the bootstrap predictive root $X_{T+h}^*- \widehat{X}^{*}_{T+h}$  to estimate the conditional distribution of the predictive root $X_{T+h}- \widehat{X}_{T+h}$ in the real world. 
Then, we can use the bootstrap quantiles associated with the bootstrap predictive root $X_{T+h}^*- \widehat{X}^{*}_{T+h}$ as
bounds for the true root  $X_{T+h}- \widehat{X}_{T+h}$;   solving
for $X_{T+h}$ yields the desired PI. The fact that this PI is actually
pertinent can be justified under additional conditions;
see Definition 2.4 of \cite{pan2016bootstrap} for a formal discussion. 
\begin{algorithm}[htbp]
\caption{$h$-step ahead bootstrap PPI of $X_{T+h}$ for a general NLAR}
\centering
\label{algori2}
    \begin{tabularx}{\textwidth}{lX}     
    Step 1 & Same as Step 1 of \cref{algori1}.
\\
    Step 2 & Apply Step 2-3 of \cref{algori1} to construct $\widehat{X}_{T+h}$, the  optimal point predictor of $X_{T+h}$ (based on  $L_1$ or $L_2$ loss).  \\
    Step 3 & (a) Resample the residuals $\{\hat{\epsilon}^{*}_{t}\}_{t=1}^{T}$ and $\{\hat{\epsilon}^{*}_{t}\}_{t=T+1}^{T+h}$ as i.i.d. from $\widehat{F}_{\epsilon}$ to create pseudo-errors. \\
    &(b) Let $(X_{-p+1}^*,\ldots, X_{0}^*)'=({X}_{0+I},\cdots,{X}_{p+I-1})'$ 
where $I$ is generated as a discrete random variable uniformly on the values $-p+1,\ldots, T-p+1$. Then, use the fitted general autoregressive model of Step 1 and the generated $\{\hat{\epsilon}^{*}_{t}\}_{t=1}^{T}$ in Step 3 (a) to get 
 bootstrap pseudo-series $\{X_t^*\}_{t = 1}^{T}$ in a recursive manner, i.e., $X_i^{*} = \widehat{G}(\bm{X}_{i-1}^{*},\hat{\epsilon}^{*}_{i})$ for $i=1,\ldots,T$.  \\
 & (c) Based on the bootstrap series $\{X^{*}_{t}\}_{t=-p+1}^{T}$, 
re-estimate the general model to obtain $\widehat{G}^{*}(\cdot, \cdot)$.\\
 & (d) Guided by the idea of forward bootstrap, re-define the last $p$ values of the bootstrap data to match the original data, i.e., re-define $X_t^*=X_t$ for $t=T-p+1, \dots, T$.\\
& (e) Use $\widehat{G}(\cdot,\cdot)$, the bootstrap data $\{X_t^*\}_{t = -p+1}^{T}$, and the pseudo-errors $\{\hat{\epsilon}_{t}^*\}_{t = T+1}^{T+h}$ to generate recursively the future bootstrap data $X_{T+1}^*,\ldots,X_{T+h}^{*}$, i.e., $X^{*}_{T+i} = \widehat{G}(\bm{X}_{T+i-1}^{*},\hat{\epsilon}^{*}_{T+i})$ for $i=1,\ldots,h$.
\\
  & (f) With bootstrap data $\{X^*_{t}\}_{t=-p+1}^{T}$ and the re-estimated model $\widehat{G}^{*}(\cdot, \cdot)$, utilize \cref{algori1} to compute the bootstrap predictor denoted by $\widehat{X}^*_{T+h}$, with the same loss criterion as in Step 2.\\
& (g) Construct the bootstrap predictive root: $X_{T+h}^*- \widehat{X}^*_{T+h}$. \\
Step 4 & Repeat Step 3 $K$ times. The $K$ bootstrap root replicates are collected in the form of an empirical distribution whose $\beta$-quantile is denoted $q(\beta)$.
The $(1-\alpha)100\%$ equal-tailed prediction interval for ${X}_{T+h}$ centered at $\widehat{X}_{T+h}$ is then approximated by
 $[\widehat{X}_{T+h}+q(\alpha/2), \widehat{X} _{T+h} +q(1-\alpha/2)].$
  \end{tabularx}
\end{algorithm}
\begin{Remark}\label{Remark2.1}
Step 3 part (f) of \cref{algori2} in the bootstrap world is an analogy with Step 2 in the real world. For Step 3 part (f), we actually make the bootstrap prediction again in the bootstrap world. The reason behind this step is that the effects of innovation can not be reduced in future predictions even if the mean of the innovation is 0, which is different from the situation for the linear time series. This double-bootstrap step is worthy and necessary if we want to get a PI that is asymptotically centered at the mean or median of the future value.   
\end{Remark}

As discussed in \cref{Sec:1}, to improve the finite-sample CVR of PIs we may use predictive (instead of fitted) residuals in the bootstrap procedure.
We will give details in the following section where we focus on a specific NLAR of interest. We should clarify that the success of \cref{algori1,algori2} and subsequent algorithms heavily depend on the estimation accuracy of $G(\cdot,\cdot)$. If the estimator $\widehat{G}(\cdot,\cdot)$ is asymptotically consistent to the true model. Besides, with uniform convergence between empirical residual and true innovation distributions, the consistency of bootstrap predictions to the exactly optimal point predictors is possible. With further conditions, the PPI is also achievable. In other words, if we can find such desirable estimations, these above prediction inferences will be valid.
\section{Prediction inference for a specific NLAR of interest}\label{Sec:3}
Here and in all that follows, we are exclusively interested in analyzing the NLAR model of the specific form:
\begin{equation}\label{Eq9}
    X_t = \phi(\bm{X_{t-1}}) + \sigma(\bm{X_{t-1}})\epsilon_t,
\end{equation}
where $\{\epsilon_t\}$ are i.i.d. innovations. \cref{Eq9} is a special case of \cref{Eq1} that decomposes the NLAR function $G  (\cdot,\cdot)$  to $\phi(\cdot)$
---that represents the conditional mean--- plus the so-called
variance function multiplying the innovations  $\epsilon_t$.
Under \cref{Eq9}, the innovations 
are more explicitly defined and thus easily estimable as residuals
after model fitting. If $\sigma(\bm{X_{t-1}})\equiv 1$,
then we have an NLAR with homoscedastic errors. For simplifying the notation, we assume that mean and variance functions have the same order $p$. We first impose some standard assumptions and suppose they are met throughout this paper:
\begin{itemize}
    \item[A1] $\phi(\cdot)$ and $\sigma(\cdot)$ are continuous functions from $\mathbb{R}^{p}$ to $\mathbb{R}$, and $\sigma(\cdot)$ is positive and bounded. Moreover, for quantifying the boundness of $X_t$ in probability to serve the proof, we assume that there are $C_{M}<\infty$ with $\mathbb{E}(|\sigma(X_0)\epsilon_1|^{M})\leq C_{M}$ for all $M<\infty$, where $X_0$ is the starting point of the time series.
    \item[A2] $\{\epsilon_t\}$ are $i.i.d.$ with
  distribution $F_{\epsilon}$, satisfying
$\mathbb{E}(\epsilon_t) = 0$ and 
  $E(\epsilon_t^2) = 1$. However, if 
 $\sigma(\bm{X_{t-1}})\equiv 1$ (homoscedastic errors case), then
$E(\epsilon_t^2) $ is not restricted to equal one, but needs to be finite.
    \item[A3] For all $t$, $\epsilon_t$ is independent of $\{X_{t-h},h>0\}$.
\end{itemize}

\subsection{Sufficient conditions for geometric ergodicity}\label{Sec:3.1}
To conduct statistical inference for non-linear time series data in the following sections, we need to find a tool to quantify the degree of asymptotic independence of time series. Popular choices are various mixing conditions. For simplifying proofs and relying on existing results, in this paper, we focus on time series with geometrically ergodic property which is equivalent to $\beta$-mixing condition with at least exponentially fast mixing rate; see \cite{bradley2005basic} made a detailed introduction of different mixing conditions and ergodicity. Thus, the first question we face is how we can make sure the geometric ergodicity of model \cref{Eq9} is satisfied.

Checking geometric ergodicity for a LAR model is simple; it is well known that the LAR model is stationary and geometrically ergodic as long as the corresponding characteristic polynomial does not have zero roots inside or on the unit circle. However, this check criterion depends on the linearity assumption, and can not be extended to serve for NLAR model directly \citep{an1996geometrical}. Thus, 
practitioners rely on Markov chain techniques to explore conditions under which the NLAR model is geometrically ergodic. The motivation is that the NLAR model can be described as a Markov chain in a general state space; the extensive discussions and literature related to Markov chains can guide the development of criteria to check the ergodicity of NLAR models. 

One of the earliest criteria developed to guarantee the ergodicity of a Markov chain is Doeblin's condition  given by \cite{doob1953stochastic}.  Later, \cite{tweedie1975sufficient} proposed a more generalized condition,   the so-called \textit{Drift criterion}. This criterion gives a sufficient condition for an aperiodic and irreducible Markov chain to be geometrically ergodic. For  completeness,
  we present this criterion using the version applied by \cite{an1996geometrical}:
\begin{Lemma}[Drift criterion]
\label{Lemma3.1}
Let $\{X_t\}$ be an aperiodic and irreducible Markov chain. Suppose that there exists a small set $C$, a non-negative measurable function $k$, positive constants $c_1$, $c_2$ and $\rho<1$ such that:
\begin{equation}
\begin{split}
    &\mathbb{E}(k(\bm{X_{t+1}})|\bm{X_{t}} = \bm{x}) \leq \rho k(\bm{x}) - c_1,~\text{for any}~\bm{x}\notin C,\\
    &\mathbb{E}(k(\bm{X_{t+1}})|\bm{X_{t}} = \bm{x}) \leq c_2,~\text{for any}~\bm{x}\in C.
\end{split}
\end{equation}
The function $k(\cdot)$ is called the test function in the literature. For the formal definition of a `small set'; see   \cite{tjostheim1990non} or \cite{tong1990non}. We will  soon see that the small set can be taken as a compact set in some situations. 
\end{Lemma}
Thus, to ensure the ergodicity of an NLAR model, people can check if
$\{X_t\}$ is aperiodic and irreducible and \cref{Lemma3.1} holds. Along with this idea, \cite{an1996geometrical} give several kinds of sufficient conditions for \cref{Eq9} with $\sigma(\bm{X_{t-1}})\equiv 1$ (homoscedastic errors case) to be geometrically ergodic. Note that the test function $k(\cdot)$ and the specific small set will change according to which condition is based. Then, \cite{min1999probabilistic} extend these results to the region of NLAR models with heteroscedastic errors. Based on this body of work, if we assume:
\begin{itemize}
    \item[A4] The probability density function of innovation  $f_{\epsilon}(\cdot)$ is continuous and everywhere positive.
    \item[A5] The conditional mean and volatility functions satisfy the  inequalities:
\begin{equation}
    \sup_{||\bm{x}||_2\leq K} |\phi(\bm{x})| < \infty~;~ \sup_{||\bm{x}||_2\leq K} |\sigma(\bm{x})| < \infty,~\text{for each}~K>0,
\end{equation}
where $\bm{x}\in\mathbb{R}^{p}$, and $||\cdot||_2$ is the Euclidean norm.
\end{itemize} 
we can obtain a useful Lemma:
\begin{Lemma}
\label{Lemma3.2}
Let $\{X_t\}$ satisfy the model \cref{Eq9}. Suppose A4 and A5 are fulfilled. Then $\{X_t\}$ is aperiodic and irreducible with respect to $\mu$ which is the Lebesgue measure. Moreover, the $\mu$-non-null compact sets are small sets.
\end{Lemma}
\begin{Remark}
The proof of \cref{Lemma3.2} can be found in \cite{min1999probabilistic}. The original proof only requires the density function to be lower semi-continuous. In this paper, since the estimation of the innovation distribution and the $L_1$ optimal predictor will be discussed, we require a stronger condition for the density function. Besides, we should notice that we can check the classical properties defined by the Markov chain to verify the aperiodicity and irreducibility. Nevertheless, it is more natural to apply \cref{Lemma3.2} for analyzing NLAR models. For example, when we consider a homoscedastic NLAR model with $p=1$, the drift criterion and the boundedness of the conditional mean function on a suitable compact set can be satisfied by a simple condition: $|\phi(x)|\leq C_2|x| + C_1$, for all $x$ and some $C_1<\infty, C_2 < 1$; this is exactly Assumption 3 (i) of \cite{franke2002properties}. In their work, they mention that the everywhere-positive assumption (A4) on the density function is unnecessarily restrictive. Thus, they replace it with their Assumption 3 (iii). However, for simplifying the proofs of our paper, we still require the everywhere-positive property. Furthermore, we will apply a convolutional technique to acquire an  everywhere-positive ``innovation distribution'' in the bootstrap world to satisfy this strong assumption.  
\end{Remark}
The condition which ensures the Drift criterion is satisfied for a 
 homoscedastic NLAR models can be drawn from \cite{an1996geometrical}. If we assume:
\begin{itemize}
    \item[A6] There exists a positive number $\lambda<1$ and a constant C such that the conditional mean function satisfies:
\begin{equation}
    |\phi(\bm{x})|\leq \lambda\max\{|x_1|,\ldots,|x_p|\} +C,
\end{equation}
\end{itemize}
then we can get the below lemma:
\begin{Lemma}[Theorem 3.2 of \cite{an1996geometrical}]
\label{Lemma3.3}
Let $\{X_t\}$ satisfy the homoscedastic version of the model \cref{Eq9}. If A4 and A6 are fulfilled, then this NLAR model is geometrically ergodic. 
\end{Lemma}
Similarly, we can obtain the condition for heteroscedastic NLAR models to be geometrically ergodic by imposing an additional assumption on the variance function:
\begin{itemize}
    \item[A7] The conditional variance function satisfies:
\begin{equation}
    \lim_{||\bm{x}||_2\overset{}{\to}\infty}\frac{\sigma(\bm{x})}{||\bm{x}||_2} = 0.
\end{equation}
\end{itemize}
Then the   Lemma holds:
\begin{Lemma}[Theorem 3.5 of \cite{min1999probabilistic}]
\label{Lemma3.4}
Let $\{X_t\}$ satisfy the model \cref{Eq9}. Suppose the conditional variance function satisfies A5 and A7. In addition, A6 holds true for the conditional mean function, and A4 holds true for the probability density function. Then, this heteroscedastic NLAR model is geometrically ergodic. 
\end{Lemma}

\subsection{Prediction inference for NLAR models with known form}
Through \cref{Lemma3.1} to \cref{Lemma3.4}, we have seen sufficient conditions to guarantee the geometric ergodicity of NLAR models. Before   considering the case when the NLAR model and innovation distribution $F_{\epsilon}$ are unknown, we first show how can we conclude prediction inference for NLAR models of known form. Although this case is usually unrealistic, it serves as a useful illustration before we investigate NLAR model cases. To simplify notation, we only consider the homoscedastic NLAR model in the main text, the analogous algorithms and theorems serve for NLAR models with heteroscedastic errors can be shown similarly; see more discussions in the Appendix. 

According to the discussion in \cref{Sec:2}, we can deploy the Monte Carlo simulation to approximate the exactly optimal point prediction or PI conditional on the past history. The procedure is summarized in \cref{algori3}.
\begin{algorithm}[H]
\caption{$h$-step ahead prediction of $X_{T+h}$ under homoscedastic \cref{Eq9} of known form}
\centering
\label{algori3}
  \begin{tabularx}{\textwidth}{lX}   
    Step 1 &  Note that the homoscedastic \cref{Eq9} can be written as
$X_{T+1} = \phi(X_{T },\ldots,X_{T +1-p})+\epsilon_{T+1}$; this can
be iterated to find an expression for $X_{T+h}$.  For example, $X_{T+2} = \phi(\phi(X_{T},\ldots,X_{T+1-p})+\epsilon_{T+1},X_{T},\ldots,X_{T+2-p})+\epsilon_{T+2}$. We can write $X_{T+h}$ as:
    \begin{equation}\label{Eq15}
        X_{T+h} = \mathcal{G}(X_{T},\ldots,X_{T-p+1};\epsilon_{T+1},\ldots,\epsilon_{T+h}),
    \end{equation}
    where we used the notation $\mathcal{G}(X_{T},\ldots,X_{T-p+1};\epsilon_{T+1},\ldots,\epsilon_{T+h})$ to specify that $X_{T+h}$
 depends on $X_{T},\ldots,X_{T-p+1}$ and   $\{\epsilon_i\}_{i=T+1}^{T+h}$.\\
    Step 2 & Simulate $\{\epsilon^{*}_{T+1},\ldots,\epsilon^{*}_{T+h}\}$ i.i.d. from $F_{\epsilon}$. \\
    Step 3 & Plug the $\{\epsilon^{*}_t\}_{t=T+1}^{T+h}$ from Step 2 and $\{X_{T-p+1},\ldots,X_{T} \}$ into \cref{Eq15} to obtain a pseudo-value of $X_{T+h}$ given by $\mathcal{G}(X_T,\ldots,X_{T-p+1};\epsilon^*_{T+1},\ldots,\epsilon^*_{T+h})$.\\
    Step 4 & Repeat Steps 2 and 3 $M$ times to get $M$ pseudo values $\{X_{T+h}^{(1)},\ldots,X_{T+h}^{(M)}\}$. The $L_2$ and $L_1$ optimal predictor can be approximated by $\frac{1}{M}\sum_{i = 1}^{M}X_{T+h}^{(i)}$ and $\text{Median}(X_{T+h}^{(1)},\ldots,X_{T+h}^{(M)})$, respectively. Furthermore, a QPI  
can be built by taking corresponding quantiles of the empirical distribution of 
$\{X_{T+h}^{(1)},\ldots,X_{T+h}^{(M)}\}$.  
  \end{tabularx}
\end{algorithm}
\begin{Remark}\label{Remark:3.2}
The procedure implied by \cref{algori3} can be easily extended to NLAR models with heteroscedastic errors. The only difference is that we compute $X_{T+k} = \phi(X_{T+k-1},\ldots,X_{T+k-p})+\sigma(X_{T+k-1},\ldots,X_{T+k-p})\epsilon_{T+k}$ iteratively for $k = 1,\ldots,h.$
\end{Remark}
We first show that the mean of pseudo values derived from \cref{algori3} can be consistent to the exactly $L_2$ optimal predictor. This is formalized in \cref{Theo3.1}.
\begin{Theorem}\label{Theo3.1}
Under assumptions A1-A6, the point predictor of homoscedastic \cref{Eq9} as $\widehat{X^{L_2}_{T+h}} = \frac{1}{M}\sum_{i = 1}^{M}X_{T+h}^{(i)}$ converges to the exactly $L_2$ optimal predictor almost sure as $M$ tends to infinity. Here, $X_{T+h}^{(i)} = \mathcal{G}(X_T,\ldots,X_{T-p+1};\epsilon^{(i)}_{T+1},$ $\ldots,\epsilon^{(i)}_{T+h})$; $\{\epsilon_{T+1}^{(i)}, \ldots,\epsilon_{T+h}^{(i)}\}$ are i.i.d. with common distribution $F_{\epsilon}$ for all $i=1,\ldots,M$.
\end{Theorem}
Next, inspired by the proof of \cite{guo1999multi}, we can show the median of pseudo values in \cref{algori3} is also consistent to the exactly $L_1$ optimal predictor. We can also build a PI that is asymptotically valid with any arbitrary CVR. To achieve this goal, we need additional one mild assumption:
\begin{itemize}
    \item[A8] The mean function $\phi(\cdot)$ is uniformly continuous.
\end{itemize}
This will lead to \cref{Theo3.2}:
\begin{Theorem}\label{Theo3.2}
Under assumptions A1-A6, if we take the point predictor as $\widehat{X^{L_1}_{T+h}} = \text{Median}$ $(\{X_{T+h}^{(1)},$$\ldots,X_{T+h}^{(M)}\})$, it is consistent to the exactly $L_1$ optimal predictor when $M$ converges to infinity. Here, $X_{T+h}^{(i)}$ and $\{\epsilon_{T+1}^{(i)}, \ldots,\epsilon_{T+h}^{(i)}\}$ have the same definitions with \cref{Theo3.1}. We can further show that the QPI is asymptotically valid with any arbitrary CVR.  
\end{Theorem}

\section{Prediction inference for parametric NLAR models of unknown form}\label{Sec:4}
In practice, it is not realistic to assume that we know the NLAR model \cref{Eq9} and the corresponding innovation distribution. Therefore, it is necessary to research if the aforementioned algorithms stand meaningful when the NLAR model is unknown. In this section, we will consider the case that the NLAR model has a known parametric specification but the parameter values are unknown. After estimating the model by Least Square (LS) technique, we show that the prediction inference can also be built with standard assumptions.
\subsection{Consistent point prediction and QPI}
In this subsection, we assume  
that the NLAR model \cref{Eq9} has the parametric specification
\begin{equation}\label{Eq16}
      X_t = \phi(\bm{X}_{t-1},\theta_1) + \sigma(\bm{X}_{t-1},\theta_2)\epsilon_t,
\end{equation}
where the functional form of $\phi( \cdot, \cdot)$
and $\sigma( \cdot, \cdot)$ is known but the real-valued parameters $\theta_1$ and $\theta_2$ are unknown. 

For carrying out prediction inference, we need to estimate $\theta_1$ and $\theta_2$ first; denote $\widehat{\theta}_1,\widehat{\theta}_2 $
the estimators. The consistency of parameter estimation is crucial for getting a consistent prediction. Thus, we suppose that we have a consistent parameter estimation and other mild conditions:
\begin{itemize}
    \item[A9] The parameter estimator $\widehat{\theta}_1$ and $\widehat{\theta}_2$ are consistent to $\theta_1$ and $\theta_2$ respectively. 
    \item[A10] For all $\bm{x}$ in the support $\mathcal{X}$ of $\bm{X}_t$, the non-linear functions $\phi(\bm{x},\cdot)$ and $\sigma(\bm{x},\cdot)$ are both Lipschitz continuous 
with respect to their 2nd argument.
    \item[A11] The probability density of innovation $f_{\epsilon}(x)$ satisfies $\sup_{x}f_{\epsilon}(x)<\infty$ and $\int|f_{\epsilon}(x) - f_{\epsilon}(x+c) |dx  = O(c)$ for finite $c$. 
\end{itemize}
Based on these assumptions, we will show how to get consistent point predictors and asymptotically valid QPI. The assumption A9 can be satisfied by applying the non-linear LS (NLS) estimation method. The reason for choosing this method and the procedure of estimation will be discussed in \cref{Sec:4.3.1}. First, we want to show the Cumulative Distribution Function (CDF) of innovations can be approximated by the empirical CDF of residuals. Their relationship can be summarized in \cref{Lemma4.1}.
\begin{Lemma}
\label{Lemma4.1}
Under A1--A7, A9--A11, the CDF of innovation $F_{\epsilon}(x)$ can be approximated by the empirical CDF of residuals $\widehat{F}_{\epsilon}(x)$ in a way:
\begin{equation}\label{Eq17}
    \sup_{x}|\widehat{F}_{\epsilon}(x) - F_{\epsilon}(x)|\overset{p}{\to} 0 ,
\end{equation}
where $\widehat{F}_{\epsilon}(x) := \frac{1}{T} \sum_{i=1}^{T}\mathbbm{1}_{\hat{\epsilon}_{i}\leq x}$; $\mathbbm{1}(\cdot)$ is the indicator function, and we define the residual $\hat{\epsilon}_i = (X_{i} - \phi(\bm{X}_{i-1},\widehat{\theta}_1))/\sigma(\bm{X}_{i-1},\widehat{\theta}_2)$ for $i = 1,\ldots,T$.
\end{Lemma}
With \cref{Lemma4.1}, we can build a QPI or find approximations of optimal $L_1$ and $L_2$ predictors. Of course, for this case, we need to apply the bootstrap technique. In short, being similar to the \cref{algori1}, we summarize the approach to obtain prediction inference for unknown parametric NLAR models \cref{Eq16} in \cref{algori4}. For simplifying notation, we focus on the case with homoscedastic errors.
\begin{algorithm}[H]
\caption{$h$-step  ahead prediction of $X_{T+h}$ under homoscedastic \cref{Eq16}}
\centering
\label{algori4}
  \begin{tabularx}{\textwidth}{lX}   
    Step 1 & Fit the homoscedastic \cref{Eq16} model based on
 $\{X_{-p+1},$ $\ldots,X_{T}\}$ to obtain parameter estimator $\widehat{\theta}_1$ that satisfies A9. Furthermore, compute and record all centered residuals $\hat{\epsilon}_t - T^{-1}\sum_{i=1}^{T}\hat{\epsilon}_i$ for $t = 1, \ldots,T$ and construct their empirical distribution
$\widehat{F}_{\epsilon}$.\\
    Step 2 & 
Similarly to  \cref{Eq15}, we see that $X_{T+h}$ depends
on $X_{T},\ldots,X_{T-p+1}$, on ${\epsilon}_{T+1} ,\ldots, {\epsilon}_{T+h}$,
and on ${\theta}_1$. We can denote this as
$$  X_{T+h} = \mathcal{G}(X_{T},\ldots,X_{T-p+1}; {\epsilon}_{T+1} ,\ldots, {\epsilon}_{T+h} ; {\theta}_1).$$
Hence, we can generate a pseudo-value of $X_{T+h}^*$ as:
    \begin{equation}\label{Eq18}
        X^*_{T+h} = \mathcal{G}(X_{T},\ldots,X_{T-p+1};\hat{\epsilon}_{T+1}^*,\ldots,\hat{\epsilon}_{T+h}^*;\hat{\theta}_1),
    \end{equation}
where $\{\hat{\epsilon}^{*}_{j}\}_{j=T+1}^{T+h}$ are drawn i.i.d. from $\widehat{F}_{\epsilon}$.

\\
    Step 3 & Repeat Step 2 $M$ times to get $\{X_{T+h}^{(1)},\ldots,X_{T+h}^{(M)}\}$. Then, $L_2$ and $L_1$ optimal predictors can be approximated by $\frac{1}{M}\sum_{i = 1}^{M}X_{T+h}^{(i)}$ and $\text{Median of }\{X_{T+h}^{(1)},\ldots,X_{T+h}^{(M)}\}$, respectively. Furthermore,  a QPI 
can be built by taking corresponding quantiles.  
  \end{tabularx}
\end{algorithm}
The analogous algorithm for NLAR model \cref{Eq16} with heteroscedastic errors is easy to be built. The asymptotic validity of QPI and consistency of optimal $L_1$ or $L_2$ point prediction are guaranteed by \cref{Theo4.1} under the additional assumption of mean and volatility functions given below:
\begin{itemize}
    \item[A12] For all parameter values in their respective domains, the non-linear functions $\phi(\cdot,\cdot)$ and $h(\cdot,\cdot)$ are continuous w.r.t their first argument.
\end{itemize}
\begin{Theorem}
\label{Theo4.1}
Under A1-A7, A9–A12, let $\{X_t\}$ satisfy  \cref{Eq16}. For $h\geq 1$ we have:
\begin{equation}
    \sup_{|x|\leq c_T}\left|F_{X^*_{T+h}|X_{T},\ldots, X_{-p+1}}(x) - F_{X_{T+h}|X_{T},\ldots, X_{T-p+1}}(x)\right|\overset{p}{\to} 0,
\end{equation}
where $X^*_{T+h} = \mathcal{G}(X_T,\ldots,X_{T-p+1};\hat{\epsilon}^*_{T+1},\ldots,\hat{\epsilon}^*_{T+h};\widehat{\theta})$;
this is computed as $X^*_{T+k} = \phi(\bm{X}^*_{T+k-1},\widehat{\theta}_1)+\sigma(\bm{X}_{T+k-1}^*,\widehat{\theta}_2)\hat{\epsilon}^*_{T+k}$ iteratively for $k = 1,\ldots,h$; $\{\hat{\epsilon}^*_i\}_{i=T+1}^{T+h} \text{are}~i.i.d. \sim \widehat{F}_{\epsilon}$; $c_T$ is an appropriate sequence converges to infinity as $T$ converges to infinity; $F_{X^*_{T+h}|X_{T},\ldots, X_{-p+1}}(x)$ is the distribution of h-step ahead future value in the bootstrap world, i.e., conditional on all observed data.
\end{Theorem}

As we discussed in the \cref{Sec:1}, instead of adopting the fitted (traditional) residuals in \cref{algori4}, we can apply the predictive residuals to compute QPI. To acquire such predictive residuals which are denoted $\hat{\epsilon}_{t}^{p}$ hereafter, we need to estimate models based on delete-$X_t$, i.e., the available data for the scatter plot of $X_i$ vs. $\{X_{i-p},\ldots, X_{i-1}\}$ excludes the single point at $i = t$. Evaluate and collect the estimation residual at this point and repeat it for $t=1,\ldots,T$, we obtain all predictive residuals $\{\hat{\epsilon}_{t}^{p}\}_{t=1}^{T}$. The corresponding prediction procedure with predictive residuals is written in \cref{algori5}.

\begin{algorithm}[H]
\caption{$h$-step ahead prediction of $X_{T+h}$ for unknown homoscedastic \cref{Eq16} with predictive residuals}
\centering
\label{algori5}
  \begin{tabularx}{\textwidth}{lX}   
    Step 1 & Fit the homoscedastic \cref{Eq16} model based on $\{X_{-p+1},$ $\ldots,X_{T}\}$ to get parameter estimation $\widehat{\theta}_1$ which satisfies A9. Furthermore, for $t = 1,\ldots,T$, re-estimate the model based on delete-$X_t$ dataset to get estimation $\widehat{\theta}^{p}_{1}$. Compute all predictive residual $\hat{\epsilon}_t^{p} = X_t - \phi(\bm{X}_{t-1},\widehat{\theta}_{1}^{p})$. Center all predictive residuals by manipulation $\hat{\epsilon}_t^{p} - T^{-1}\sum_{i = 1}^{T}\hat{\epsilon}_i^{p}$ for $t = 1, \ldots,T$ and record the result as the empirical predictive residual distribution $\widehat{F}_{\epsilon}^{p}$.\\
    Steps 2-3 & Change the residual distribution $\widehat{F}_{\epsilon}$ to $\widehat{F}_{\epsilon}^{p}$, the rest is the same as \cref{algori4}.
  \end{tabularx}
\end{algorithm}
\begin{Remark}\label{Remark:4.1}
When $T$ tends 
to infinity, the effects of leaving out
one data pair $X_t \text{vs}~\{X_{t-1},\ldots,X_{t-p}\}$ is negligible.
Hence, for large $T$, the predictive residual $\hat{\epsilon}_t^{p}$ is approximately the same as the fitted residual $\hat{\epsilon}_{t}$. Therefore, \cref{Lemma4.1} and \cref{Theo4.1} are still true with predictive residuals. Similarly, for NLAR models with heteroscedastic errors, we can define the predictive residual $\hat{\epsilon}_t^{p} = \frac{X_t - \phi(\bm{X}_{t-1},\widehat{\theta}_{1}^{p})}{\sigma(\bm{X}_{t-1},\widehat{\theta}_{2}^{p})}$, where $\widehat{\theta}_{2}^{p}$ is the parameter estimation of the variance function with the delete-$X_t$ dataset. In \cref{Sec:5}, we will show the positive effects brought by applying predictive residuals on the empirical CVR of QPI.
\end{Remark}

\FloatBarrier
 
\subsection{Pertinent PIs}\label{Sec:4.2}
As shown in \cref{Theo4.1}, it is straightforward to build a QPI for $X_{T+h}$. However, 
 this type of prediction interval can not capture the variability arising from the model estimation. Thus, we propose a new bootstrap procedure based on the general \cref{algori2} and summarize it in \cref{algori6}.
\begin{algorithm}[htbp]
\caption{$h$-steps ahead PPI of $X_{T+h}$ for unknown homoscedastic \cref{Eq16} with fitted residuals}
\centering
\label{algori6}
  \begin{tabularx}{\textwidth}{lX}   
    Step 1 & Fit the homoscedastic \cref{Eq16} model based on $\{X_{-p+1},$ $\ldots,X_{T}\}$ to get parameter estimation $\widehat{\theta}_1$ which satisfies A9. Furthermore, compute and record $\hat{\epsilon}_t - T^{-1}\sum_{i=1}^{T}\hat{\epsilon}_i$ for $t = 1, \ldots,T$ to get $\widehat{F}_{\epsilon}$.\\
    Step 2 & Find the prediction $\widehat{X}_{T+h}$ based on \cref{algori4}.\\
    Step 3 & (a) Resample (with replacement) the residuals from $\widehat{F}_{\epsilon}$ to create pseudo-errors $\{\hat{\epsilon}^{*}_{t}\}_{t=p+1}^{T}$ and $\{\hat{\epsilon}^{*}_{t}\}_{t=T+1}^{T+h}$.\\
    &(b) Let $(X_{-p+1}^*,\ldots, X_{0}^*)'=({X}_{0+I},\cdots,{X}_{p+I-1})'$ where $I$ is generated as a discrete random variable uniformly on the values $-p+1,\ldots, T-p+1$. Then, use the fitted homoscedastic NLAR model of Step 1 and the generated $\{\hat{\epsilon}^{*}_{t}\}_{t=p+1}^{T}$ in Step 3 (a) to create 
 bootstrap pseudo-data $\{X_t^*\}_{t = 1}^{T}$ in a recursive manner, i.e., compute $X_{k}^* = \phi(\bm{X}_{k-1}^*,\widehat{\theta}_1) + \hat{\epsilon}_{1}^*$ for $k=1,\ldots,T$. \\
 & (c) Based on the bootstrap data $\{X^{*}_{t}\}_{t=-p+1}^{T}$, re-estimate the homoscedastic NLAR model to get $\widehat{\theta}_1^{*}$.\\
 & (d) Guided by the idea of forward bootstrap, re-define the last $p$ values of the  bootstrap data to match the original, i.e., re-define $X_t^*=X_t$ for $t=T-p+1, \dots, T$.\\
& (e) With parameter estimation $\widehat{\theta}_1$, the bootstrap data $\{X_t^*\}_{t = -p+1}^{T}$, and the pseudo-errors $\{\hat{\epsilon}_{t}^*\}_{t = T+1}^{T+h}$
  to generate recursively the future bootstrap data $X_{T+1}^*,\ldots,X_{T+h}^{*}$.\\
  & (f) With bootstrap data $\{X^*_{t}\}_{t=-p+1}^{T}$ and parameter estimation $\widehat{\theta}^{*}_1$, utilize \cref{algori4} to compute the bootstrap predictor which is denoted by $\widehat{X}^*_{T+h}$. For generating innovations, we still use $\widehat{F}_{\epsilon}$.\\
& (g) Determine the bootstrap root: $X_{T+h}^*- \widehat{X}^*_{T+h}$. \\ 
Step 4 & Repeat Step 3 $K$ times; 
the $K$ bootstrap root replicates are collected in the form of an empirical distribution whose $\beta$-quantile is denoted $q(\beta)$.
The $(1-\alpha)100\%$ equal-tailed prediction interval for ${X}_{T+h}$ centered at $\widehat{X}_{T+h}$ is then estimated by
 $[\widehat{X}_{T+h}+q(\alpha/2), \widehat{X} _{T+h}+q(1-\alpha/2)].$
  \end{tabularx}
\end{algorithm}
Analogously to the development of QPI with predictive residuals, we also propose the variant of \cref{algori6} with predictive residuals to improve the CVR in finite sample cases; this is given in \cref{algori7}. \cref{algori6} and \cref{algori7}
 focus on a homoscedastic NLAR satisfying \cref{Eq16};  
  corresponding algorithms for an NLAR  with heteroscedastic error can be built similarly. 

\begin{algorithm}[htbp]
\caption{\small{$h$-steps ahead PPI of $X_{T+h}$ for unknown homoscedastic \cref{Eq16} with predictive residuals}}
\centering
\label{algori7}
  \begin{tabularx}{\textwidth}{lX}   
    Step 1 & Fit the homoscedastic \cref{Eq16} model to get parameter estimation $\widehat{\theta}_{1}$ which satisfies A9. Furthermore, for $t = 1,\ldots,T$, re-estimate the model based on delete-$X_t$ dataset to compute the estimate $\widehat{\theta}^{p}_{1}$. Compute the predictive residual $\hat{\epsilon}_t^{p} = X_t - \phi(\bm{X}_{t-1},\widehat{\theta}_{1}^{p})$. Then, take $\hat{\epsilon}^p_t - T^{-1}\sum_{i=1}^{T}\hat{\epsilon}^p_i$ for $t = 1, \ldots,T$ as the empirical predictive residual distribution $\widehat{F}_{\epsilon}^{p}$.\\
    Step 2 & Find the prediction $\widehat{X}_{T+h}$ based on \cref{algori5}.\\
    Steps 3-4 & Chang the residual distribution $\widehat{F}_{\epsilon}$ to $\widehat{F}^{p}_{\epsilon}$ and change the application of \cref{algori4} to \cref{algori5}; the rest is the same as \cref{algori6}. \\
  \end{tabularx}
\end{algorithm}
\FloatBarrier
\begin{Remark}
    To build the PPI for \cref{Eq16} with heteroscedastic error, we need to make predictions and generate bootstrap series including variance function; see \cref{Remark:3.2}. In addition, since we assume the innovation $\epsilon_1$ has variance 1 if $\sigma(\cdot)\not\equiv 1$, we also need to normalize the variance of fitted/predictive residuals to 1 in \cref{algori6} and \cref{algori7}, i.e, take $\frac{\hat{\epsilon}_t - T^{-1}\sum_{i=1}^{T}\hat{\epsilon}_i}{\sqrt{\sum_{t=1}^T(\hat{\epsilon}_t - T^{-1}\sum_{i=1}^{T}\hat{\epsilon}_i)^2/T}}$ to get $\widehat{F}_{\epsilon}$ and take $\frac{\hat{\epsilon}^p_t - T^{-1}\sum_{i=1}^{T}\hat{\epsilon}^p_i}{\sqrt{\sum_{t=1}^T(\hat{\epsilon}^p_t - T^{-1}\sum_{i=1}^{T}\hat{\epsilon}^p_i)^2/T}}$  to get $\widehat{F}^p_{\epsilon}$ for $t = 1, \ldots,T$. This additional manipulation for \cref{Eq16} with heteroscedastic error simplifies the theoretical proof. Moreover, from a practical issue, the length of PPI will decrease with this additional step. 
\end{Remark}
\FloatBarrier

\subsection{Asymptotic validity of the PPIs}\label{Sec:4.3}

The idea that underlies \cref{algori6} and \cref{algori7} is approximating the distribution of predictive root $X_{T+h} - \widehat{X}_{T+h}$ by its bootstrap version ${X}_{T+h}^{*} - \widehat{X}_{T+h}^{*}$. From a theoretical view, as we have clarified in \cref{Remark:4.1}, the difference between applying fitted residuals and predictive residuals lies in the available dataset. The difference in parameter estimations based on fitted residuals and predictive residuals  asymptotically converges to zero in probability
under standard assumptions. In other words, the difference between fitted residuals and predictive residuals also converges to zero in probability. Thus, in what follows, we just analyze the
asymptotic performance of a PPI with fitted residuals since it is
the same as  the asymptotic performance of a PPI with predictive residuals. However, the latter invariably has a larger finite-sample CVR.
 
For $\widehat{X}_{T+h}^{*}$, if we insist on doing the prediction in the bootstrap world parallelly with the procedure in the real world, we need to use the residual distribution $\Hat{\Hat{F}}_{\epsilon}$, which is obtained from re-estimating the model in the bootstrap world. However, in the bootstrap world, the underlying ``true innovation'' distribution is the $\widehat{F}_{\epsilon}$ and we can expect that $\Hat{\Hat{F}}_{\epsilon}$ converges to $\widehat{F}_{\epsilon}$, so we apply $\widehat{F}_{\epsilon}$ to compute $\widehat{X}_{T+h}^{*}$ directly. All in all, we want to compare two predictive roots $X_{T+h} -  \widehat{X}_{T+h}$ and $X_{T+h}^{*} - \widehat{X}_{T+h}^{*}$. If we can show
\begin{equation}\label{Eq20}
    \sup_{|x|\leq c_T}\left|\mathbb{P}\left(X_{T+h}^{*} -\widehat{X}_{T+h}^{*} \leq x|X_T,\ldots,X_{-p+1}\right) - \mathbb{P}\left(X_{T+h} -  \widehat{X}_{T+h}\leq x|X_T,\ldots,X_{T-p+1}\right)  \right| \overset{p}{\to} 0,
\end{equation}
then we can utilize the distribution of $X_{T+h}^{*} -\widehat{X}_{T+h}^{*}$ to consistently estimate the distribution of $X_{T+h} -  \widehat{X}_{T+h}$;
as a result, the PPI has asymptotic validity within $c_T$. 

  \cref{Eq20} can be shown based on \cref{Theo4.1} with one additional condition,  namely that  $\widehat{\theta}^{*} = (\widehat{\theta}^{*}_1, \widehat{\theta}^{*}_2)$ is consistent to $\widehat{\theta} =(\widehat{\theta}_1,\widehat{\theta}_2)$. Before going into detail about this property, we first discuss the conditions under which $\widehat{\theta}$ is consistent to $\theta = (\theta_1,\theta_2)$.

\FloatBarrier
\subsection{The consistency of $\widehat{\theta}$ to $\theta$}\label{Sec:4.3.1}
In this paper, we adopt the NLS technique to 
perform parameter estimation; the reason is that NLS is based entirely
on the scatter plot of $X_t$ vs.~$(X_{t-1},\ldots,  X_{t-p})$ so that
predictive residuals can be easily defined.
 First, we consider a homoscedastic version of \cref{Eq16}. The heteroscedastic version will be handled by a two-step estimation approach later. To simplify notation in the proofs, we consider an NLAR model with order one, and
we attempt to minimize a quadratic empirical loss function as given below:
\begin{equation}\label{Eq21}
    \widehat{\theta}_1 = \arg\min_{\vartheta\in\Theta_1}L_{T}(\vartheta) 
\ \ \mbox{where} \ \ 
L_{T}(\vartheta) =\frac{1}{T}\sum_{t=1}^{T}(X_{t} - \phi(X_{t-1},\vartheta))^2,
\end{equation}
where $\Theta_1$ is the domain of possible values of $\theta_1$. With correctly specified model, the true parameter $\theta_1$ satisfies that:
\begin{equation}
    \theta_1 = \arg\min_{\vartheta\in\Theta_1}L(\vartheta) 
\ \ \mbox{where} \ \ 
L(\vartheta) =\mathbb{E}(X_{t} - \phi(X_{t-1},\vartheta))^2.
\end{equation}
The consistency of the non-linear least squares estimator $\widehat{\theta}_1$ to $\theta_1$ can be guaranteed with below additional assumptions:
\begin{itemize}
    \item[A13] $ \Theta_1$ is bounded, closed and with finite dimension. 
    \item[A14] $\theta_1$ uniquely minimizes $L(\vartheta)$ over $\vartheta\in\Theta_1$.
\end{itemize}
If we can not correctly specify the model, we call $\theta_1$ the optimal parameter in the sense of minimizing $L(\vartheta)$. We can still build the consistency relationship between $\widehat{\theta}_1$ and $\theta_1$ if we assume A14. As we have clarified at the beginning of \cref{Sec:1}, we focus on the case where we can correctly specify the model. The model misspecification case can be analyzed similarly.
\begin{Lemma}\label{Lemma4.2}Under assumptions A1-A7 and A10, A13-A14, if $\{X_t\}$  satisfies a homoscedastic \cref{Eq16}, the non-linear least squares estimation $\widehat{\theta}_1$ converges to the true parameter $\theta_1$ in probability, i.e., for any $\epsilon>0$, 
\begin{equation}
    \mathbb{P}(|\widehat{\theta}_1-\theta_1 |>\epsilon) \overset{}{\to} 0.
\end{equation}
\end{Lemma}

To handle the heteroscedastic \cref{Eq16}, we still estimate $\theta_1$   by minimizing the empirical risk \cref{Eq21}. The corresponding true risk 
with respect to $\theta_1$ is:
\begin{equation}\label{Eq24}
\begin{split}
    \mathbb{E}(X_{t} - \phi(X_{t-1},\vartheta))^2 &= \mathbb{E}(\phi(X_{t-1},\theta_1) + \sigma(X_{t-1},\theta_2)\epsilon_{t} - \phi(X_{t-1},\vartheta))^2 \\
    & = \mathbb{E}(\phi(X_{t-1},\theta_1) - \phi(X_{t-1},\vartheta))^2 + \mathbb{E}(\sigma(X_{t-1},\theta_2)^2),
\end{split}
\end{equation}
which implies that the minimizer of risk \cref{Eq24} is equal to the true $\theta_1$. Thus, the minimizer of empirical risk will still converge to the true parameter in probability. After securing this consistent estimation
of $\theta_1$, we proceed  to estimate the $\theta_2$ by
minimizing the below empirical risk:
\begin{equation}
     \widehat{\theta}_2 = \arg\min_{\vartheta\in\Theta_2}K_{T}(\vartheta,\widehat{\theta}_1) =  \arg\min_{\vartheta\in\Theta_2}\left|\frac{1}{T}\sum_{t=1}^{T}\left(\frac{X_{t} - \phi(X_{t-1},\widehat{\theta}_1)}{h(X_{t-1},\vartheta)}\right)^2 - 1 \right|.
\end{equation}
The corresponding true risk should be:
\begin{equation}
\begin{split}
         \vartheta_2 = \arg\min_{\vartheta\in\Theta_2}K(\vartheta,\theta_1) &=  \arg\min_{\vartheta\in\Theta_2}\left|\mathbb{E}\left(\frac{X_{t} - \phi(X_{t-1},\theta_1)}{h(X_{t-1},\vartheta)}\right)^2 - 1 \right|\\
         & = \arg\min_{\vartheta\in\Theta_2}\left|\mathbb{E}\left(\frac{\phi(X_{t-1},\theta_1) + h(X_{t-1},\theta_2)\epsilon_t - \phi(X_{t-1},\theta_1)}{h(X_{t-1},\vartheta)}\right)^2 - 1 \right|\\
         & = \arg\min_{\vartheta\in\Theta_2}\left|\mathbb{E}\left(\frac{ h(X_{t-1},\theta_2)}{h(X_{t-1},\vartheta)}\right)^2 - 1 \right|,
\end{split}
\end{equation}
which implies $\vartheta_2 = \theta_2$. Under the additional  assumptions:
\begin{itemize}
    \item[A15] $ \Theta_2$ is bounded, closed and with finite dimension. 
    \item[A16] $\theta_2$ uniquely minimizes $K(\vartheta,\theta_1)$ over $\vartheta\in\Theta_2$,
\end{itemize}
we can derive the lemma below to ensure the consistency of $\widehat{\theta}_2$ to $\theta_2$:
\begin{Lemma}\label{Lemma4.3}
Under assumptions A1-A7 and A10, A13-A16, if $\{X_t\}$ 
satisfies a heteroscedastic \cref{Eq16}, the NLS estimators $\widehat{\theta}_1$ and $\widehat{\theta}_2$ converge  respectively
to the true parameters $\theta_1$ and $\theta_2$ in probability.
\end{Lemma}

In fact, under the consistency of parameter estimations, we can impose more conditions on the mean and variance functions to develop estimation inference of $\widehat{\theta}_1$ and $\widehat{\theta}_2$, i.e., building confidence interval for $\widehat{\theta}_1$ and $\widehat{\theta}_2$, respectively. If we can further approximate the confidence interval by bootstrap, the PPIs derived from \cref{algori6,algori7} are indeed pertinent. We first discuss the consistency property of $\widehat{\theta}^*$; then the discussions of estimation inference on $\widehat{\theta}^*$ and $\widehat{\theta}$ are put in \cref{Subsec:estimationinfe}.

\subsection{The consistency of $\widehat{\theta}^{*}$ to $\widehat{\theta}$
in the bootstrap world}\label{Subsec:consistencybootrealest}
From the last subsection, we have seen the non-linear least squares can return a satisfied estimation but this is still not enough for us to derive the
asymptotic validity of the PPI. As we discussed in \cref{Sec:4.3}, the necessary component is the consistency between $(\widehat{\theta}^{*}_1,\widehat{\theta}^{*}_2)$ and $(\widehat{\theta}_1,\widehat{\theta}_2)$. We first investigate the relationship between $\widehat{\theta}^{*}_1$ and $\theta_1$. Once this relationship is determined, the consistency between $\widehat{\theta}_1^{*}$ and $\widehat{\theta}_1$ is trivial. In the work of \cite{franke2000bootstrapping}, a similar problem is considered for the regression case. In short, this consistency can be proved by showing that analogous $L_{T}^{*}(\vartheta)$ converges uniformly to $L(\vartheta)$. In our case, $L_{T}^{*}(\vartheta)$ has the form as below:
\begin{equation}
    L_{T}^{*}(\vartheta) = \frac{1}{T}\sum_{t=1}^{T}(X^{*}_{t} - \phi(X^{*}_{t-1},\vartheta))^2.
\end{equation}
Here, $\{X_t^*\}$ is the bootstrap series, which mimics the property of the true series. It can be created based on \cref{algori6} or \cref{algori7} with fitted or predictive residuals, respectively. And $\widehat{\theta}_1^{*}$ satisfies that:
\begin{equation}\label{eq54}
    \widehat{\theta}_1^{*} = \arg\min_{\vartheta\in\Theta_1}L^{*}_{T}(\vartheta) =  \arg\min_{\vartheta\in\Theta_1}\frac{1}{T}\sum_{t=1}^{T}(X^{*}_{t} - \phi(X^{*}_{t-1},\vartheta))^2.
\end{equation}
As discussed in \cref{Sec:4.3.1}, it is necessary that we have the additional condition: the bootstrap series is also geometrically ergodic. Then, with close enough empirical residual distribution and true innovation distribution, we may show the uniform convergence of $L_{T}^{*}(\vartheta)$ and $L(\vartheta)$. Then, the consistency of $\widehat{\theta}_1^{*}$ to $\theta_1$ is easily found.

The first problem we face is that the probability density of the fitted residual $\hat{\epsilon}$ is not continuous and positive everywhere which means the basic assumption A4 needed to show the ergodicity of the bootstrap series is not met. In a similar situation,  \cite{franke2002properties} apply the kernel technique to build a probability density of $\hat{\epsilon}$. Here, we take a convolution approach to make the density function of empirical residual continuous and positive everywhere, i.e., we define another random variable $\widetilde{\epsilon}_i$ which is the sum of empirical fitted residual $\hat{\epsilon}_i$ and an independent normal random variable with mean 0 and suitable variance $\xi(T)$, i.e., let:
\begin{equation}\label{Eq29}
    \widetilde{\epsilon}_i= \hat{\epsilon}_i + z_i,~\text{for}~i=1,\ldots,T,
\end{equation}
where $z_i\sim N(0,\xi(T))$, where $\xi(T)$ converges to 0 as $T\overset{}{\to} \infty$ with a suitable rate. Then, we create bootstrap residuals
by drawing i.i.d. from $\widetilde{F}_{\epsilon}$, the CDF of $\{\widetilde{\epsilon}_i\}$, in order to build a bootstrap series $\{\widetilde{X}^{*}_{t}\}$ as we did in \cref{algori6}. Subsequently, we re-estimate the parameter of NLAR based on the bootstrap series $\{\widetilde{X}^{*}_{t}\}$. Since the convergence in mean squares implies the convergence in probability, we can easily see that \cref{Lemma4.1} still stands true for $\widetilde{F}_{\epsilon}$. Thus, if we replace $\widehat{F}_\epsilon$ by $\widetilde{F}_{\epsilon}$ in \cref{algori4} and \cref{algori6}, all claims will still be true. 

\begin{Remark}\label{Remarkdensity}
Here, we take a convolution approach to create residuals that possess a continuous probability density function. We should notice that this approach is equivalent to the kernel density estimator taken by \cite{franke2002bootstrap,franke2002properties} with a Gaussian kernel. More specifically, the variance $\xi(T)$ plays the same role as the bandwidth $h$ in the Gaussian kernel. Thus, if we take $\xi(T) = O(T^{-\delta})$ for some constant $\delta>0$, we can show that the probability density of $\widetilde{\epsilon}$, $\widetilde{f}_{\epsilon}$ converges uniformly to $f_{\epsilon}$,i.e., $||\widetilde{f}_{\epsilon} - f_{\epsilon} ||_{\infty} = op(1)$, see the proof of Lemma 4 from \cite{franke2002properties} for a reference. In addition, this convolution technique can also be applied to predictive residuals. As we have discussed in \cref{Sec:4.3}, the predictive residual is equivalent to the fitted residual asymptotically. 
\end{Remark}

Although we will use $\widetilde{F}_{\epsilon}$ and the corresponding density function $\widetilde{f}_{\epsilon}$ to develop subsequent theorems, in practice we still apply $\widehat{F}_{\epsilon}$ since effects stemming from
$z_i$ are negligible when we sample innovations from the empirical residual distribution. For simplifying notations, we keep using $\widehat{F}_{\epsilon}$ and $\widehat{f}_{\epsilon}$, though their representation may change according to the context. Similarly with the deduction of \cref{Lemma4.2}, we can get:
\begin{equation}
     \mathbb{P}(|\widehat{\theta}_1^{*}-\theta_1 |>\epsilon) \leq \mathbb{P}(2\sup_{\vartheta\in\Theta_1}|L^{*}_{T}(\vartheta) - L(\vartheta)|>C ),
\end{equation}
for some constant $C>0$. Focusing on analyzing $\sup_{\vartheta\in\Theta_1}|L^{*}_{T}(\vartheta) - L(\vartheta)|$, we can partition the parameter space into different balls, i.e., make a $\varepsilon$-covering of $\Theta_1$. Let the $\varepsilon$-covering number of $\Theta$ be $C_{N} = N(\varepsilon;\Theta_1;||\cdot||)$ which means for every $\vartheta\in\Theta_1$, $\exists~i\in\{1,2,\ldots, C_{N}\}$ s.t. $||\vartheta - \theta^{i} ||\leq \varepsilon$ for $\forall \varepsilon>0$. Define $\Xi_{\theta} \in \{\theta^{1},\ldots,\theta^{C_{N}}\}$, we can consider:
\begin{equation}\label{Eq31}
\begin{split}
    &\sup_{\vartheta\in\Theta_1}|L^{*}_{T}(\vartheta) - L(\vartheta)|\\
    & = \sup_{\vartheta\in\Theta_1}|L^{*}_{T}(\vartheta) - L(\vartheta) + L^{*}_{T}(\Xi_{\theta}) -  L^{*}_{T}(\Xi_{\theta})  + L(\Xi_{\theta}) - L(\Xi_{\theta})|\\
    &\leq \max_{\Xi_{\theta}\in\{\vartheta^{1},\ldots,\vartheta^{C_{N}}\}}| L^{*}_{T}(\Xi_{\theta})  - L(\Xi_{\theta})| + \sup_{\vartheta\in\Theta_1}| L^{*}_{T}(\vartheta)  - L^{*}_{T}(\Xi_{\theta}) | + \sup_{\vartheta\in\Theta_1}|L(\vartheta) -  L(\Xi_{\theta})|.
\end{split}
\end{equation}
Consider the second term of the r.h.s. of the above inequality. From Lipschitz continuous assumption of $\phi(\cdot,\cdot)$ 
with respect to $\vartheta$, we can get:
\begin{equation}
    \sup_{\vartheta\in\Theta_1}| L^{*}_{T}(\vartheta)  - L^{*}_{T}(\Xi_{\theta}) | \leq \sup_{\vartheta\in\Theta_1}L||\vartheta - \Xi_{\theta} ||\leq L\cdot\varepsilon \overset{}{\to} 0.
\end{equation}
Similarly, we can find the $\sup_{\vartheta\in\Theta}|L(\vartheta) -  L(\Xi_{\theta})|\overset{}{\to} 0$. For the first term of the r.h.s of \cref{Eq31}, if we can show the bootstrap series is also ergodic when the parameter is fixed, then we actually have $L_T^*(\Xi_\vartheta) \overset{p}{\to} \mathbb{E}^{*}[X_1^* - \phi(X_0^*,\Xi_\vartheta)]$, such a similar result is also implied by Theorem 5 of \cite{franke2002properties}, here $\mathbb{E}^{*}(\cdot)$ stands for the conditional expectation in the bootstrap world. Therefore, for getting the uniform convergence of $L^{*}_{T}(\vartheta)$ to $L(\vartheta)$ in probability, it is enough to show:
\begin{equation}\label{Eq33}
    \mathbb{E}^{*}[X_{1}^{*} - \phi(\bm{X}_{0}^{*},\Xi_{\theta})]^2 \overset{p}{\to} \mathbb{E}[X_{1} - \phi(\bm{X}_{0},\Xi_{\theta})]^2,~\text{for each}~\Xi_{\theta}.
\end{equation}

For notational simplicity, we consider an NLAR(1) model; then, the l.h.s. of \cref{Eq33} equals:
\begin{equation}\label{Eq34}
    \int_{\mathbb{R}^{2}}(x_{1} - \phi(x_{0},\Xi_{\theta}))^2\widehat{f}_{\epsilon}(x_1 - \phi(x_0,\widehat{\theta}_1))\pi^{*}(x_0)dx_1dx_0,
\end{equation}
where $\pi^{*}(\cdot)$ stands for the marginal stationary density function of the bootstrap series. As we can see, the uniform convergence of \cref{Eq31} in probability depends on the ergodic property of the bootstrap series and the closeness of $\pi^{*}(\cdot)$ and $\pi(\cdot)$ which is the true stationary density function of the real series. In other words, the ergodic property of bootstrap series is not enough to get our desired result. Actually, such ergodic property alone is of little use since the main goal of performing bootstrap is to mimic features of the true process, i.e., we require the stationary distribution of the bootstrap series and real series should be close enough to show \cref{Eq33}. Here, we develop a theorem to illustrate the required conditions. 
\begin{Theorem}\label{Theom: meanbootreal}
Suppose that the data generating process obeys \cref{Eq16} and the bootstrap time series $\{X_t^*\}_{t=-p+1}^{T}$ are generated according to \cref{algori6} or \cref{algori7}. Under A1-A7, A9-A12, 
we have:
\begin{equation}\label{Eq35}
    \mathbb{E}^{*}[X_{1}^{*} - \phi(X_{0}^{*},\Xi_{\theta})]^2 \overset{p}{\to} \mathbb{E}[X_{1} - \phi(X_{0},\Xi_{\theta})]^2.
\end{equation}
\end{Theorem}
Then, we can show that  \cref{Eq31} converges to zero in probability, and that $\widehat{\theta}_1^{*}$ converges to $\widehat{\theta}_1$ in probability. In addition, applying the two-stage estimation strategy taken in \cref{Sec:4.3.1}, the consistency between $\widehat{\theta}_2^{*}$ and $\widehat{\theta}_2$ can also be shown. Thus, $\widehat{\theta}^{*}$ is consistent to $\widehat{\theta}$, meeting the requirement of proving \cref{Eq20}. For showing the asymptotic validity of PPI with predictive residuals, we can apply the same strategy to make predictive residuals have a smooth probability density function as we did in \cref{Eq29}. Then, the consistency between the parameter estimation in the bootstrap world and the true parameter can also be shown as proof of applying fitted residuals. Although the PPI with fitted residuals and predictive residuals are the same asymptotically, the performance of PPI with predictive residuals is better in finite sample cases. Examples can be found in the next section. 

All in all, the prediction intervals defined in \cref{algori6,algori7} are asymptotically valid, since we can get consistent parameter estimations in the bootstrap world. \cref{Coro:4.1} below summarizes this result:
\begin{Corollary}\label{Coro:4.1}
Under assumptions A1-A7, A9-A12, $\widehat{\theta}^{*}$ is consistent to $\widehat{\theta}$ with both fitted and predictive residuals, which substantiates \cref{Eq20}. Thus, the conditional distribution of $X_{T+h} -  \widehat{X}_{T+h}$ can be asymptotically approximated by the conditional distribution of $X^{*}_{T+h} -  \widehat{X}^{*}_{T+h}$ which guarantees the validity of PI derived by \cref{algori6} or \cref{algori7} with both homoscedastic or heteroscedastic errors asymptotically. 
\end{Corollary}

\subsection{The estimation inference of $\widehat{\theta}$ and $\widehat{\theta}^*$}\label{Subsec:estimationinfe}
From the above sections, the asymptotic validity of PPI is checked. However, with a more complicated prediction procedure in \cref{algori6,algori7}, we expect to get a stronger PI, i.e, pertinent PI. The crucial part behind the pertinence is that we can approximate the distribution of $\widehat{\theta}$ by the distribution of $\widehat{\theta}^*$. In other words, the estimation variability can be captured by the bootstrap-based PI. To derive the estimation inference, we need stronger assumptions than A10 on the mean and variance function. We assume:
\begin{itemize}
    \item[A17] For all $\bm{x}$ in the support $\mathcal{X}$ of $\bm{X}_t$, $\phi(\bm{x},\cdot)$ and $\sigma(\bm{x},\cdot)$ are twice differentiable w.r.t. parameters in some neighborhood of true parameters. 
    
    \item[A18] If we write $L_{T}(\vartheta)$ as $\frac{1}{T}\sum_{t=1}^{T}q_{t}(\vartheta)$, we need $\mathbb{E}\nabla^2$ $q_{1}(\theta_1)$ is non-singular; If we write $K_{T}(\vartheta,\theta_1)$ as $\frac{1}{T}\sum_{t=1}^{T}q_{t}(\vartheta,\theta_1)$, we need $\mathbb{E}\nabla^2$ $q_{1}(\theta_2,\theta_{1})$ is non-singular.  
    \item[A19] The first order condition of minimizing the empirical risk function satisfies that $\nabla L_{T}(\widehat{\theta}_1) = op(T^{-1/2})$ and $\nabla K_{T}(\widehat{\theta}_2,\widehat{\theta}_1) = op(T^{-1/2})$. Similarly, we assume we can achieve such accuracy for optimization in the bootstrap world.

\end{itemize}
A19 implies that the first-order conditions for minimizing criterion functions hold approximately since it may be hard to find exactly $\widehat{\theta}_1$ or $\widehat{\theta}_2$. Then, we first develop the estimation inference of $\widehat{\theta}_1$ and $\widehat{\theta}_2$ in the \cref{Thm:estinfereal}. 

\begin{Theorem}\label{Thm:estinfereal}
Under A1-A7, A13-A19, with consistent parameter estimations $\widehat{\theta}_1$ and $\widehat{\theta}_2$, we have:
\begin{equation}
    \sqrt{T}(\widehat{\theta}_1 - \theta_1) \overset{d}{\to} N(0,B_1^{-1}\Omega_1 B_1^{-1}),
\end{equation}
where $\Omega_1 = 4\cdot\mathbb{E}(\sigma(X_{0},\theta_2)R_{1}\sigma(X_{0},\theta_2))$; $B_1 =2\cdot\mathbb{E}\left(\nabla\phi(X_0,\theta_1)(\nabla\phi(X_0,\theta_1))^{\top}\right)$; $R_{1} = \nabla \phi(X_{0},\theta_{1})$ $\nabla \phi(X_{0},\theta_{1})^{\top}$; $\nabla$ is the gradient operator w.r.t. $\theta_1$. Similarly, we can analyze the distribution of parameter estimation $\widehat{\theta}_2$:
\begin{equation}
    \sqrt{T}(\widehat{\theta}_2 - \theta_2) \overset{d}{\to} N(0,B_2^{-1}\Omega_2 B_2^{-1}),
\end{equation}
where $\Omega_2 = 4\cdot\mathbb{E}(B_3R_2B_3^{\top})$; $B_3 = \mathbb{E}(\nabla g(X_{1},X_{0},\theta_2,\theta_1))$; $R_2 = (g(X_1,X_0,\theta_2, \theta_1) -1)^2$; $B_2 = 2\cdot(\mathbb{E}(\nabla g(X_{1},$ $X_{0},\theta_2,\theta_1))\cdot (\mathbb{E}(\nabla g(X_{1},X_{0},\theta_2,\theta_1))^{\top}$; $ g(X_{1},X_{0},\theta_2,\theta_1)=$ $\left(\frac{X_{1} - \phi(X_{0},\theta_1)}{\sigma(X_{0},\theta_2)}\right)^2$; here $\nabla$ is the gradient operator w.r.t. $\theta_2$. 
\end{Theorem}
\begin{Remark}
  From here, we can see the distribution of $\widehat{\theta}_1$ depends on the time series structure. If we do not assume that we can specify the correct model format, the covariance matrix of the parameter's asymptotic distribution will depend on the whole structure of the time series. This is the reason why we need the forward bootstrap to generate time series and do estimation in the bootstrap world, otherwise, we can not approximate the covariance term well.   
\end{Remark}

In the bootstrap world, we can perform a similar parameter estimation procedure as we did in \cref{Subsec:consistencybootrealest}. As we have seen in the proof of \cref{Theom: meanbootreal}, for $(X_{-p+1},\ldots,X_{T})\in \Omega_{T}$, where $\Omega_{T}\subseteq \mathbb{R}^{T+p}$ and $\mathbb{P}((X_{-p+1},\ldots,X_{T})\notin \Omega_{T}) = o(1)$, under the consistency of parameter estimation in the real world, the bootstrap series is also ergodic in the sense of $\beta$-mixing. Thus, we can have below consistency results in the bootstrap world:
\begin{equation}
\begin{split}
    &\nabla^2 L^*_{T}(\widetilde{\theta}^*_1) = \frac{1}{T}\nabla^2 \sum_{t=1}^{T}\left(X^*_{t} - \phi(X^*_{t-1},\widetilde{\theta}^*_1)  \right)^2 \overset{p}{\to} 2\cdot\mathbb{E}^*\left(\nabla\phi(X^*_0,\widehat{\theta}_1)(\nabla\phi(X^*_0,\widehat{\theta}_1))^{\top}\right) = B_1^* ;\\
&\nabla^2 K^*_{T}(\widetilde{\theta}^*_2,\widehat{\theta}^*_1) = 2\cdot\left(\frac{1}{T}\sum_{t=1}^{T}\nabla g^*(X^*_t,X^*_{t-1},\widetilde{\theta}^*_2, \widehat{\theta}^*_1) \right)\cdot\left(\frac{1}{T}\sum_{t=1}^{T}\nabla g^*(X^*_t,X^*_{t-1},\widetilde{\theta}^*_2, \widehat{\theta}^*_1) \right)^{\top}  \\
      & + 2\cdot\left(\frac{1}{T}\sum_{t=1}^{T}g^*(X^*_t,X^*_{t-1},\widetilde{\theta}^*_2, \widehat{\theta}^*_1) -1 \right)\cdot \left(\frac{1}{T}\sum_{t=1}^{T}\nabla^2 g^*(X^*_t,X^*_{t-1},\widetilde{\theta}^*_2, \widehat{\theta}^*_1) \right)\\
      & \overset{p}{\to}2\cdot\mathbb{E}^*\left(\nabla g^*(X^*_1,X^*_{0},\widehat{\theta}_2, \widehat{\theta}_1)\right) \mathbb{E}^*\left(\nabla g^*(X^*_1,X^*_{0},\widehat{\theta}_2, \widehat{\theta}_1)\right)^{\top} = B_2^*,
\end{split}
\end{equation}
where $\widetilde{\theta}^*_1$ is between $\widehat{\theta}^*_1$ and $\widehat{\theta}_1$; $\widetilde{\theta}^*_2$ is between $\widehat{\theta}^*_2$ and $\widehat{\theta}_2$; hence $\widetilde{\theta}^*_1$ and $\widetilde{\theta}^*_2$ also converge to $\theta_1$ and $\theta_2$in probability, respectively. It is easily to find $B_1^* \overset{}{\to} B_1$ and $B_2^*\overset{}{\to} B_2$ for $(X_{-p+1},\ldots,X_{T})\in \Omega_{T}$. To simplify the result about $\nabla^2 K^*_{T}(\widetilde{\theta}^*_2,\widehat{\theta}^*_1)$, we need the variance of $\epsilon^*_{1}$ to be one to remove the second term. This is guaranteed since we normalize the variance of the residuals to 1 when we perform the bootstrap prediction algorithms for models with heteroscedastic errors. Another advantage of this manipulation comes from analyzing $\nabla K^*_{T}(\widehat{\theta}_2,\widehat{\theta}_1) = 2\cdot\left(\frac{1}{T}\sum_{t=1}^{T}g^*(X^*_t,X^*_{t-1},\widehat{\theta}_2, \widehat{\theta}_1) -1 \right)\cdot\left(\frac{1}{T}\sum_{t=1}^{T}\nabla g^*(X^*_t,X^*_{t-1},\widehat{\theta}_2, \widehat{\theta}_1) \right)$. With this additional manipulation, $\mathbb{E}^*(g^*(X^*_t,X^*_{t-1},\widehat{\theta}_2, \widehat{\theta}_1) -1)$ is 0, which implies that the asymptotic distribution of $\nabla K^*_{T}(\widehat{\theta}_2,\widehat{\theta}^*_1)$ has mean 0. By the CLT for a triangular array of strongly mixing series given in \cite{politis1997subsampling}, we can further show:
\begin{equation}
\begin{split}
&  \sqrt{T}\nabla L^*_{T}(\widehat{\theta}_1) \overset{d}{\to} N(0,\Omega_1); \\
    &\sqrt{T}\nabla K^*_{T}(\widehat{\theta}_2,\widehat{\theta}_1^*) \overset{d}{\to} N(0,\Omega_2).
\end{split}
\end{equation}
The required assumptions can be checked in the same way shown in Theorem 5 of \cite{franke2002properties}. All in all, we can develop estimation inference for parameter estimation in the bootstrap world, i.e, \cref{Corollary:estinfboot} as below:
\begin{Corollary}\label{Corollary:estinfboot}
If we restrict on observed data $\{X_{-p+1},\ldots,X_T\}\in \Omega_T$, where $\mathbb{P}((X_{-p+1},\ldots,X_T)\in \Omega_T) = o(1)$ as $T\to \infty$, under assumptions of \cref{Thm:estinfereal}, we can further build the estimation inference of parameter estimations in the bootstrap world, i.e, we have:
\begin{equation}\label{estimationinfboot}
\begin{split}
     &\sqrt{T}(\widehat{\theta}^*_1 - \widehat{\theta}_1) \overset{d}{\to} N(0,B_1^{-1}\Omega_1 B_1^{-1});\\
     &\sqrt{T}(\widehat{\theta}^*_2 - \widehat{\theta}_2) \overset{d}{\to} N(0,B_2^{-1}\Omega_2 B_2^{-1}).
\end{split}
\end{equation}
\end{Corollary}
\cref{Thm:estinfereal} and \cref{Corollary:estinfboot} together guarantee the pertinence of PPI returned by \cref{algori6,algori7} with \textit{high probability}. The notable advantage of this type of PI will be illustrated in \cref{Sec:5}. 

\section{Simulations}\label{Sec:5}
In this section, we deploy simulations to check the performance of our bootstrap point predictions and the asymptotic validity of corresponding PIs in $R$ platform for finite sample size. 
We first consider a simple case: NLAR model with order one and homoscedastic errors. We present the model as below:
\begin{equation}\label{Eq36}
    X_t = a + \log(b+ |X_{t-1}|) + \epsilon_t,  b>0,
\end{equation}
where $\epsilon_t$ satisfies A2. We can easily check that the model \cref{Eq36} satisfies the condition to make sure the time series is geometrically ergodic. Assuming that we have observed series $\{X_1,\ldots,X_T\}$, we want to predict the value of $X_{T+h}$. 
 As   pointed out before, the exactly $L_2$ optimal predictor is the conditional mean of $X_{T+h}$:
\begin{equation}\label{Eq37}
    \mathbb{E}(X_{t+h}|X_1,\ldots,X_T) = \int\cdots\int\mathcal{G}(X_T,\epsilon_{T+1},\ldots,\epsilon_{T+h})dF_{\epsilon_{T+1}}\cdots dF_{\epsilon_{T+h}},
\end{equation}
where $\mathcal{G}(X_T,\epsilon_{T+1},\ldots,\epsilon_{T+h})$ represents the analytic formula of $X_{T+h}$ which can be obtained by computing $X_{T+k} = a + \log(b+|X_{T+k-1}|) + \epsilon_{T+k}$ for $k = 1,\ldots,h$ iteratively. When we know the NLAR model and the innovation distribution, \cref{Eq37} can be computed by multiple-integration directly. However, it is very time-consuming or even impossible to perform such computation for a long prediction horizon. Thus, throughout this section, we assume that the forecasting result returned by the simulation approach of \cref{algori3} represents the exactly optimal $L_1$ or $L_2$ prediction. We take the simulation repeating number $M = 1000$ to get a satisfying approximation. For a fair comparison, we also do $1000$ times bootstrap with \cref{algori4} or \cref{algori5} to get the bootstrap prediction when the model and innovation are unknown. Starting from a simple example, we consider $a = 0.2, b = 0.5$ and $\{\epsilon_i\}\sim N(0,1)$. For the prediction horizon, we consider $h = 1,2,\ldots,5$ for the moment. To generate the data of \cref{Eq36}, we take $X_0\sim \text{Uniform}(-1,1)$, then generate a series with size $B+T$. Here, $B$ is a large enough burn-in number to remove the effects of the initial distribution of $X_0$. 
\begin{Remark}\label{Remark: 5.1}
In the model of \cref{Eq36}, we include the constant parameter $a$ 
in order to safely ensure that the innovation distribution has mean zero;
this is important for the consistency of NLS.
\end{Remark}

\begin{figure}[!h]
    \centering
    \includegraphics[scale=0.8]{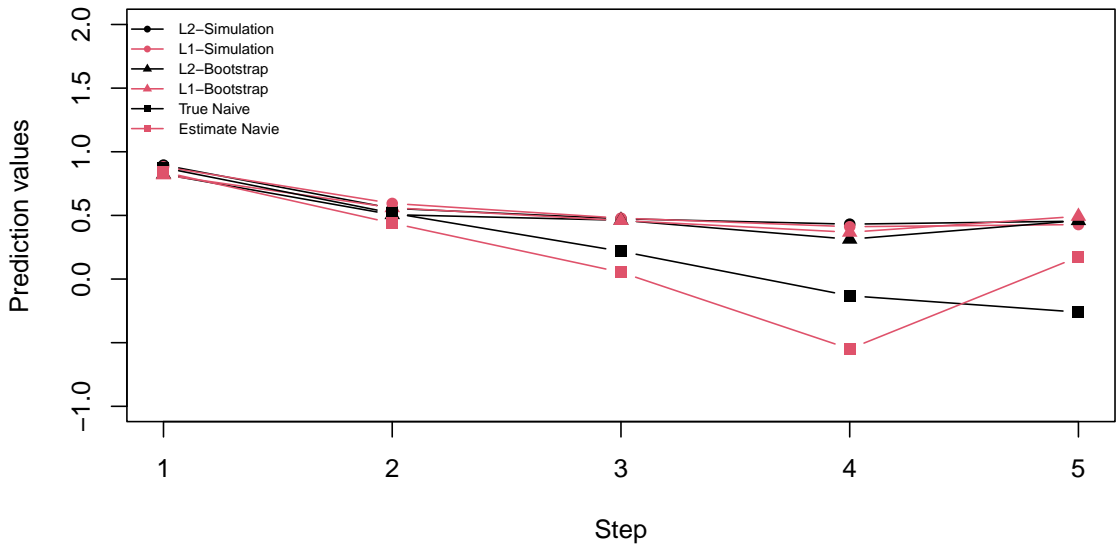}
    \caption{One experiment of performing four prediction methods based on \cref{Eq36}; the figure shows the prediction value
of $h$-step ahead predictors vs.~the step $h$.}
    \label{fig1}
\end{figure}

To see the crucial difference between the prediction of LAR and NLAR models, we apply two naive prediction methods which predict $X_{T+h}$ of \cref{Eq36} by computing  $X_{T+k} = a + \log(b + |{X}_{T+k-1}|)$ or $X_{T+k} = \widehat{a} + \log(\widehat{b} + |{X}_{T+k-1}|)$ repeatedly for $k = 1,\ldots,h$. 
In total, we compare four methods to do prediction. We call them (1) Simulation, with a known model and innovation; (2) Bootstrap, with an unknown model and innovation; (3) True Naive Prediction---naive prediction with the known model; (4) Estimated Naive Prediction---naive prediction with the estimated model. We call (3) and (4) naive since the variability in model estimation is not captured by these two methods. Compared to bootstrap-based methods (2) and (4), the simulation and true naive prediction methods are ``oracles'' since we assume that model and innovation information are known to us. Setting the burn-in number $B = 1000$ and $T = 400$, we present one simulation experiment result in \cref{fig1} showing the $h$-step ahead optimal prediction value vs.~the step $h$.

For the optimal prediction of $X_{T+1}$, all four methods return almost the same value as expected. This is in our expectation since the innovation is assumed to have 0 mean. In addition, the normal distribution is symmetric. Thus, two naive predictions, $L_2$ and $L_1$ optimal predictions from simulation and bootstrap approaches coincide. For prediction horizon beyond 1-step, we can still see the approximations of $L_1$ and $L_2$ optimal predictors based on bootstrap are very close to the oracle $L_1$ and $L_2$ optimal predictors return by the simulation method. On the other hand, true and estimated naive prediction methods give significantly different and even opposite results. We should notice that the only difference between these two naive prediction methods is using true parameters or estimated parameters. Under the consistent relationship between true parameters and estimated parameters which is indicated by the closeness of simulation and bootstrap approaches, the importance of including innovation in prediction gets emphasized. Otherwise, the multi-step ahead prediction of the NLAR model will be spoiled by the error accumulation issue.

Obviously, we can not get a conclusion based on only one experiment. For comparing the relationship between simulation and bootstrap methods, we repeat the above experiment $N = 5000$ times. Then, we approximate the Mean Squared Difference (MSD) between oracle (simulation-based) and bootstrap-based predictions according to the below equation:
\begin{equation}
    \text{MSD of the}~h\text{-th step ahead prediction} = \frac{1}{N}\sum_{n=1}^{N}(S_{n,h} - B_{n,h})^2,~\text{for}~h=1,\ldots,5,
\end{equation}
where $S_{n,h}$ and $B_{n,h}$ are $h$-th step ahead oracle and bootstrap predictions in the $n$-th replication, respectively. We summarize all MSD different prediction steps in \cref{Tab:1}. For all prediction horizons, the MSD is very small, which supports the conclusion that the bootstrap-based optimal prediction is consistent to the simulation-based optimal prediction.
\begin{table}[htbp]
\centering
  \caption{The MSD of bootstrap-based $L_1$ and $L_2$ predictions under Model \cref{Eq36}}
  \label{Tab:1}
\begin{tabular}{lccccc}
  \toprule 
Prediction Horizon & 1 & 2 & 3 & 4 & 5 \\
  \midrule
    $L_2$-Bootstrap  & $4.52\times10^{-3}$ & $3.98\times10^{-3}$ & $3.88\times10^{-3}$ & $3.87\times10^{-3}$ & $3.84\times10^{-3}$ \\
    $L_1$-Bootstrap  & $5.71\times10^{-3}$ & $4.84\times10^{-3}$ & $4.77\times10^{-3}$ & $4.71\times10^{-3}$ & $4.70\times10^{-3}$ \\
       \bottomrule
    \end{tabular}\\
\end{table}

In addition to comparing oracle and bootstrap-based predictions, we are also interested in the Mean Squared Prediction Error (MSPE) of all predictions. MSPE of predictions is approximated based on the below formula.
\begin{equation}
    \text{MSPE of the}~h \text{-th ahead prediction} = \frac{1}{N}\sum_{n=1}^{N}(X_{n,h} - P_{n,h})^2,~\text{for}~h=1,\ldots,5,
\end{equation}
where $P_{n,h}$ represents $h$-th step ahead predictions implied by four approaches and $X_{n,h}$ stands for the true future value in the $n$-th replication. All MSPE are recorded in \cref{Tab:2}.
\begin{table}[htbp]
\centering
  \caption{The MSPE of all prediction methods with $N(0,1)$ innovation under Model \cref{Eq36}}
  \label{Tab:2}
\begin{tabular}{lccccc}
  \toprule 
Prediction Horizon & 1 & 2 & 3 & 4 & 5 \\
  \midrule
    $L_2$-Simulation & 0.9595 & 1.2357 & 1.2101 & 1.1905 & 1.2153 \\
    $L_1$-Simulation & 0.9594 & 1.2360 & 1.2107 & 1.1901 & 1.2156\\
    $L_2$-Bootstrap  & 0.9639 & 1.2390 & 1.2144 & 1.1958 & 1.2181 \\
    $L_1$-Bootstrap  & 0.9640 & 1.2406 & 1.2158 &1.1960 & 1.2193 \\
    True Naive  & 0.9596   & 1.3748  &  1.4894 & 1.5581 & 1.6309\\
    Estimated naive & 0.9641 & 1.3826 & 1.4910 & 1.5518 & 1.6084\\
       \bottomrule
    \end{tabular}\\
\end{table}

We can find the MSPE of simulation- and bootstrap-based $L_1$ or $L_2$ optimal predictions are also very close to each other, respectively. Since the bootstrap optimal prediction is obtained with an estimated model and innovation distribution, it is not surprising that the MSPE is slightly larger than the simulation-based (oracle) optimal prediction, no matter if $L_2$ or $L_1$ is the loss criterion. In our expectation, the MSPE of simulation- and bootstrap-based prediction are all smaller than the MSPE of two naive predictions. Again, the importance of including the innovation effect in NLAR prediction is highlighted. 

Rather than applying the standard normal innovation distribution, we also research the MSPE of different methods with a skewed innovation distribution. Take $\epsilon_t\sim \chi^2(3) - 3$, here, we subtract 3 from $\chi^2(3)$ for getting a mean-zero innovation distribution. Repeating the above procedure, we summarize the MSPE of all prediction approaches in \cref{Tab:3}.
\begin{table}[htbp]
\centering
  \caption{The MSPE of all prediction methods with $\chi^2(3) - 3$ innovation under Model \cref{Eq36}}
  \label{Tab:3}
\begin{tabular}{lccccc}
  \toprule 
Prediction Horizon & 1 & 2 & 3 & 4 & 5 \\
  \midrule
    $L_2$-Simulation &5.7351	&6.1817	&6.3648	&6.8767	&6.3593 \\
    $L_1$-Simulation &6.1200	&6.4845	&6.6904	&7.2933	&6.7039 \\
    $L_2$-Bootstrap  &5.7885	&6.1704	&6.3679	&6.8900	&6.3749 \\
    $L_1$-Bootstrap  &6.1695	&6.4643	&6.6953	&7.2931	&6.7036 \\
    True Naive  & 5.7398	&6.5561	&6.8901	&7.7830	&7.4002\\
    Estimated naive & 5.7798	&6.5294	&6.8703	&7.7172	&7.2002 \\
       \bottomrule
    \end{tabular}\\
\end{table}
From there, the relative performance of different prediction methods is consistent to results implied by \cref{Tab:2}. Simulation-based methods are the best
as expected;  bootstrap-based methods give just slightly higher MSPE compared to the simulation-based method. Both computationally heavy methods outperform the  naive methods. 

Another notable phenomenon indicated by \cref{Tab:3} is that the MSPE of $L_2$ optimal prediction is always less than its corresponding $L_1$ optimal prediction. This is reasonable since the $L_2$ optimal prediction is determined based on $L_2$ loss which coincides with the mean squared errors. However, for the simulation result in \cref{Tab:2}, this phenomenon is not remarkable since the innovation distribution is symmetric.

As we have clarified in \cref{Remark: 5.1}, the constant parameter in \cref{Eq36} is important and can cover cases in which the mean of the innovation distribution is not zero. We present the MSPE of all prediction methods with $\chi^2(3)$ in \cref{Tab:4}. Here, we can see the MSPE of bootstrap-based methods is still very close to the corresponding MSPE of simulation-based methods. The interesting finding is that the estimated naive approach also gives great MSPE values. The reason behind this phenomenon is that the estimated parameter value $a$ already includes the mean effects of the innovation distribution. On the other hand, the true naive estimation approach returns terrible MSPE, since the effects of innovation distribution are discarded. 
\begin{table}[htbp]
\centering
  \caption{The MSPE of all prediction methods with $\chi^2(3)$ innovation under Model \cref{Eq36}}
  \label{Tab:4}
\begin{tabular}{lccccc}
  \toprule 
Prediction Horizon & 1 & 2 & 3 & 4 & 5 \\
  \midrule
    $L_2$-Simulation &6.329292	&6.066646 &6.040300	&6.358251&5.988886 \\
    $L_1$-Simulation &6.831067	&6.445046	&6.413609	&6.754066	&6.352465 \\
    $L_2$-Bootstrap  &6.346281	&6.099135	&6.075808	&6.378988	&6.014661 \\
    $L_1$-Bootstrap  &6.836730	&6.469492	&6.440389	&6.775695	&6.366932 \\
    True Naive  &15.818671	&20.260416	&23.381642	&26.636934	&29.088646\\
    Estimated naive &6.335469	&6.097773	&6.080526	&6.381737	&6.021346 \\
       \bottomrule
    \end{tabular}\\
\end{table}

Beyond analyzing the performance of point prediction, we are also interested in measuring prediction accuracy by building PIs. As discussed before, we can build two types of bootstrap-based prediction intervals, (1) Quantile PI; (2) Pertinent PI. The advantage of pertinent PI is that it can be centered in the optimal $L_2$ or $L_1$ predictors. Moreover, it includes the estimation error of parameters into consideration, which means a superior empirical CVR, especially in short data size situations. For deriving pertinent PI, we take $K = 1000$ in \cref{algori6}. We repeat experiment $N=5000$ times and set significance level $\alpha = 0.05$. Then, we compute empirical CVR of bootstrap-based QPI and PPI for $h = 1,\ldots,5$ step ahead predictions with the below formula:
\begin{equation}\label{empircvr}
    \text{CVR of the}~h \text{-th ahead prediction} = \frac{1}{N}\sum_{n=1}^{N}\mathbbm{1}_{X_{n,h}\in [L_{n,h},U_{n,h}]}, \text{for}~h=1,\ldots,5,
\end{equation}
where $[L_{n,h},U_{n,h}]$ and $X_{n,h}$ represent $h$-th step ahead prediction intervals and the true future value in the $n$-th replication, respectively. Recall that the PPI can be centered at $L_1$ or $L_2$ optimal point predictor, thus we have three types of PIs based on bootstrap. Moreover, we can build PIs with fitted or predictive residuals. Besides, if all model information is known to us, we can use \cref{algori3} to get oracle QPI based on simulations. Thus, we totally have seven different PIs in hand. We denote them by (1) QPI-f, QPI with fitted residuals; (2) QPI-p, QPI with predictive residuals; (3) $L_2$-PPI-f, PPI centered at $L_2$ optimal predictor with fitted residuals; (4) $L_2$-PPI-p, PPI centered at $L_2$ optimal predictor with predictive residuals; (5) $L_1$-PPI-f, PPI centered at $L_1$ optimal predictor with fitted residuals; (6) $L_1$-PPI-p, PPI centered at $L_1$ optimal predictor with predictive residuals; (7) SPI, which is QPI based on simulations. In addition, since building a valid PI is more challenging work, we totally consider seven different models to check the feasibility of our methods:
\begin{itemize}
    \item Model 1: $X_t = (0.1\cdot X_{t-1}) I(X_{t-1}\leq 0) + (0.8\cdot X_{t-1}) I(X_{t-1}>0)$ + $\epsilon_t$.
    \item Model 2: $X_t = (0.5\cdot X_{t-1} + 0.2\cdot X_{t-2} + 0.1\cdot X_{t-3})I(X_{t-1}\leq 0) + (0.8X_{t-1})I(X_{t-1}>0) $ + $\epsilon_t$.
    \item Model 3: $X_t = (0.1\cdot X_{t-1} + 0.5\cdot e^{-X_{t-1}^2}\epsilon_t) I(X_{t-1}\leq 0) + (0.8\cdot X_{t-1} +0.5\cdot e^{-X_{t-1}^2}\epsilon_t) I(X_{t-1}>0)$.
    \item Model 4: $X_t = 0.2 + \log(0.5+ |X_{t-1}|) + \epsilon_t$.
    \item Model 5: $X_{t} = 2\cdot\log(X_{t-1}^2)+\epsilon_t$.
    \item Model 6: $X_t = \log(10 + 5\cdot e^{0.9\cdot X_{t-1}}) + \epsilon_t$.
    \item Model 7: $X_t = \log(4\cdot e^{0.9\cdot X_{t-2}} + 5\cdot e^{0.9\cdot X_{t-1}} + 6\cdot e^{0.9\cdot X_{t-3}}) + \epsilon_t$,
\end{itemize}
where $\epsilon_t\sim N(0,1)$ and $I(X_{t-1}\leq 0)$ is the indicator function which equals to 0 when $X_{t-1}\leq 0$ and 1 otherwise. Models 1-2 are two different threshold models in the form $X_{t} = \sum_{i=1}^{s}$ $\{\alpha_{i0}+$ $ \sum_{j=1}^{p}\alpha_{ij}X_{t-j}\}$ $I(X_{t-d}\in I_{i})$ $+\epsilon_t$, where $I_{i}$ are different non-overlapping intervals. The ergodicity can be guaranteed if $\max_{i}\sum_{j=1}^{p}{|\alpha_{ij}|}<1$, which is satisfied by Models 1-2. In addition, for Model 3 with heteroscedastic errors, A5 and A7 are also met so that the ergodicity is achieved. Model 4 is according to \cref{Eq36}. The Models 5 is according to the form $X_{t} = \alpha*\log(X_{t-1}^2)$ for positive $\alpha$. It is also in NLAR form and ergodic by checking the first condition of \cite{an1996geometrical}, but they are simpler compared to Model 4. For Models 6 and 7, they have a general form: $X_t = \log(\alpha_0+\alpha_1e^{\beta X_{t-1}} + \alpha_2e^{\beta X_{t-2}}+\cdots+\alpha_p e^{\beta X_{t-p}})$ for $\beta > 0, \alpha_0,\ldots,\alpha_p \geq 0$. The ergodicity of this model is guaranteed by checking A6.  All simulation results of CVR are summarized in \cref{Tab:5,Tab:6,Tab:7,Tab:8,Tab:9,Tab:10,Tab:11}. Besides the analyses of CVR, we are also concerned about the average length of PIs of different methods. In practice, a wide PI is less useful even though it has great coverage probability. We define the average length (LEN) of PIs as below:
\begin{equation}
        \text{LEN of the}~h \text{-th ahead PI} = \frac{1}{N}\sum_{n=1}^{N}(U_{n,h}-L_{n,h}), \text{for}~h=1,\ldots,5,
\end{equation}
where $U_{n,h}$ and $L_{n,h}$ are higher and lower bounds of the $h$-th step ahead PI in the $n$-th replication, respectively. Accordingly, we present LEN of different PIs along with CVR in \cref{Tab:5,Tab:6,Tab:7,Tab:8,Tab:9,Tab:10,Tab:11}.
\begin{Remark}
    We should clarify that the CVR computed by \cref{empircvr} is the unconditional coverage rate of $X_{T+k}$ since it is an average of the conditional coverage of $X_{T+k}$ for all replications. Also, when the sample size is small, we may get parameter estimations that make the time series close to be unstationary, especially for estimating different regions of a threshold model where the sample size further decreases. This will destroy our prediction process when multi-step ahead predictions are required. Thus, we redo the simulation once we find such abnormal larger or small predictions.  
\end{Remark}

From these simulations, the first thing we can notice is that all CVR for SPI are great and close to the oracle coverage level even for short data, which implies the simulation-based approach works well once we know the true model and innovation distribution. For $T=400$, we can find all PPIs work well and are even competitive compared to the SPI. On the other hand, the QPI with fitted residuals is the worst one, especially for complicated Models 6 and 7. By applying predictive residuals, the CVR gets improved for QPI. For $T=100$, no matter with fitted or predictive residuals, PPIs dominate QPIs. Generally speaking, PI with predictive residuals works better than the corresponding one with fitted residuals. The naive QPI with fitted residuals is still the worst method among all choices. For $T = 50$, the gap between applying fitted residuals and predictive residuals gets amplified. All PIs with predictive residuals work better than the corresponding ones with fitted residuals. Moreover, the gap between QPI and PPI also gets amplified. The simple QPI with fitted residuals only returns around 85$\%$ and 89$\%$ CVR for 5-step ahead predictions of Model 6 and Model 2, respectively. On the other hand, the PPI with predictive residuals achieves more than 90$\%$ and 93$\%$ CVR, respectively. For the LEN of different PIs, we can find that LENs of SPIs are barely changed for a specific model with various sample sizes. This is in our expectation since SPI is oracle and we can approximate the true quantile region of future value by simulation even with 50 sample points. Also, compared to the LEN of PPI and QPI, SPI tends to have smaller LEN, but possesses more accurate CVR due to the oracle property again. For PPI, although its LEN tends to be slightly larger than the LEN of SPI and QPI, it is the best bootstrap-type PI according to the CVR. Based on these simulation results, we summarize some important conclusions below:
\begin{itemize}
    \item If we know the parameters of the model and innovation distribution, SPI can work well and give accurate CVR even for short data, but it is usually unrealistic in practice.
    \item If we do not have model information and the data is short, PPI with predictive residuals is the best method, which can give competitive performance compared to SPI for simple models. However, the QPI can not cover future values well and its CVR is severely lower than oracle level. 
    \item If we do not have model information and the data in hand is large enough, both QPI and PPI work well.
\item Since in practice we can not judge whether the data in hand is large enough
for the problem at hand, using the PPI (with predictive residuals) is recommendable. 
\end{itemize}

\begin{Remark}
To perform bootstrap-based prediction, we ran simulations in a parallel
fashion using 30 Xeon(R) E5-2630 CPUs. Besides, we should notice that the constant parameter inside the $\log$ function of Model 6 is the hardest one to be estimated, since the low change rate of the partial derivative. This may be the reason for the relatively poor performance of bootstrap-based prediction methods on Model 6.
\end{Remark}

\begin{table}[htbp]
\centering
  \caption{The CVR and LEN of PIs for Model 1}
  \vspace{2pt}
  \label{Tab:5}
\begin{tabular}{lcccccccccc}
  \toprule 
 Model 1: & \multicolumn{10}{c}{$X_t = (0.1\cdot X_{t-1})I(X_{t-1}\leq 0) + (0.8\cdot X_{t-1})I(X_{t-1}>0)$ + $\epsilon_t$} \\
 \midrule
  & \multicolumn{5}{c}{CVR for each step} & \multicolumn{5}{c}{LEN for each step}\\
    $T = 400$  & 1     & 2     & 3     & 4     & 5 & 1 & 2 & 3 & 4 & 5  \\[3pt]
    QPI-f      & 0.9456 & 0.9472 & 0.9478 & 0.9496 & 0.9484 & 3.89 & 4.60 & 4.87 & 5.01 & 5.07 \\
    QPI-p       & 0.9470 & 0.9468 & 0.9478 & 0.9502 & 0.9500 & 3.91 & 4.62 & 4.90 & 5.03 & 5.09  \\
    $L_2$-PPI-f      & 0.9474 & 0.9454 & 0.9486 & 0.9500 & 0.9514 & 3.90 & 4.61 & 4.89 & 5.03 & 5.09 \\
    $L_2$-PPI-p     & 0.9474 & 0.9480 & 0.9480 & 0.9510 & 0.9526 & 3.92 & 4.63 & 4.92 & 5.05 & 5.12 \\
    $L_1$-PPI-f   & 0.9468 & 0.9456 & 0.9494 & 0.9494 & 0.9520 & 3.90 & 4.61 & 4.89 & 5.03 & 5.09   \\
    $L_1$-PPI-p      & 0.9464 & 0.9468 & 0.9484 & 0.9512 & 0.9530 & 3.92 & 4.63 & 4.92 & 5.05 & 5.12\\
    SPI       & 0.9484 & 0.9474 & 0.9500 & 0.9502 & 0.9546 & 3.90 & 4.62 & 4.91 & 5.04 & 5.10 \\[3pt]
    $T = 100$        &       &       &       &       &  \\[3pt]
    QPI-f       & 0.9388 & 0.9408 & 0.9362 & 0.9328 & 0.9330 & 3.86 & 4.52 & 4.78 & 4.91 & 4.97 \\
    QPI-p       & 0.9438 & 0.9446 & 0.9394 & 0.9366 & 0.9352 & 3.94 & 4.61 & 4.88 & 5.00 & 5.07\\
    $L_2$-PPI-f       & 0.9416 & 0.9424 & 0.9382 & 0.9350 & 0.9358 & 3.91 & 4.58 & 4.85 & 4.98 & 5.05 \\
    $L_2$-PPI-p         & 0.9478 & 0.9478 & 0.9442 & 0.9396 & 0.9402 & 3.99 & 4.67 & 4.94 & 5.08 & 5.15  \\
    $L_1$-PPI-f        & 0.9428 & 0.9430 & 0.9376 & 0.9346 & 0.9358 & 3.91 & 4.58 & 4.85 & 4.97 & 5.04  \\
    $L_1$-PPI-p        & 0.9476 & 0.9482 & 0.9442 & 0.9402 & 0.9404 & 3.99 & 4.67 & 4.94 & 5.07 & 5.14   \\
    SPI         & 0.9502 & 0.9482 & 0.9464 & 0.9468 & 0.9460 & 3.90 & 4.61 & 4.89 & 5.03 & 5.09 \\[3pt]
    $T = 50$        &       &       &       &       &  \\[3pt]
    QPI-f        & 0.9168 & 0.9248 & 0.9204 & 0.9106 & 0.9218 & 3.74 & 4.44 & 4.69 & 4.81 & 4.87  \\
    QPI-p        & 0.9296 & 0.9360 & 0.9334 & 0.9238 & 0.9324 & 3.91 & 4.63 & 4.90 & 5.02 & 5.09  \\
    $L_2$-PPI-f       & 0.9306 & 0.9318 & 0.9268 & 0.9176 & 0.9306 & 3.91 & 4.57 & 4.83 & 4.96 & 5.04\\
    $L_2$-PPI-p        & 0.9402 & 0.9438 & 0.9392 & 0.9292 & 0.9390 & 4.07 & 4.76 & 5.04 & 5.18 & 5.26  \\
    $L_1$-PPI-f        & 0.9302 & 0.9314 & 0.9264 & 0.9170 & 0.9300 & 3.91 & 4.56 & 4.82 & 4.95 & 5.02 \\
    $L_1$-PPI-p       & 0.9390 & 0.9438 & 0.9366 & 0.9290 & 0.9364 & 4.08 & 4.75 & 5.03 & 5.16 & 5.24  \\
    SPI        & 0.9486 & 0.9492 & 0.9508 & 0.9452 & 0.9464 & 3.90 & 4.61 & 4.90 & 5.03 & 5.09 \\
       \bottomrule
    \end{tabular}\\
\end{table}

\begin{table}[htbp]
\centering
  \caption{The CVR and LEN of PIs for Model 2}
  \vspace{2pt}
  \label{Tab:6}
\begin{tabular}{ll>{\centering\arraybackslash}p{1cm}>{\centering\arraybackslash}p{1cm}>{\centering\arraybackslash}p{1cm}>{\centering\arraybackslash}p{1cm}>{\centering\arraybackslash}p{1cm}>{\centering\arraybackslash}p{1cm}>{\centering\arraybackslash}p{1cm}>{\centering\arraybackslash}p{1cm}>{\centering\arraybackslash}p{1cm}>{\centering\arraybackslash}p{1cm}}
  \toprule 
 Model 2: & \multicolumn{10}{c}{{\small $X_t = (0.5\cdot X_{t-1} + 0.2\cdot X_{t-2} + 0.1\cdot X_{t-3})I(X_{t-1}\leq 0) + (0.8\cdot X_{t-1})I(X_{t-1}>0) $ + $\epsilon_t$}} \\
 \midrule
    & \multicolumn{5}{c}{CVR for each step} & \multicolumn{5}{c}{LEN for each step}\\
    $T = 400$  & 1     & 2     & 3     & 4     & 5 & 1     & 2     & 3     & 4     & 5 \\[3pt]
    QPI-f       & 0.9420 & 0.9506 & 0.9468 & 0.9444 & 0.9372 & 3.88 & 4.68 & 5.11 & 5.40 & 5.58  \\
    QPI-p      & 0.9462 & 0.9512 & 0.9502 & 0.9474 & 0.9428 & 3.92 & 4.72 & 5.16 & 5.45 & 5.64 \\
    $L_2$-PPI-f      & 0.9446 & 0.9510 & 0.9486 & 0.9470 & 0.9408 & 3.90 & 4.71 & 5.15 & 5.44 & 5.63 \\
    $L_2$-PPI-p     & 0.9466 & 0.9542 & 0.9516 & 0.9494 & 0.9434 & 3.94 & 4.75 & 5.20 & 5.49 & 5.69 \\
    $L_1$-PPI-f   & 0.9448 & 0.9518 & 0.9478 & 0.9468 & 0.9402 & 3.90 & 4.71 & 5.15 & 5.44 & 5.62   \\
    $L_1$-PPI-p      & 0.9470 & 0.9544 & 0.9500 & 0.9486 & 0.9436 & 3.94 & 4.75 & 5.20 & 5.49 & 5.68 \\
    SPI       & 0.9446 & 0.9534 & 0.9508 & 0.9510 & 0.9454 & 3.90 & 4.71 & 5.16 & 5.46 & 5.65  \\[3pt]
    $T = 100$        &       &       &       &       &  \\[3pt]
    QPI-f       & 0.9270 & 0.9304 & 0.9294 & 0.9272 & 0.9250 & 3.81 & 4.57 & 4.98 & 5.23 & 5.40  \\
    QPI-p        & 0.9370 & 0.9412 & 0.9368 & 0.9372 & 0.9372 & 3.98 & 4.76 & 5.19 & 5.46 & 5.63    \\
    $L_2$-PPI-f       & 0.9358 & 0.9352 & 0.9338 & 0.9314 & 0.9298 & 3.95 & 4.71 & 5.13 & 5.40 & 5.59 \\
    $L_2$-PPI-p         & 0.9454 & 0.9454 & 0.9444 & 0.9430 & 0.9418 & 4.10 & 4.90 & 5.34 & 5.63 & 5.83 \\
    $L_1$-PPI-f        & 0.9364 & 0.9360 & 0.9336 & 0.9310 & 0.9304 & 3.95 & 4.71 & 5.13 & 5.39 & 5.58 \\
    $L_1$-PPI-p        & 0.9450 & 0.9456 & 0.9432 & 0.9422 & 0.9412 & 4.11 & 4.90 & 5.33 & 5.62 & 5.81 \\
    SPI         & 0.9446 & 0.9472 & 0.9498 & 0.9474 & 0.9478 & 3.90 & 4.71 & 5.16 & 5.46 & 5.65 \\[3pt]
    $T = 50$        &       &       &       &       &  \\[3pt]
    QPI-f        & 0.8980 & 0.9054 & 0.9018 & 0.8950 & 0.8926  & 3.66 & 4.47 & 4.87 & 5.14 & 5.38 \\
    QPI-p        & 0.9260 & 0.9314 & 0.9272 & 0.9218 & 0.9212 & 4.05 & 4.97 & 5.42 & 5.74 & 5.99  \\
    $L_2$-PPI-f       & 0.9340 & 0.9268 & 0.9214 & 0.9164 & 0.9152 & 4.22 & 5.10 & 5.86 & 6.89 & 8.97\\
    $L_2$-PPI-p        & 0.9522 & 0.9478 & 0.9404 & 0.9400 & 0.9376  & 4.60 & 5.57 & 6.36 & 7.33 & 9.03 \\
    $L_1$-PPI-f       & 0.9338 & 0.9268 & 0.9194 & 0.9144 & 0.9130 & 4.23 & 5.09 & 5.82 & 6.79 & 8.71   \\
    $L_1$-PPI-p       & 0.9522 & 0.9482 & 0.9384 & 0.9378 & 0.9356 & 4.61 & 5.55 & 6.30 & 7.20 & 8.71 \\
    SPI         & 0.9494 & 0.9448 & 0.9464 & 0.9458 & 0.9462 & 3.90 & 4.71 & 5.16 & 5.46 & 5.65 \\
       \bottomrule
    \end{tabular}\\
\end{table}

\begin{table}[htbp]
\centering
  \caption{The CVR and LEN of PIs for Model 3}
  \vspace{2pt}
  \label{Tab:7}
\begin{tabular}{ll>{\centering\arraybackslash}p{1cm}>{\centering\arraybackslash}p{1cm}>{\centering\arraybackslash}p{1cm}>{\centering\arraybackslash}p{1cm}>{\centering\arraybackslash}p{1cm}>{\centering\arraybackslash}p{1cm}>{\centering\arraybackslash}p{1cm}>{\centering\arraybackslash}p{1cm}>{\centering\arraybackslash}p{1cm}>{\centering\arraybackslash}p{1cm}}
  \toprule 
 Model 3: & \multicolumn{10}{c}{{\small $X_t = (0.1\cdot X_{t-1}+ 0.5\cdot e^{-X_{t-1}^2}\cdot\epsilon_t)I(X_{t-1}\leq 0) + (0.8\cdot X_{t-1} +0.5\cdot e^{-X_{t-1}^2}\cdot\epsilon_t)I(X_{t-1}>0)$}} \\
 \midrule
     & \multicolumn{5}{c}{CVR for each step} & \multicolumn{5}{c}{LEN for each step}\\
    $T = 400$  & 1     & 2     & 3     & 4     & 5  & 1     & 2     & 3     & 4     & 5\\[3pt]
    QPI-f       & 0.9478 & 0.9442 & 0.9526 & 0.9444 & 0.9418  & 1.47 & 1.74 & 1.82 & 1.84 & 1.85  \\
    QPI-p      & 0.9474 & 0.9486 & 0.9504 & 0.9444 & 0.9432 & 1.47 & 1.74 & 1.82 & 1.84 & 1.85\\
    $L_2$-PPI-f      & 0.9520 & 0.9488 & 0.9542 & 0.9436 & 0.9434 & 1.59 & 2.22 & 2.24 & 2.30 & 2.29 \\
    $L_2$-PPI-p     & 0.9510 & 0.9486 & 0.9522 & 0.9454 & 0.9446 &  1.64 & 2.37 & 2.37 & 2.44 & 2.42  \\
    $L_1$-PPI-f   & 0.9514 & 0.9480 & 0.9540 & 0.9448 & 0.9440 & 1.63 & 1.88 & 2.10 & 2.17 & 2.18   \\
    $L_1$-PPI-p     & 0.9480 & 0.9514 & 0.9530 & 0.9474 & 0.9448 & 1.68 & 1.92 & 2.19 & 2.27 & 2.28 \\
    SPI      & 0.9500 & 0.9500 & 0.9516 & 0.9444 & 0.9442 & 1.47  & 1.74 & 1.82 & 1.84 & 1.85\\[3pt]
    $T = 100$        &       &       &       &       &  \\[3pt]
    QPI-f       & 0.9344 & 0.9388 & 0.9420 & 0.9390 & 0.9372 & 1.47 & 1.73 & 1.81 & 1.84 & 1.85  \\
    QPI-p        & 0.9318 & 0.9348 & 0.9404 & 0.9392 & 0.9378  & 1.47 & 1.73 & 1.82 & 1.84 & 1.86   \\
    $L_2$-PPI-f       & 0.9406 & 0.9418 & 0.9452 & 0.9418 & 0.9434  & 1.55 & 2.08 & 2.11 & 2.13 & 2.11  \\
    $L_2$-PPI-p        & 0.9424 & 0.9422 & 0.9452 & 0.9410 & 0.9452 & 1.64 & 2.40 & 2.38 & 2.40 & 2.36 \\
    $L_1$-PPI-f        & 0.9400 & 0.9426 & 0.9464 & 0.9430 & 0.9440 & 1.60 & 1.91 & 2.01 & 2.02 & 2.03 \\
    $L_1$-PPI-p        & 0.9398 & 0.9440 & 0.9466 & 0.9412 & 0.9454 & 1.72 & 2.04 & 2.16 & 2.17 & 2.18\\
    SPI         & 0.9482 & 0.9474 & 0.9506 & 0.9518 & 0.9456 & 1.47 & 1.74 & 1.82 & 1.84 & 1.85 \\[3pt]
    $T = 50$        &       &       &       &       &  \\[3pt]
    QPI-f       & 0.9060 & 0.9268 & 0.9266 & 0.9222 & 0.9302 & 1.43 & 1.71 & 1.80 & 1.83 & 1.84 \\
    QPI-p        & 0.9030 & 0.9286 & 0.9262 & 0.9206 & 0.9312  & 1.44 & 1.73 & 1.82 & 1.85 & 1.87  \\
    $L_2$-PPI-f       & 0.9300 & 0.9394 & 0.9350 & 0.9338 & 0.9408 & 1.55 & 3.37 & 3.34 & 3.23 & 3.11 \\
    $L_2$-PPI-p        & 0.9302 & 0.9414 & 0.9358 & 0.9372 & 0.9398 & 1.64 & 3.85 & 3.74 & 3.60 & 3.44  \\
    $L_1$-PPI-f        & 0.9302 & 0.9410 & 0.9358 & 0.9356 & 0.9412 & 1.61 & 2.39 & 2.58 & 2.54 & 2.50 \\
    $L_1$-PPI-p        & 0.9308 & 0.9422 & 0.9364 & 0.9384 & 0.9412 & 1.79 & 2.79 & 2.81 & 2.74 & 2.68  \\
    SPI     & 0.9486 & 0.9520 & 0.9486 & 0.9444 & 0.9518 & 1.47 & 1.74 & 1.81 & 1.84 & 1.85\\
       \bottomrule
    \end{tabular}\\
\end{table}

\begin{table}[htbp]
\centering
  \caption{The CVR and LEN of PIs for Model 4}
  \vspace{2pt}
  \label{Tab:8}
\begin{tabular}{lccccccccccc}
  \toprule 
 Model 4: & \multicolumn{6}{c}{$ X_t = 0.2 + \log(0.5 + |X_{t-1}|) + \epsilon_t$} \\
 \midrule
 & \multicolumn{5}{c}{CVR for each step} & \multicolumn{5}{c}{LEN for each step}\\
    $T = 400$  & 1     & 2     & 3     & 4     & 5  & 1     & 2     & 3     & 4     & 5 \\[3pt]
    QPI-f       & 0.9498 & 0.9446 & 0.9482 & 0.9444 & 0.9444  & 3.89 & 4.30 & 4.33 & 4.33 & 4.33 \\
    QPI-p       & 0.9486 & 0.9462 & 0.9512 & 0.9450 & 0.9464 & 3.91 & 4.33 & 4.35 & 4.35 & 4.35 \\
    $L_2$-PPI-f      & 0.9492 & 0.9454 & 0.9480 & 0.9440 & 0.9466 & 3.90 & 4.32 & 4.34 & 4.34 & 4.34 \\
    $L_2$-PPI-p      & 0.9508 & 0.9442 & 0.9504 & 0.9458 & 0.9466 & 3.92 & 4.33 & 4.35 & 4.36 & 4.36 \\
    $L_1$-PPI-f   &0.9496 & 0.9460 & 0.9486 & 0.9454 & 0.9468 & 3.90 & 4.32 & 4.34 & 4.34 & 4.34  \\
    $L_1$-PPI-p      &0.9510 & 0.9452 & 0.9502 & 0.9462 & 0.9472 & 3.93 & 4.34 & 4.35 & 4.36 & 4.36 \\
    SPI       & 0.9502 & 0.9456 & 0.9492 & 0.9460 & 0.9504  & 3.90 & 4.32 & 4.34 & 4.34 & 4.34\\[3pt]
    $T = 100$        &       &       &       &       &  \\[3pt]
    QPI-f       &0.9350 &0.9440 &0.9358 &0.9412 &0.9348 & 3.85 & 4.27 & 4.29 & 4.29 & 4.29 \\
    QPI-p        &0.9412 &0.9482 &0.9404 &0.9456 &0.9412 & 3.93 & 4.34 & 4.37 & 4.37 & 4.37  \\
    $L_2$-PPI-f        &0.9376 &0.9442 &0.9370 &0.9438 &0.9362 & 3.90 & 4.30 & 4.32 & 4.32 & 4.32 \\
    $L_2$-PPI-p         &0.9412 &0.9504 &0.9406 &0.9478 &0.9426 & 3.98 & 4.38 & 4.39 & 4.40 & 4.40  \\
    $L_1$-PPI-f        &0.9386 &0.9446 &0.9378 &0.9448 &0.9364 & 3.90 & 4.30 & 4.32 & 4.32 & 4.32 \\
    $L_1$-PPI-p        &0.9412 &0.9502 &0.9404 &0.9470 &0.9418 & 3.98 & 4.38 & 4.40 & 4.40 & 4.41 \\
    SPI         &0.9480 &0.9502 &0.9426 &0.9504 &0.9466 & 3.90 & 4.32 & 4.34 & 4.34 & 4.34  \\[3pt]
    $T = 50$        &       &       &       &       &  \\[3pt]
    QPI-f       &0.9280 &0.9288 &0.9328 &0.9326 &0.9312  & 3.75 & 4.24 & 4.27 & 4.26 & 4.27 \\
    QPI-p        &0.9386 &0.9396 &0.9420 &0.9422 &0.9428 & 3.92 & 4.40 & 4.43 & 4.43 & 4.43 \\
    $L_2$-PPI-f       &0.9404 &0.9316 &0.9372 &0.9372 &0.9352 & 3.89 & 4.30 & 4.32 & 4.32 & 4.33 \\
    $L_2$-PPI-p        &0.9496 &0.9398 &0.9452 &0.9448 &0.9438 & 4.06 & 4.46 & 4.48 & 4.49 & 4.49 \\
    $L_1$-PPI-f        &0.9410 &0.9308 &0.9376 &0.9384 &0.9350 & 3.90 & 4.30 & 4.33 & 4.33 & 4.33 \\
    $L_1$-PPI-p       &0.9504 &0.9398 &0.9452 &0.9458 &0.9438 & 4.06 & 4.47 & 4.49 & 4.50 & 4.49  \\
    SPI        &0.9530 &0.9462 &0.9456 &0.9444 &0.9428 & 3.90 & 4.32 & 4.34 & 4.34 & 4.34\\
       \bottomrule
    \end{tabular}\\
\end{table}

\begin{table}[htbp]
\centering
  \caption{The CVR and LEN of PIs for Model 5}
  \vspace{2pt}
  \label{Tab:9}
\begin{tabular}{lcccccccccc}
  \toprule 
 Model 5: & \multicolumn{10}{c}{$ X_t = 2\cdot \log(X_{t-1}^2) + \epsilon_t$} \\
 \midrule
  & \multicolumn{5}{c}{CVR for each step} & \multicolumn{5}{c}{LEN for each step}\\
    $T = 400$  & 1     & 2     & 3     & 4     & 5 & 1     & 2     & 3     & 4     & 5 \\[3pt]
    QPI-f      & 0.9432 & 0.9502 & 0.9476 & 0.9478 & 0.9510 & 3.90 & 4.32 & 4.42 & 4.44 & 4.45 \\
    QPI-p       & 0.9440 & 0.9524 & 0.9492 & 0.9480 & 0.9484 & 3.91 & 4.33 & 4.43 & 4.46 & 4.46 \\
    $L_2$-PPI-f       & 0.9468 & 0.9502 & 0.9516 & 0.9494 & 0.9518 & 3.90 & 4.34 & 4.43 & 4.46 & 4.47\\
    $L_2$-PPI-p       & 0.9448 & 0.9536 & 0.9500 & 0.9478 & 0.9498 & 3.92 & 4.35 & 4.45 & 4.47 & 4.48 \\
    $L_1$-PPI-f   & 0.9448 & 0.9510 & 0.9506 & 0.9496 & 0.9504 & 3.91 & 4.34 & 4.44 & 4.46 & 4.47\\
    $L_1$-PPI-p      & 0.9440 & 0.9532 & 0.9494 & 0.9480 & 0.9498 & 3.92 & 4.35 & 4.45 & 4.47 & 4.48\\
    SPI        & 0.9462 & 0.9538 & 0.9510 & 0.9480 & 0.9472 & 3.90 & 4.33 & 4.42 & 4.45 & 4.45 \\[3pt]
    $T = 100$        &       &       &       &       &  \\[3pt]
    QPI-f      & 0.9484 & 0.9406 & 0.9392 & 0.9452 & 0.9418 & 3.87 & 4.29 & 4.39 & 4.41 & 4.42  \\
    QPI-p         & 0.9498 & 0.9450 & 0.9406 & 0.9480 & 0.9450 & 3.92 & 4.34 & 4.44 & 4.46 & 4.46  \\
    $L_2$-PPI-f       & 0.9512 & 0.9436 & 0.9418 & 0.9488 & 0.9456 & 3.90 & 4.33 & 4.44 & 4.47 & 4.48 \\
    $L_2$-PPI-p         & 0.9526 & 0.9468 & 0.9436 & 0.9480 & 0.9470 & 3.94 & 4.38 & 4.48 & 4.52 & 4.53 \\
    $L_1$-PPI-f        & 0.9524 & 0.9440 & 0.9420 & 0.9476 & 0.9454 & 3.90 & 4.33 & 4.44 & 4.47 & 4.48 \\
    $L_1$-PPI-p       & 0.9530 & 0.9470 & 0.9438 & 0.9488 & 0.9476 & 3.94 & 4.38 & 4.48 & 4.52 & 4.53 \\
    SPI        & 0.9562 & 0.9514 & 0.9470 & 0.9512 & 0.9496 & 3.90 & 4.33 & 4.42 & 4.45 & 4.45  \\[3pt]
    $T = 50$        &       &       &       &       &  \\[3pt]
    QPI-f        & 0.9246 & 0.9314 & 0.9326 & 0.9342 & 0.9376 & 3.79 & 4.27 & 4.36 & 4.38 & 4.39 \\
    QPI-p        & 0.9300 & 0.9390 & 0.9390 & 0.9394 & 0.9432 & 3.88 & 4.36 & 4.45 & 4.48 & 4.48 \\
    $L_2$-PPI-f      & 0.9336 & 0.9366 & 0.9398 & 0.9404 & 0.9452 & 3.89 & 4.34 & 4.46 & 4.50 & 4.51\\
    $L_2$-PPI-p       & 0.9362 & 0.9418 & 0.9426 & 0.9444 & 0.9492 & 3.96 & 4.43 & 4.55 & 4.59 & 4.60 \\
    $L_1$-PPI-f        & 0.9332 & 0.9374 & 0.9392 & 0.9398 & 0.9444 & 3.89 & 4.35 & 4.46 & 4.50 & 4.51 \\
    $L_1$-PPI-p      & 0.9364 & 0.9426 & 0.9432 & 0.9436 & 0.9480 & 3.97 & 4.44 & 4.55 & 4.59 & 4.60 \\
    SPI        & 0.9508 & 0.9498 & 0.9480 & 0.9454 & 0.9516 & 3.90 & 4.33 & 4.43 & 4.45 & 4.45\\
       \bottomrule
    \end{tabular}\\
\end{table}
\begin{table}[htbp]
\centering
  \caption{The CVR and LEN of PIs for Model 6}
  \vspace{2pt}
  \label{Tab:10}
\begin{tabular}{lcccccccccc}
  \toprule 
 Model 6: & \multicolumn{10}{c}{ $ X_t = \log(10 + 5\cdot e^{0.9\cdot X_{t-1}}) + \epsilon_t$} \\
 \midrule
   & \multicolumn{5}{c}{CVR for each step} & \multicolumn{5}{c}{LEN for each step}\\
    $T = 400$  & 1     & 2     & 3     & 4     & 5 & 1     & 2     & 3     & 4     & 5 \\[3pt]
    QPI-f       & 0.9506 & 0.9452 & 0.9440 & 0.9420 & 0.9388 & 3.88 & 5.18 & 6.01 & 6.60 & 7.03\\
    QPI-p        & 0.9528 & 0.9472 & 0.9454 & 0.9414 & 0.9378 & 3.90 & 5.22 & 6.05 & 6.64 & 7.07\\
    $L_2$-PPI-f      & 0.9532 & 0.9488 & 0.9476 & 0.9434 & 0.9418 & 3.90 & 5.22 & 6.07 & 6.67 & 7.11\\
    $L_2$-PPI-p       & 0.9506 & 0.9500 & 0.9478 & 0.9456 & 0.9412 & 3.92 & 5.26 & 6.12 & 6.71 & 7.16\\
    $L_1$-PPI-f  & 0.9536 & 0.9488 & 0.9470 & 0.9442 & 0.9424 & 3.90 & 5.22 & 6.07 & 6.67 & 7.12 \\
    $L_1$-PPI-p     & 0.9514 & 0.9510 & 0.9482 & 0.9456 & 0.9424 & 3.93 & 5.26 & 6.12 & 6.72 & 7.17\\
    SPI       & 0.9532 & 0.9508 & 0.9508 & 0.9490 & 0.9498 & 3.90 & 5.25 & 6.13 & 6.76 & 7.23 \\[3pt]
    $T = 100$        &       &       &       &       &  \\[3pt]
    QPI-f   &0.9350 & 0.9244 & 0.9216 & 0.9108 & 0.9038 & 3.83 & 5.02 & 5.77 & 6.28 & 6.67 \\
    QPI-p  &0.9404 & 0.9302 & 0.9298 & 0.9176 & 0.9116 & 3.94 & 5.17 & 5.93 & 6.46 & 6.86\\
    $L_2$-PPI-f  &0.9422 & 0.9332 & 0.9330 & 0.9224 & 0.9136  & 3.94 & 5.23 & 6.04 & 6.62 & 7.06\\
    $L_2$-PPI-p  &0.9498 & 0.9388 & 0.9392 & 0.9324 & 0.9218 & 4.05 & 5.37 & 6.21 & 6.81 & 7.25\\
    $L_1$-PPI-f  &0.9424 & 0.9340 & 0.9340 & 0.9224 & 0.9146  & 3.95 & 5.23 & 6.04 & 6.63 & 7.07 \\
    $L_1$-PPI-p  &0.9498 & 0.9400 & 0.9384 & 0.9318 & 0.9214 & 4.05 & 5.38 & 6.21 & 6.81 & 7.26\\
    SPI  & 0.9498 & 0.9496 & 0.9504 & 0.9458 & 0.9494  & 3.90 & 5.25 & 6.13 & 6.76 & 7.23\\[3pt]
    $T = 50$        &       &       &       &       &  \\[3pt]
    QPI-f      & 0.9056 & 0.8930 & 0.8796 & 0.8640 & 0.8526 & 3.72 & 4.85 & 5.50 & 5.97 & 6.34\\
    QPI-p       & 0.9200 & 0.9102 & 0.8934 & 0.8872 & 0.8716 & 3.93 & 5.13 & 5.83 & 6.33 & 6.71\\
    $L_2$-PPI-f     & 0.9276 & 0.9172 & 0.9032 & 0.8984 & 0.8860 & 4.04 & 5.32 & 6.17 & 6.81 & 7.36\\
    $L_2$-PPI-p      & 0.9412 & 0.9302 & 0.9188 & 0.9160 & 0.9042 & 4.27 & 5.62 & 6.50 & 7.16 & 7.72\\
    $L_1$-PPI-f   & 0.9290 & 0.9170 & 0.9034 & 0.8982 & 0.8856 & 4.04 & 5.33 & 6.17 & 6.81 & 7.36\\
    $L_1$-PPI-p     & 0.9412 & 0.9300 & 0.9192 & 0.9166 & 0.9034  & 4.27 & 5.62 & 6.50 & 7.16 & 7.72\\
    SPI       & 0.9508 & 0.9460 & 0.9432 & 0.9484 & 0.9472& 3.90 & 5.25 & 6.13 & 6.75 & 7.23 \\
       \bottomrule
    \end{tabular}\\
\end{table}
\begin{table}[htbp]
\centering
  \caption{The CVR and LEN of PIs for Model 7}
  \vspace{2pt}
  \label{Tab:11}
\begin{tabular}{lcccccccccc}
  \toprule 
 Model 7: & \multicolumn{10}{c}{$ X_t = \log(4\cdot e^{0.9\cdot X_{t-2}} + 5\cdot e^{0.9\cdot X_{t-1}} + 6\cdot e^{0.9\cdot X_{t-3}}) + \epsilon_t$} \\
 \midrule
    & \multicolumn{5}{c}{CVR for each step} & \multicolumn{5}{c}{LEN for each step}\\
    $T = 400$  & 1     & 2     & 3     & 4     & 5 & 1     & 2     & 3     & 4     & 5\\[3pt]
    QPI-f       & 0.9450 & 0.9426 & 0.9442 & 0.9384 & 0.9376 & 3.88 & 4.02 & 4.28 & 4.70 & 4.87\\
    QPI-p        & 0.9484 & 0.9444 & 0.9470 & 0.9410 & 0.9400 & 3.92 & 4.06 & 4.33 & 4.75 & 4.92 \\
    $L_2$-PPI-f       & 0.9472 & 0.9434 & 0.9476 & 0.9394 & 0.9406 & 3.91 & 4.05 & 4.31 & 4.75 & 4.93\\
    $L_2$-PPI-p       & 0.9498 & 0.9460 & 0.9498 & 0.9442 & 0.9428 & 3.94 & 4.09 & 4.36 & 4.80 & 4.98 \\
    $L_1$-PPI-f   & 0.9462 & 0.9430 & 0.9474 & 0.9408 & 0.9406 & 3.91 & 4.05 & 4.31 & 4.75 & 4.93\\
    $L_1$-PPI-p      & 0.9484 & 0.9466 & 0.9486 & 0.9444 & 0.9434 & 3.95 & 4.09 & 4.36 & 4.80 & 4.98\\
    SPI        & 0.9462 & 0.9452 & 0.9480 & 0.9446 & 0.9432 & 3.90 & 4.04 & 4.31 & 4.76 & 4.94\\[3pt]
    $T = 100$        &       &       &       &       &  \\[3pt]
    QPI-f       & 0.9330 & 0.9404 & 0.9298 & 0.9238 & 0.9278 & 3.82 & 3.96 & 4.19 & 4.56 & 4.70\\
    QPI-p        & 0.9436 & 0.9496 & 0.9440 & 0.9354 & 0.9390 & 3.99 & 4.14 & 4.38 & 4.76 & 4.91\\
    $L_2$-PPI-f       & 0.9400 & 0.9464 & 0.9404 & 0.9332 & 0.9408 & 3.94 & 4.09 & 4.35 & 4.77 & 4.95\\
    $L_2$-PPI-p      & 0.9498 & 0.9536 & 0.9504 & 0.9430 & 0.9508 & 4.10 & 4.26 & 4.53 & 4.98 & 5.16\\
    $L_1$-PPI-f   & 0.9396 & 0.9466 & 0.9404 & 0.9312 & 0.9406 & 3.94 & 4.09 & 4.35 & 4.77 & 4.95 \\
    $L_1$-PPI-p     & 0.9504 & 0.9548 & 0.9508 & 0.9422 & 0.9504 & 4.10 & 4.26 & 4.54 & 4.98 & 5.16\\
    SPI        & 0.9502 & 0.9542 & 0.9484 & 0.9468 & 0.9550 & 3.90 & 4.04 & 4.31 & 4.76 & 4.94\\[3pt]
    $T = 50$        &       &       &       &       &  \\[3pt]
    QPI-f        & 0.9152 & 0.9132 & 0.9202 & 0.9044 & 0.9020 & 3.71 & 3.89 & 4.12 & 4.46 & 4.60 \\
    QPI-p       & 0.9366 & 0.9402 & 0.9428 & 0.9340 & 0.9284 & 4.04 & 4.25 & 4.50 & 4.87 & 5.03 \\
    $L_2$-PPI-f       & 0.9344 & 0.9312 & 0.9366 & 0.9254 & 0.9236 & 3.97 & 4.13 & 4.42 & 4.88 & 5.08\\
    $L_2$-PPI-p      & 0.9518 & 0.9506 & 0.9570 & 0.9460 & 0.9432 & 4.31 & 4.49 & 4.82 & 5.31 & 5.54 \\
    $L_1$-PPI-f   & 0.9340 & 0.9310 & 0.9360 & 0.9264 & 0.9248 & 3.97 & 4.13 & 4.42 & 4.88 & 5.08  \\
    $L_1$-PPI-p    & 0.9528 & 0.9494 & 0.9554 & 0.9450 & 0.9434 & 4.31 & 4.50 & 4.82 & 5.32 & 5.54 \\
    SPI       & 0.9442 & 0.9464 & 0.9520 & 0.9486 & 0.9508 & 3.90 & 4.05 & 4.31 & 4.76 & 4.94\\
       \bottomrule
    \end{tabular}\\
\end{table}

\FloatBarrier

\clearpage
\section{Conclusions and discussions}\label{Sec:6}
In the paper at hand, we first propose comprehensive algorithms which serve for optimal multi-step ahead point predictions and prediction intervals with general NLAR models. Depending on whether the model information and the innovation distribution are known to us, we can rely on simulation- or bootstrap-based methods to perform predictions. 

Theoretically, we analyzed prediction inference with a specific form of NLAR model which possesses separate mean and variance functions. When we know the model and innovation information, we show that the simulation-based approach can return consistent predictions. When we only know the form of parametric NLAR models, the bootstrap-based prediction is also shown to be consistent to true optimal future values. Moreover, we can obtain asymptotically valid or pertinent prediction intervals. 

In addition, we propose the idea of combining predictive residuals with the bootstrap-based NLAR prediction. The simulation studies verify the efficiency of our methods. Constructing pertinent prediction intervals with predictive residuals can improve the empirical CVR, especially for short data.


\section*{Acknowledgements}
 The authors are thankful to Professor Yunyi Zhang for his valuable suggestions. 
The research of the first author was partially supported by the Richard Libby Graduate Research Award. The research of the second author was partially supported by NSF grant DMS 19-14556.

\clearpage
\appendix

\section*{\textsc{Appendix: Proofs}}\label{Appendix:Proof}
\begin{proof}[\textbf{\textsc{Proof of Theorem 3.1}}]
Denote the exactly $L_2$ optimal predictor of $X_{T+h}$ by $X_{T+h}^{L_2}$. We know it is a conditional mean given observed data:
\begin{equation}
    X_{T+h}^{L_2} = \mathbb{E}(X_{T+h}|X_T,\ldots,X_{T-p+1}).
\end{equation}
Due to the strong stationary property, we can rewrite \cref{Eq15} as $X_{T+h} = f(\bm{Y},\bm{\epsilon_{h}})$, where $\bm{Y}$ and $\bm{\epsilon_{h}}$ represent $\{X_{t}\}_{t = 1}^{p}$ and $\{\epsilon_t\}_{t=T+1}^{T+h}$, respectively. Moreover, by the causality assumption, we get $\bm{Y}$ and $\bm{\epsilon_{h}}$ are independent. In addition, $\{\bm{\epsilon_h}^{(i)}\}_{i=1}^{M}$ are also $i.i.d.$. Thus, $\{X_{T+h}^{(i)}\}_{i=1}^{M}$ are conditionally $i.i.d.$ given $\bm{Y}$. Based on Theorem 4.2 of \cite{majerek2005conditional}, the conditional version of the strong law of large numbers implies that:
\begin{equation}
    \widehat{X^{L_2}_{T+h}} = \frac{1}{M}\sum_{i = 1}^{M}X_{T+h}^{(i)} \overset{a.s.}{\to} X_{T+h}^{L_2}, \text{assuming that~} X_{T+h}^{L_2}~ \text{exits}.
\end{equation}
The existence of $X_{T+h}^{L_2}$ is guaranteed by assumptions A2 and A5 - A7. This proof can be directly extended to the NLAR model with heteroscedastic errors, since the relationship $X_{T+h} = f(\bm{Y},\bm{\epsilon_{h}})$ is also satisfied, so $\{X_{T+h}^{(i)}\}_{i=1}^{M}$ are still conditionally $i.i.d.$ given $\bm{Y}$. Therefore, without changing other parts of this proof, we can show the analogous theorem for NLAR models with heteroscedastic error cases. 
\end{proof}
\vspace{5pt}

\begin{proof}[\textbf{\textsc{Proof of Theorem 3.2}}]
We actually want to show that under the ergodic property:
\begin{equation}
    \widehat{X^{L_1}_{T+h}} \overset{p}{\to}X_{T+h}^{L_1}, \text{as~} M~\text{converges to infinity.} 
\end{equation}
We can write $\widehat{X^{L_1}_{T+h}}$ and $X_{T+h}^{L_1}$ as:
\begin{equation}
    \widehat{X^{L_1}_{T+h}} = H_{M}(\bm{Y},\bm{\mathcal{E}_h}) = \text{Median}(f(\bm{Y},\bm{\epsilon_h}^{(1)}),\ldots,f(\bm{Y},\bm{\epsilon_h}^{(M)}))~;~X_{T+h}^{L_1} = H(\bm{Y},\bm{\epsilon_h}) =  Q_{X_{T+h}|\bm{X}_{T}}(1/2),
\end{equation}
where $\bm{Y}$ and $\bm{\epsilon_{h}}$ represent $\{X_{t}\}_{t = 1}^{p}$ and $\{\epsilon_t\}_{t=T+1}^{T+h}$, respectively; $\bm{\mathcal{E}_h}$ represents the whole set $\{\bm{\epsilon_h}^{(i)}\}_{i=1}^{M}$. $Q_{X_{T+h}|\bm{X}_{T}}$ is the conditional quantile function of $X_{T+h}$. By assumption A8, $f(\cdot,\cdot)$ is also uniformly continuous in $\bm{x}$ since it is a composition of uniformly continuous functions. Thus, for a given $\eta>0$, there exists a constant $\delta>0$ such that:
\begin{equation}\label{Eq19}
    |f(\bm{y_1},\bm{\epsilon_{h}}) - f(\bm{y_2},\bm{\epsilon_{h}})| < \eta,~\text{when}~||\bm{y_1}-\bm{y_2}||\leq\delta,
\end{equation}
where $||\cdot||$ is any norm equivalent to the Euclidean norm. 

Then, we can split the $p$-dimensional ball $B(D) = \{\bm{y},||\bm{y}||\leq D\}$ into some disjoint subsets $S_{j},~j = 1,\ldots,k$. Let $||\bm{y}-\bm{s_{j}}||<\delta$ for $\bm{s_{j}}\in S_{j}$ and any point $\bm{y}\in B(D)$ such that:
\begin{equation}
    |f(\bm{y},\bm{\epsilon_{h}}) - f(\bm{s_i},\bm{\epsilon_{h}})| < \eta, \forall \eta > 0.
\end{equation}
Thus, $|f(\bm{y},\bm{\epsilon_{h}}^{(i)}) - f(\bm{s_j},\bm{\epsilon_{h}}^{(i)})|<\eta$, $\forall i\in{1,\ldots,M}$. It is possible to fix a small enough $\eta$ to make sure that the order of $\{f(\bm{y},\bm{\epsilon_h}^{(1)}),\ldots,f(\bm{y},\bm{\epsilon_h}^{(M)}) \}$ is same with the order of $\{f(\bm{s_j},\bm{\epsilon_h}^{(1)}),\ldots,f(\bm{s_j},\bm{\epsilon_h}^{(M)}) \}$. In other words, we have:
\begin{equation}\label{Eq47}
    \sum_{j=1}^{k}I_{j}|H_{M}(\bm{Y},\bm{\mathcal{E}_h}) - H_{M}(\bm{s_{j}},\bm{\mathcal{E}_h})|<\eta,
\end{equation}
where $I_{j}$ represents the indicator function $I(\bm{Y}\in S_{j}),~j=1,\ldots,k.$ In addition, we also have:
\begin{equation}\label{Eq48}
    \sum_{j=1}^{k}I_{j}|H(\bm{Y},\bm{\epsilon_h)} - H(\bm{s_{j}},\bm{\epsilon_h})|<\eta.
\end{equation}
Therefore, define $I_{0} := I(\bm{Y}\notin B(D))$, by combining \cref{Eq47,Eq48}, we can get:
\begin{equation}
\begin{split}
    &|H_{M}(\bm{Y},\bm{\mathcal{E}_h}) - H(\bm{Y},\bm{\epsilon_h})|\\
    &= \sum_{j=1}^{k}I_j \left|H_{M}(\bm{Y},\bm{\mathcal{E}_h}) - H_{M}(\bm{s_j},\bm{\mathcal{E}_h}) + H_{M}(\bm{s_j},\bm{\mathcal{E}_h}) -  H(\bm{s_j},\bm{\epsilon_h}) +  H(\bm{s_j},\bm{\epsilon_h}) -  H(\bm{Y},\bm{\epsilon_h})\right| + \\
    &I_0 |H_{M}(\bm{Y},\bm{\mathcal{E}_h}) - H_{M}(\bm{s_0},\bm{\mathcal{E}_h}) + H_{M}(\bm{s_0},\bm{\mathcal{E}_h}) -  H(\bm{s_0},\bm{\epsilon_h}) +  H(\bm{s_0},\bm{\epsilon_h}) -  H(\bm{Y},\bm{\epsilon_h})|
    \\
    & \leq \sum_{j=1}^{k}I_{j}\left(|H_{M}(\bm{Y},\bm{\mathcal{E}_h}) - H_{M}(\bm{s_j},\bm{\mathcal{E}_h})| + |H_{M}(\bm{s_j},\bm{\mathcal{E}_h}) -  H(\bm{s_j},\bm{\epsilon_h})| + |H(\bm{s_j},\bm{\epsilon_h}) -  H(\bm{Y},\bm{\epsilon_h})|\right) + I_0\cdot C \\
    &\leq 2\eta + I_0\cdot C + \sum_{j=1}^{k}I_{j}|H_{M}(\bm{s_j},\bm{\mathcal{E}_h}) -  H(\bm{s_j},\bm{\epsilon_h})|.
\end{split}
\end{equation}
Comparing $H_{M}(\bm{s_j},\bm{\mathcal{E}_h})$ and $H(\bm{s_j},\bm{\epsilon_h})$, where $\bm{s_j}$ is a fixed point, by applying the CLT on the sample median for the ergodic series, we can get $H_{M}(\bm{s_j},\bm{\mathcal{E}_h})$ converges to $H(\bm{s_j},\bm{\epsilon_h})$ in probability. The non-zero property of the probability density of $X_{n+h}$ at the median is guaranteed by the everywhere positive density function of innovation. Thus, we have:
\begin{equation}
    \mathbb{P}(|H_{M}(\bm{Y},\bm{\mathcal{E}_h}) - H(\bm{Y},\bm{\epsilon_h})| \leq 2\eta + I_0\cdot C)\overset{}{\to}1.
\end{equation}
Besides, $D$ can be arbitrarily large. Also, $\eta$ can be arbitrarily small. Finally, we get:
\begin{equation}
    \mathbb{P}(|H_{M}(\bm{Y},\bm{\mathcal{E}_h}) - H(\bm{Y},\bm{\epsilon_h})|\leq \varepsilon)\overset{}{\to} 1.
\end{equation}
The above proof can be extended to other quantile estimators. Thus, we can build asymptotically valid QPI with any CVR. For extending such proof to NLAR models with heteroscedastic errors, we need that the variance function is also uniformly continuous. Then, $f(\cdot,\cdot)$ is still uniformly continuous in $\bm{x}$. Therefore, without changing other parts of this proof, we can show the analogous theorem for NLAR models with heteroscedastic error cases. 
\end{proof}

\begin{proof}[\textbf{\textsc{Proof of Lemma 4.1}}]
For simplifying the notation, we just consider the NLAR model in \cref{Eq16} with order 1 and homoscedastic errors. The case with higher order and heteroscedastic errors can be proved similarly. 

The proof is inspired by the work of \cite{boldin1983estimation}. For connecting $F_{\epsilon}(x)$ and $\widehat{F}_{\epsilon}(x)$, we should notice that the empirical distribution $F_{T}(x)$ based on $\{\epsilon_{i}\}_{i=1}^{T}$ can be a bridge, i.e., we have:
\begin{equation}
\begin{split}
    \sup_{x}|\widehat{F}_{\epsilon}(x) - F_{\epsilon}(x)|&= \sup_{x}|\widehat{F}_{\epsilon}(x) - F_{T}(x) + F_{T}(x) -  F_{\epsilon}(x)|\\
    &\leq \sup_{x}|\widehat{F}_{\epsilon}(x) - F_{T}(x)| + \sup_{x}|F_{T}(x) - F_{\epsilon}(x)|.
\end{split}
\end{equation}
From the Glivenko–Cantelli theorem, we know $\sup_{x}|F_{T}(x) - F_{\epsilon}(x)|$ converges to 0 a.s.. Thus, we only need to show:
\begin{equation}
    \sup_{x}|\widehat{F}_{\epsilon}(x) - F_{T}(x)|\overset{p}{\to} 0.
\end{equation}
First, we know
\begin{equation}
    \text{if}~\Delta_i(x) = \begin{cases}
    1, & \epsilon_{i}\leq x\\
    0, & \epsilon_i> x
    \end{cases}~\text{then,}~F_{T}(x) = \frac{1}{T}\sum_{i=1}^{T}\Delta_{i}(x).
\end{equation}
Thus, we can get:
\begin{equation}\label{Eq55}
    \widehat{F}_{\epsilon}(x) = \frac{1}{T}\sum_{i=1}^{T}\Delta_i(x+\phi(X_{i-1},\widehat{\theta}_1)-\phi(X_{i-1},\theta_1)),
\end{equation}
since $\hat{\epsilon}_{i} = \phi(X_{i-1},\theta_1) - \phi(X_{i-1},\widehat{\theta}_1) + \epsilon_i$. For handling the randomness of $\phi(X_{i-1},\widehat{\theta}_1)-\phi(X_{i-1},\theta_1)$ inside $\Delta_{i}(\cdot)$ of \cref{Eq55}, we use nonrandom $\eta_{T},T=1,2,3,\ldots,$ to replace $\phi(X_{i-1},\widehat{\theta}_1)-\phi(X_{i-1},\theta_1)$. Then, we can consider the process:
\begin{equation}
    z_{T}(x,\eta_{T}) = \widehat{F}_{\epsilon}(x) - F_{T}(x) = \frac{1}{T}\sum_{i=1}^{T}\left( \Delta_{i}(x+\eta_{T}) - \Delta_{i}(x) \right).
\end{equation}
Indeed, we have:
\begin{equation}\label{eq23}
    \mathbb{P}(\sup_{x}|\widehat{F}_{\epsilon}(x) - F_{T}(x)|>\epsilon)\leq \mathbb{P}(\sup_{x}\sup_{|\eta_T|\leq T^{-\lambda}}|z_{n}(x,\eta_{T})|>\epsilon) + \mathbb{P}(|\phi(X_{i-1},\widehat{\theta}_1)-\phi(X_{i-1},\theta_1)|>T^{-\lambda}).
\end{equation}
Without loss of generality, we select an appropriate $\lambda$ to make sure the second term on the right-hand side of the above inequality converges to 0 under A9 and A10.

Then, we shall show $\mathbb{P}(\sup_{x}\sup_{|\eta_T|\leq T^{-\lambda}}|z_{T}(x,\eta_{T})|>\epsilon)$ also converges to 0. Since this term depends on the continuum of values of $x$, we can partition the real axis into $N_{T} \sim T^{1/2}$ parts by points:
\begin{equation}
    -\infty = x_{0}<x_{1}<\cdots x_{k}< \cdots<x_{N_{T}-1}<x_{N_T} = \infty,~\text{where}~F_{\epsilon}(x_k) = kN_{T}^{-1}.
\end{equation}
Hence, for $x_{r}$ and $x_{r+1}$ such that $x_{r}\leq x \leq x_{r+1}$, we have:
\begin{equation}
    x_{r} + \eta_{T} \leq x + \eta_{T} \leq x_{r+1} + \eta_{T}.
\end{equation}
In addition, since $\Delta_i(x)$ is monotonic, we obtain:
\begin{equation}
\begin{split}
     z_{T}(x,\eta_{T}) &\geq z_{T}(x_r,\eta_{T}) + \frac{1}{T}\sum_{i=1}^{T}\Delta_i(x_r) - \frac{1}{T}\sum_{i=1}^{T}\Delta_i(x_{r+1}); \\
     z_{T}(x,\eta_{T}) &\leq z_{T}(x_{r+1},\eta_{T}) + \frac{1}{T}\sum_{i=1}^{T}\Delta_i(x_{r+1}) - \frac{1}{T}\sum_{i=1}^{T}\Delta_i(x_{r}).
\end{split}
\end{equation}
Therefore, we have:
\begin{equation}\label{Eq61}
\begin{split}
    \sup_{x}\sup_{|\eta_T|\leq T^{-\lambda}}&|z_{T}(x,\eta_{T})|\\ 
    &\leq \sup_{k \leq N_{T}-1}\sup_{|\eta_T|\leq T^{-\lambda}}|z_{T}(x_{k+1},\eta_{T})| \\
    & + \sup_{k \leq N_{T}}\sup_{|\eta_T|\leq T^{-\lambda}}|z_{T}(x_{k},\eta_{T})|\\
    & + \sup_{|t_1 - t_2|\leq N_{T}^{-1}}\frac{1}{T}\left |\sum_{i=1}^{T}\left(\Delta_{i}(F_{\epsilon}^{-1}(t_1)) -  \Delta_{i}(F_{\epsilon}^{-1}(t_2)) \right)\right |.
\end{split}
\end{equation}
For the last term on the r.h.s. of \cref{Eq61}:
\begin{equation}
\begin{split}
    &\sup_{|t_1 - t_2|\leq N_{T}^{-1}}\frac{1}{T}\left |\sum_{i=1}^{T}\left(\Delta_{i}(F_{\epsilon}^{-1}(t_1)) -  \Delta_{i}(F_{\epsilon}^{-1}(t_2)) \right)\right | \\
    &= \sup_{|t_1 - t_2|\leq N_{T}^{-1}}\frac{1}{T}\left |\sum_{i=1}^{T}\left(\Delta_{i}(F_{\epsilon}^{-1}(t_1)) - t_1 +t_1 -t_2 + t_2 -  \Delta_{i}(F_{\epsilon}^{-1}(t_2)) \right)\right |\\
    & \leq \sup_{t_1, s.t. |t_1 - t_2|\leq N_{T}^{-1}}\left|\frac{1}{T}\sum_{i=1}^{T}\Delta_{i}(F_{\epsilon}^{-1}(t_1)) - t_1 \right| + \sup_{|t_1 - t_2|\leq N_{T}^{-1}}\left| t_1 -t_2  \right| +  \sup_{t_2, s.t.|t_1 - t_2|\leq N_{T}^{-1}}\left| t_2  - \frac{1}{T}\sum_{i=1}^{T} \Delta_{i}(F_{\epsilon}^{-1}(t_2))    \right|. 
\end{split}
\end{equation}
By the Glivenko–Cantelli theorem, it is obvious that $\sup_{|t_1 - t_2|\leq N_{T}^{-1}}\frac{1}{T}\left |\sum_{i=1}^{T}\left(\Delta_{i}(F_{\epsilon}^{-1}(t_1)) -  \Delta_{i}(F_{\epsilon}^{-1}(t_2)) \right)\right |$ is $o_p(1)$. Next, we consider the second term of the r.h.s of \cref{Eq61}:
\begin{equation}\label{Eq63}
\begin{split}
    &\sup_{k \leq N_{T}}\sup_{|\eta_T|\leq T^{-\lambda}}|z_{T}(x_{k},\eta_{T})|\\
    &=  \sup_{k \leq N_{T}}\sup_{|\eta_T|\leq T^{-\lambda}}\left|  \frac{1}{T}\sum_{i=1}^{T}\left( \Delta_{i}(x_{k}+\eta_{T}) - \Delta_{i}(x_k) \right) \right| \\
    &=  \sup_{k \leq N_{T}}\sup_{|\eta_T|\leq T^{-\lambda}}\left|  \frac{1}{T}\sum_{i=1}^{T} \Delta_{i}(x_{k}+\eta_{T}) - F_{\epsilon}(x_{k}+\eta_{T}) + F_{\epsilon}(x_{k}+\eta_{T}) - F_{\epsilon}(x_{k}) +  F_{\epsilon}(x_{k}) -  \frac{1}{T}\sum_{i=1}^{T} \Delta_{i}(x_{k})    \right|\\
    &\leq \sup_{k \leq N_{T}}\sup_{|\eta_T|\leq T^{-\lambda}}\left\{  \left|  \frac{1}{T}\sum_{i=1}^{T} \Delta_{i}(x_{k}+\eta_{T}) - F_{\epsilon}(x_{k}+\eta_{T}) \right| +\left| F_{\epsilon}(x_{k}+\eta_{T}) - F_{\epsilon}(x_{k}) \right| + \left| F_{\epsilon}(x_{k}) -  \frac{1}{T}\sum_{i=1}^{T} \Delta_{i}(x_{k})  \right| \right\}.
\end{split}
\end{equation}
Applying the Glivenko–Cantelli theorem again, we can find the first and third term in the r.h.s. of \cref{Eq63} converges to 0 a.s.. For the middle term:
\begin{equation}
\begin{split}
    &\sup_{k \leq N_{T}}\sup_{|\eta_T|\leq T^{-\lambda}}\left| F_{\epsilon}(x_{k}+\eta_{T}) - F_{\epsilon}(x_{k}) \right|\\
    &=  \sup_{k \leq N_{T}}\sup_{|\eta_T|\leq T^{-\lambda}}\left| F_{\epsilon}(x_k) + F_{\epsilon}^{\prime}(\Tilde{x})\eta_T - F_{\epsilon}(x_k)  \right| \\
    & \leq \sup_{x}\sup_{|\eta_T|\leq T^{-\lambda}}\left|F_{\epsilon}^{\prime}(\Tilde{x})\eta_T  \right| \overset{}{\to} 0,~\text{Under A11.}
\end{split}
\end{equation}
We can do a similar analysis for the first term of the r.h.s of \cref{Eq61}. Combining all parts together, we prove \cref{Eq17}. The proof of \cref{Eq16} with heteroscedastic errors is similar.
\end{proof}

\begin{proof}[\textbf{\textsc{Proof of Theorem 4.1}}]
For simplifying the notations, we consider the NLAR model with order 1 and homoscedastic errors. For NLAR models with higher order and heteroscedastic errors, the proof can be written similarly. We show the proof of $h=2$ as an example. The proof of higher steps or one-step prediction can be written similarly.

First, by the tower property, we can show that $F_{X_{T+2}|X_T}(x)$ is equivalent to:
\begin{equation}\label{truedis}
\begin{split}
    F_{X_{T+2}|X_T}(x) &=  \mathbb{P}(X_{T+2}\leq x|X_T)\\
      &=  \mathbb{P}(\phi(X_{T+1},\theta_1) + \epsilon_{T+2}\leq x|X_T)\\
    & = \mathbb{P}\left(\epsilon_{T+2}\leq x - \phi(\phi(X_{T},\theta_1)+\epsilon_{T+1},\theta_1)\bigg\vert X_T\right)\\
    & = \mathbb{E}\left[\mathbb{P}\left(\epsilon_{T+2}\leq x - \phi(\phi(X_{T},\theta_1)+\epsilon_{T+1},\theta_1)\bigg\vert\epsilon_{T+1},X_T\right)\bigg\vert X_T\right]\\
    & = \mathbb{E}\left[F_{\epsilon}\left(x - \phi(\phi(X_{T},\theta_1)+\epsilon_{T+1},\theta_1)\right)\bigg\vert X_T\right]\\
    & = \mathbb{E}\left[F_{\epsilon}\left( \mathcal{L}(x,X_T,\epsilon_{T+1})   \right)\bigg\vert X_T\right];
\end{split}
\end{equation}
we use $\mathcal{L}(x,X_T,\epsilon_{T+1}) $ to represent $x - \phi(\phi(X_{T},\theta_1)+\epsilon_{T+1},\theta_1)$ to simplify notations. Similarly, we can analyze $F_{X^*_{T+2}|X_T,\ldots,X_0}(x)$, it has below equivalent expressions:
\begin{equation}
\begin{split}
    F_{X^*_{T+2}|X_T,\ldots,X_0}(x) &=  \mathbb{P}(X^*_{T+2}\leq x|X_T,\ldots,X_0)\\
    & = \mathbb{E}\left[\mathbb{P}\left(\hat{\epsilon}^*_{T+2}\leq \widehat{\mathcal{L}}(x, X_T,\hat{\epsilon}^*_{T+1})\bigg\vert \hat{\epsilon}^*_{T+1},X_T,\ldots,X_0\right)\bigg\vert X_T,\ldots,X_0\right]\\
    & = \mathbb{E}^*\left[\widehat{F}_{\epsilon}\left(\widehat{\mathcal{L}}(x, X_T,\hat{\epsilon}^*_{T+1})\right)\right],
\end{split}
\end{equation}
 where $\widehat{\mathcal{L}}(x, X_T,\hat{\epsilon}^*_{T+1})$ represents $x - \phi(\phi(X_{T},\widehat{\theta}_1)+\hat{\epsilon}^*_{T+1},\widehat{\theta}_1)$ and $\mathbb{E}^*(\cdot)$ represents the expectation in the bootstrap world, i.e., $\mathbb{E}(\cdot|X_T,\ldots,X_0)$. Thus, we hope to show:
 \begin{equation}\label{equivatheorem}
     \sup_{|x|\leq c_T}\bigg\vert \mathbb{E}^*\left[\widehat{F}_{\epsilon}(\widehat{\mathcal{L}}(x, X_T,\hat{\epsilon}^*_{T+1})) \right] - \mathbb{E}\left[F_{\epsilon}\left( \mathcal{L}(x,X_T,\epsilon_{T+1})   \right)\bigg\vert X_T\right] \bigg\vert \overset{p}{\to} 0. 
 \end{equation}
From here, we first bound the region of $X_t$ by Lemma 1 of \cite{franke2004bootstrapping} under A1--A7:
 \begin{equation}\label{xTregion}
    \mathbb{P}(|X_T|>\gamma_T)\to 0,  
 \end{equation}
where $\{\gamma_T\}$ is a sequence of sets, such that $\gamma_{1}\subseteq\cdots\subseteq\gamma_{T}\subseteq\gamma_{T+1}\subseteq\cdots$ with the form $\gamma_{T} = [-T^c,T^c]$; $c$ is some appropriate constant. For deriving such a result for a time series model with heteroscedastic errors, we need the additional assumption of variance function in A1; then the proof is referred to Lemma 1 of \cite{franke2004bootstrapping}. In addition, we have a relationship:
 \begin{equation}\label{decompose}
 \begin{split}
     &\mathbb{P}\left( \sup_{|x|\leq c_T}\bigg\vert \mathbb{E}^*\left[\widehat{F}_{\epsilon}(\widehat{\mathcal{L}}(x, X_T,\hat{\epsilon}^*_{T+1})) \right] - \mathbb{E}\left[F_{\epsilon}\left( \mathcal{L}(x,X_T,\epsilon_{T+1})   \right)\bigg\vert X_T\right] \bigg\vert > \varepsilon\right) \\
     &\leq\mathbb{P}(|X_T|> \gamma_T) + \mathbb{P}\left((|X_T|\leq \gamma_T )\bigcap \left(\sup_{|x|\leq c_T}\bigg\vert \mathbb{E}^*\left[\widehat{F}_{\epsilon}(\widehat{\mathcal{L}}(x, X_T,\hat{\epsilon}^*_{T+1})) \right] - \mathbb{E}\left[F_{\epsilon}\left( \mathcal{L}(x,X_T,\epsilon_{T+1})   \right)\bigg\vert X_T\right] \bigg\vert>\varepsilon\right) \right).
 \end{split}
 \end{equation}
Thus, to verify \cref{equivatheorem}, we just need to show that the second term of the r.h.s. of \cref{decompose} converges to 0. We can take the sequence $c_T$ and $\gamma_T$ to be the same sequence which converges to infinity slowly enough. Then, it is enough for us to analyze the asymptotic probability of the below expression:
\begin{equation}\label{start}
    \sup_{|x|, |y|\leq c_T}\bigg\vert \mathbb{E}^*\left[\widehat{F}_{\epsilon}(\widehat{\mathcal{L}}(x, y,\hat{\epsilon}^*_{T+1})) \right] - \mathbb{E}\left[F_{\epsilon}\left( \mathcal{L}(x,y,\epsilon_{T+1})   \right)\right] \bigg\vert>\varepsilon.
\end{equation}
Decompose the l.h.s. of \cref{start} as:
\begin{equation}\label{intermediate}
\begin{split}
    &\sup_{|x|, |y|\leq c_T}\bigg\vert \mathbb{E}^*\left[\widehat{F}_{\epsilon}(\widehat{\mathcal{L}}(x, y,\hat{\epsilon}^*_{T+1})) \right] - \mathbb{E}\left[F_{\epsilon}\left( \mathcal{L}(x,y,\epsilon_{T+1})   \right)\right] \bigg\vert\\
    & \leq \sup_{|x|, |y|\leq c_T}\bigg\vert \mathbb{E}^*\left[\widehat{F}_{\epsilon}(\widehat{\mathcal{L}}(x, y,\hat{\epsilon}^*_{T+1})) \right] - \mathbb{E}^*\left[ F_{\epsilon}(\widehat{\mathcal{L}}(x, y,\hat{\epsilon}^*_{T+1})) \right]   \bigg\vert \\
    &+ \sup_{|x|, |y|\leq c_T}\bigg\vert  \mathbb{E}^*\left[ F_{\epsilon}(\widehat{\mathcal{L}}(x, y,\hat{\epsilon}^*_{T+1})) \right] -  \mathbb{E}\left[F_{\epsilon}\left( \mathcal{L}(x,y,\epsilon_{T+1})   \right)\right]   \bigg\vert.
\end{split}
\end{equation}
Then, we analyze two terms on the r.h.s. of \cref{intermediate} separately. For the first term, we have:
\begin{equation}\label{part1}
\begin{split}
    &\sup_{|x|, |y|\leq c_T}\bigg\vert \mathbb{E}^*\left[\widehat{F}_{\epsilon}(\widehat{\mathcal{L}}(x, y,\hat{\epsilon}^*_{T+1})) \right] - \mathbb{E}^*\left[ F_{\epsilon}(\widehat{\mathcal{L}}(x, y,\hat{\epsilon}^*_{T+1})) \right]   \bigg\vert\\
    & \leq \sup_{|x|, |y|\leq c_T}\mathbb{E}^*\bigg\vert \widehat{F}_{\epsilon}(\widehat{\mathcal{L}}(x, y,\hat{\epsilon}^*_{T+1})) -  F_{\epsilon}(\widehat{\mathcal{L}}(x, y,\hat{\epsilon}^*_{T+1}))  \bigg\vert\\
    & \leq \sup_{|x|, |y|\leq c_T, z} \bigg\vert \widehat{F}_{\epsilon}(\widehat{\mathcal{L}}(x, y,z)) -  F_{\epsilon}(\widehat{\mathcal{L}}(x, y,z))  \bigg\vert \overset{p}{\to} 0,~\text{under \cref{Lemma4.1}}.
\end{split}
\end{equation}
For the second term on the r.h.s. of \cref{intermediate}, we have:
\begin{equation}\label{part2}
\begin{split}
    &\sup_{|x|, |y|\leq c_T}\bigg\vert  \mathbb{E}^*\left[ F_{\epsilon}(\widehat{\mathcal{L}}(x, y,\hat{\epsilon}^*_{T+1})) \right] -  \mathbb{E}\left[F_{\epsilon}\left( \mathcal{L}(x,y,\epsilon_{T+1})   \right)\right]   \bigg\vert \\
    &\leq \sup_{|x|, |y|\leq c_T}\bigg\vert \frac{1}{T}\sum_{i=1}^{T}F_{\epsilon}(\widehat{\mathcal{L}}(x,y,\hat{\epsilon}_i)) -  \frac{1}{T}\sum_{i=1}^{T}F_{\epsilon}(\mathcal{L}(x,y,\epsilon_i))    \bigg\vert\\
    &+ \sup_{|x|, |y|\leq c_T}\bigg\vert \frac{1}{T}\sum_{i=1}^{T}F_{\epsilon}(\mathcal{L}(x,y,\epsilon_i)) - \mathbb{E}\left[F_{\epsilon}\left( \mathcal{L}(x,y,\epsilon_{T+1})   \right)\right]      \bigg\vert, 
\end{split}
\end{equation}
where $\{\epsilon_j\}_{j=1}^{T}$ are taken as $X_i - \phi(X_{i-1},\theta_1)$ for $i=1,\ldots,T$. We can show:
\begin{equation}\label{epandhatep}
\begin{split}
    &\mathbb{P}\left(\max_{i=1,\ldots,T}\bigg\vert\epsilon_i - \hat{\epsilon}_i\bigg\vert>\varepsilon\right)\\
    & = \mathbb{P}\left(\max_{i=1,\ldots,T}\bigg\vert X_i - \phi(X_{i-1},\theta_1) - X_i + \phi(X_{i-1},\widehat{\theta}_1)          \bigg\vert>\varepsilon\right)\\
    & \leq \mathbb{P}\left((\max_{i=1,\ldots,T}|X_{i-1}|>c_T)\right) \\
    &+ \mathbb{P}\left(  \left(\max_{i=1,\ldots,T}|X_{i-1}|<c_T\right) \bigcap \left(\max_{i=1,\ldots,T}\bigg\vert  \phi(X_{i-1},\widehat{\theta}_1) - \phi(X_{i-1},\theta_1)    \bigg\vert>\varepsilon\right) \right)\\
    & \leq o(1) + \mathbb{P}\left(\sup_{|x| \leq c_T}\bigg\vert \phi(x,\widehat{\theta}_1) - \phi(x,\theta_1)   \bigg\vert >\varepsilon\right) \\
    & \to 0,~\text{under A9 and A10.}
\end{split}
\end{equation}
We further consider two terms on the r.h.s. of \cref{part2} separately. For the first term, by Taylor expansion, we have:
\begin{equation}\label{part2part1}
\begin{split}
    &\sup_{|x|, |y|\leq c_T}\bigg\vert \frac{1}{T}\sum_{i=1}^{T}F_{\epsilon}(\widehat{\mathcal{L}}(x,y,\hat{\epsilon}_i)) -  \frac{1}{T}\sum_{i=1}^{T}F_{\epsilon}(\mathcal{L}(x,y,\epsilon_i))    \bigg\vert\\
    & = \sup_{|x|, |y|\leq c_T}\bigg\vert \frac{1}{T}\sum_{i=1}^{T}\left( F_{\epsilon}(\mathcal{L}(x,y,\epsilon_i)) + f_{\epsilon}(o_i)(\widehat{\mathcal{L}}(x,y,\hat{\epsilon}_i) - \mathcal{L}(x,y,\epsilon_i))  \right)  -  \frac{1}{T}\sum_{i=1}^{T}F_{\epsilon}(\mathcal{L}(x,y,\epsilon_i)) \bigg\vert \\
    & = \sup_{|x|, |y|\leq c_T}\bigg\vert \frac{1}{T}\sum_{i=1}^{T} f_{\epsilon}(o_i)(\widehat{\mathcal{L}}(x,y,\hat{\epsilon}_i) - \mathcal{L}(x,y,\epsilon_i))     \bigg\vert \\
    & \leq \sup_{|x|, |y|\leq c_T} \frac{1}{T}\sum_{i=1}^{T}\bigg\vert  f_{\epsilon}(o_i)(\widehat{\mathcal{L}}(x,y,\hat{\epsilon}_i) - \mathcal{L}(x,y,\epsilon_i))\bigg\vert  \\ 
    &\leq \sup_{|x|, |y|\leq c_T} \sup_{z}|f_{\epsilon}(z)|\cdot\frac{1}{T}\sum_{i=1}^{T}\bigg\vert  \widehat{\mathcal{L}}(x,y,\hat{\epsilon}_i) - \mathcal{L}(x,y,\epsilon_i)\bigg\vert\\
    & \leq  \sup_{|x|, |y|\leq c_T} C\cdot\frac{1}{T}\sum_{i=1}^{T}\bigg\vert  \widehat{\mathcal{L}}(x,y,\hat{\epsilon}_i) - \mathcal{L}(x,y,\epsilon_i)    \bigg\vert~\text{(under A11)}\\
    & \leq \sup_{|x|, |y|\leq c_T,j\in\{1,\ldots,T\}} C\cdot\bigg\vert  \widehat{\mathcal{L}}(x,y,\hat{\epsilon}_j) - \mathcal{L}(x,y,\epsilon_j)   \bigg\vert.
\end{split} 
\end{equation}
 From \cref{epandhatep} and A9--A12, we have \cref{part2part1} converges to 0 in probability. For the second term on the r.h.s. of \cref{part2}, by the uniform law of large numbers, we have: 
 \begin{equation}
     \sup_{|x|, |y|\leq c_T}\bigg\vert \frac{1}{T}\sum_{i=1}^{T}F_{\epsilon}(\mathcal{L}(x,y,\epsilon_i)) - \mathbb{E}\left[F_{\epsilon}\left( \mathcal{L}(x,y,\epsilon_{T+1})   \right)\right]      \bigg\vert \overset{p}{\to} 0. 
 \end{equation}
 Combine all pieces, \cref{start} converges to 0 in probability, which implies \cref{Theo4.1}. 
\end{proof}

\begin{proof}[\textbf{\textsc{Proof of Lemma 4.2}}]
Under A4-A7, the time series $\{X_t\}$ is ergodic. Besides, under A2 and A10, we can show $L(\vartheta)$ is uniformly finite for $\vartheta\in\Theta_1$. With A13, by the uniform law of larger numbers for the ergodic series, see Theorem 6 of \cite{kirch2012testing} for a reference, we have:
\begin{equation}\label{Eq75}
    \sup_{\vartheta\in\Theta_1}|L_{T}(\vartheta) - L(\vartheta)| \overset{p}{\to} 0.
\end{equation}
Under A13 and A14, we can easily show that:
\begin{equation}\label{Eq76}
    \inf_{|\vartheta - \theta_1|>\varepsilon}L(\vartheta)>L(\theta_1),\text{for}~\forall\epsilon,
\end{equation}
\cref{Eq76} implies that given $\forall\varepsilon>0$, $\exists C>0$ such that $|\vartheta-\theta_1|>\epsilon \Rightarrow L(\vartheta) - L(\theta_1)\geq C >0$, thus we have:
\begin{equation}\label{Eq77}
\begin{split}
     \mathbb{P}(|\widehat{\theta}_1-\theta_1 |>\epsilon) &\leq \mathbb{P}(L(\widehat{\theta}_1) - L(\theta_1)\geq C )\\
     & = \mathbb{P}(L(\widehat{\theta}_1) - L_T(\widehat{\theta}_1) + L_T(\widehat{\theta}_1) -  L(\theta_1)\geq C)\\
     &\leq \mathbb{P}( L(\widehat{\theta}_1) - L_T(\widehat{\theta}_1) + L_T(\theta_1) -  L(\theta_1)\geq C   )\\
     &\leq \mathbb{P}(2\sup_{\vartheta\in\Theta_1}|L_{T}(\vartheta) - L(\vartheta)|>C ) \overset{}{\to} 0.
\end{split}
\end{equation}
The last inequality of \cref{Eq77} is guaranteed by \cref{Eq75}. 
\end{proof}

\begin{proof}[\textbf{\textsc{Proof of Lemma 4.3}}]
The proof of the consistency of $\widehat{\theta}_1$ to $\theta_1$ is the same as the proof of \cref{Lemma4.2}. Similar with the proof of \cref{Lemma4.2}, by the ergodic property of the series $\{X_t\}$, we know:
\begin{equation}\label{Eq78}
    \sup_{\vartheta\in\Theta_2}\left|\frac{1}{T}\sum_{t=1}^{T}\left(\frac{X_{t} - \phi(X_{t-1},\theta_1)}{h(X_{t-1},\vartheta)}\right)^2 - \mathbb{E}\left(\frac{X_{t} - \phi(X_{t-1},\theta_1)}{h(X_{t-1},\vartheta)}\right)^2\right|\overset{p}{\to} 0.
\end{equation}
Actually, after the first step estimation, we have got the $\widehat{\theta}_1$ which is consistent to the true parameter $\theta_1$. Thus, we need to find the below convergence relationship:
\begin{equation}\label{Eq79}
     \sup_{\vartheta\in\Theta_2}\left|\frac{1}{T}\sum_{t=1}^{T}\left(\frac{X_{t} - \phi(X_{t-1},\widehat{\theta}_1)}{h(X_{t-1},\vartheta)}\right)^2 - \mathbb{E}\left(\frac{X_{t} - \phi(X_{t-1},\theta_1)}{h(X_{t-1},\vartheta)}\right)^2\right|\overset{p}{\to} 0. 
\end{equation}
Since $\widehat{\theta}_1$ converges to $\theta_1$ in probability, it is easily to find:
\begin{equation}
    \sup_{\vartheta\in\Theta_2}\left|\frac{1}{T}\sum_{t=1}^{T}\left(\frac{X_{t} - \phi(X_{t-1},\widehat{\theta}_1)}{h(X_{t-1},\vartheta)}\right)^2 - \frac{1}{T}\sum_{t=1}^{T}\left(\frac{X_{t} - \phi(X_{t-1},\theta_1)}{h(X_{t-1},\vartheta)}\right)^2   \right|\overset{p}{\to} 0,
\end{equation}
which implies \cref{Eq79} for a compact set $\Theta_2$ in conjunction with \cref{Eq78}. Then, since $|~\cdot~ - a|$ is a uniformly continuous function for a constant $a$, by applying the uniform continuous mapping theorem which is the Theorem 1 of \cite{kasy2019uniformity} on \cref{Eq79}, we can get:
\begin{equation}\label{Eq81}
       \sup_{\vartheta\in\Theta_2}|K_{T}(\vartheta,\widehat{\theta}_1) - K(\vartheta,\theta_1)| \overset{p}{\to} 0 ,
\end{equation}
where $K(\vartheta,\theta_1)$ and $K_T(\vartheta,\widehat{\theta}_1)$ are:
\begin{equation}
    K(\vartheta,\theta_1) = \left|\mathbb{E}\left(\frac{X_{t} - \phi(X_{t-1},\theta_1)}{h(X_{t-1},\vartheta)}\right)^2 - 1 \right|~;~ K_T(\vartheta,\widehat{\theta}_1) = \left|\frac{1}{T}\sum_{t=1}^{T}\left(\frac{X_{t} - \phi(X_{t-1},\widehat{\theta}_1)}{h(X_{t-1},\vartheta)}\right)^2 - 1 \right|.
\end{equation}
Then, we can repeat the procedure in the proof of \cref{Lemma4.2} to show:
\begin{equation}
    \mathbb{P}(|\widehat{\theta}_2-\theta_2 |>\varepsilon)\overset{}{\to} 0,~\forall \varepsilon>0.
\end{equation}
\end{proof}

\begin{proof}[\textbf{\textsc{Proof of Theorem 4.2}}]
Throughout this proof, we focus on a sequence of sets $\Omega_{T+p}\subseteq \mathbb{R}^{T+p}$, such that $\mathbb{P}\left((X_{-p+1},\ldots,X_T)\in\Omega_{T+p}\right) = o(1)$.We first explain the truth that the bootstrap series is ergodic for $(X_{-p+1},\ldots,X_T)\in\Omega_{T+p}$. Particularly, we want to check analogous A4-A7 in the bootstrap world. With the consistency property of parameter estimators, i.e., A9 and the continuous density function of residuals after the convolutional manipulation, we can easily find A4-A7 are satisfied in the bootstrap world; see Theorem 2 of \cite{franke2004bootstrapping} for more discussions. 

To show the closeness of the stationary distribution of bootstrap and real series. Based on the proof of Theorem 3 of \cite{franke2002properties} and Theorem 3 of \cite{franke2004bootstrapping}, we can get:
\begin{equation}\label{eq70}
    \sup_{B}|\Pi(B) -\Pi^*(B)| = o(1),
\end{equation}
which holds for all measurable sets $B$, where $\Pi(B)$ and $\Pi^*(B)$ represent stationary distribution for real series and bootstrap series, respectively. This implies, in conjunction with \cref{Eq34} and the condition that $\mathbb{P}(X_{t}\notin\gamma_{T}) = o(1)$, we can get:
\begin{equation}
\begin{split}
        &\int_{\mathbb{R}^{2}}(x_{1} - \phi(x_{0},\Xi_{\theta}))^2\hat{f}_{\epsilon}(x_1 - \phi(x_0,\hat{\theta}))\pi^{*}(x_0)dx_1dx_0\\
        & = \int_{\gamma_T^2}(x_{1} - \phi(x_{0},\Xi_{\theta}))^2\hat{f}_{\epsilon}(x_1 - \phi(x_0,\hat{\theta}))\pi^{*}(x_0)dx_1dx_0 + o(1)\\
        & \overset{}{\to} \mathbb{E}[X_1 - \phi(X_0,\Xi_\theta)]^2,
\end{split}
\end{equation}
which implies \cref{Eq35} directly. 

\end{proof}

\begin{proof}[\textbf{\textsc{Proof of Theorem 4.3}}]
To simplify notation, we consider the case $p=q=1$. Higher order cases can be handled similarly. We first build the estimation inference for $\widehat{\theta}_1$. Starting from A19, we have:
    \begin{equation}\label{starttheta1}
        op(T^{-1/2}) = \nabla L_{T}(\widehat{\theta}_1) = \nabla L_{T}(\theta_1) + \nabla^2 L_{T}(\widetilde{\theta}_1)(\widehat{\theta}_1 - \theta_1),
    \end{equation}
where $\widetilde{\theta}_1$ is between $\widehat{\theta}_1$ and $\theta_1$; hence $\widetilde{\theta}_1$ also converges to $\theta_1$ in probability. First, we consider $\nabla^2 L_{T}(\widetilde{\theta}_1)$, which has a form as below:
\begin{equation}
    \nabla^2 L_{T}(\widetilde{\theta}_1) = \frac{1}{T}\nabla^2 \sum_{t=1}^{T}\left(X_{t} - \phi(X_{t-1},\widetilde{\theta}_1)  \right)^2 =  \frac{1}{T}\sum_{t=1}^{T} \nabla^2 q_t(\widetilde{\theta}_1).
\end{equation}
Under A18, since it is easy to check $\mathbb{E}\sup_{\vartheta\in\Theta_{1}^{0}}$ $||\nabla^2 q_{1}(\vartheta)$ $||<\infty$, by combining the uniform law of larger numbers for the ergodic series, dominated convergence theorem, the consistency between $\widetilde{\theta}_1$ and $\theta_1$ and the continuity of $L_T(\cdot)$ w.r.t. $\theta_1$, we can get:
\begin{equation}
    \nabla^2 L_{T}(\widetilde{\theta}_1) \overset{p}{\to} B_1,
\end{equation}
where $B_1 = 2\cdot\mathbb{E}\left(\nabla\phi(X_0,\theta_1)(\nabla\phi(X_0,\theta_1))^{\top}\right)$. Thus, we can multiply both side of \cref{starttheta1} by $\sqrt{T}$ to get:
\begin{equation}
    op(1) = \sqrt{T}\nabla L_{T}(\theta_1) + (B_1 + op(1))\sqrt{T}(\widehat{\theta}_1 - \theta_1),
\end{equation}
which further implies that:
\begin{equation}\label{estinf11}
    \sqrt{T}(\widehat{\theta}_1 - \theta_1) = -B_1^{-1}\sqrt{T}\nabla L_{T}(\theta_1) + op(1),
\end{equation}
where $\sqrt{T}\nabla L_{T}(\theta_1)$ has a concrete form as below:
\begin{equation}\label{normalpart}
    -\sqrt{T}\nabla L_{T}(\theta_1) = \frac{2}{\sqrt{T}}\sum_{t=1}^{T}(X_t - \phi(X_{t-1},\theta_1))\nabla \phi(X_{t-1},\theta_{1}).
\end{equation}
We need the CLT for strongly mixing processes to show the asymptotic normality of \cref{normalpart}. Based on Theorem 1.7 of \cite{bosq2012nonparametric} for univariate case, we can show the normality with the Cramér–Wold theorem. Also, \cite{kirch2012testing} applied the strong invariance principle from \cite{kuelbs1980almost} to show the asymptotic distribution of multivariate cases directly. Since we need to analyze the asymptotic distribution of parameters in the bootstrap world, we take the first approach which is more clear. We consider:
\begin{equation}
   -\sqrt{T}\bm{a}^{\top}\nabla L_{T}(\theta_1) = \frac{2}{\sqrt{T}}\sum_{t=1}^{T}(X_t - \phi(X_{t-1},\theta_1))\bm{a}^{\top}\nabla \phi(X_{t-1},\theta_{1}),
\end{equation}
where $\bm{a}^{\top}$ is any real vector that has the same dimension as $\theta_1$. Since we assume we can correctly specify the model, we have:
\begin{equation}
   -\sqrt{T}\bm{a}^{\top}\nabla L_{T}(\theta_1) \overset{d}{\to} N(0,\tau_1^2),
\end{equation}
where $\tau_1^2$ has the form as follows:
\begin{equation}
\begin{split}
    \tau_1^2 &= \sum_{i = -\infty}^{\infty}\text{Cov}\left(2\cdot(X_1 - \phi(X_{0},\theta_1))\bm{a}^{\top}\nabla \phi(X_{0},\theta_{1}), 2\cdot(X_{i+1} - \phi(X_{i},\theta_1))\bm{a}^{\top}\nabla \phi(X_{i},\theta_{1})  \right)\\
    & = 4\cdot\mathbb{E}\left((\bm{a}^{\top}\nabla \phi(X_{0},\theta_{1}))(X_1 - \phi(X_{0},\theta_1))^2(\nabla \phi(X_{0},\theta_{1})^{\top}\bm{a}) \right) \\
    & + 2\sum_{i=1}^{\infty}\text{Cov}\left(2\cdot(X_1 - \phi(X_{0},\theta_1))\bm{a}^{\top}\nabla \phi(X_{0},\theta_{1}), 2\cdot(X_{i+1} - \phi(X_{i},\theta_1))\bm{a}^{\top}\nabla \phi(X_{i},\theta_{1})  \right)\\
    & = 4\cdot\mathbb{E}(\bm{a}^{\top}\sigma(X_{0},\theta_2)\nabla \phi(X_{0},\theta_{1})\nabla \phi(X_{0},\theta_{1})^{\top}\sigma(X_{0},\theta_2)\bm{a}) \\
    & + 2\sum_{i=1}^{\infty}\text{Cov}\left(2\cdot\epsilon_1\bm{a}^{\top}\nabla \phi(X_{0},\theta_{1}),2\cdot \epsilon_{i+1}\bm{a}^{\top}\nabla \phi(X_{i},\theta_{1})  \right)\\
    & = 4\cdot\mathbb{E}(\bm{a}^{\top}\sigma(X_{0},\theta_2)R_{1}\sigma(X_{0},\theta_2)\bm{a}),
\end{split}
\end{equation}
where $R_{1} = \nabla \phi(X_{0},\theta_{1})\nabla \phi(X_{0},\theta_{1})^{\top}$. Thus, applying Cramér–Wold theorem, we have:
\begin{equation}
    -\sqrt{T}\nabla L_{T}(\theta_1) \overset{d}{\to} N(0,\Omega_1),
\end{equation}
where $\Omega_1$ is $4\cdot\mathbb{E}(\sigma(X_{0},\theta_2)R_{1}\sigma(X_{0},\theta_2))$
Thus, we can conclude from \cref{estinf11} that:
\begin{equation}
    \sqrt{T}(\widehat{\theta}_1 - \theta_1) \overset{d}{\to} N(0,B_1^{-1}\Omega_1 B_1^{-1}).
\end{equation}
Compared to the result in \cite{kirch2012testing}, they got another form of $\Omega_1$:
\begin{equation}
    \Omega_1 = \lim_{T\to\infty}\frac{1}{T}\mathbb{E}\left[\left(\nabla\sum_{t=1}^{T}\left(X_{t} - \phi(X_{t-1},\theta_1)  \right)^2\right) \left(\nabla\sum_{t=1}^{T}\left(X_{t} - \phi(X_{t-1},\theta_1)  \right)^2\right)^{T} \right].
\end{equation}
With a correctly specified model, it is equivalent to our form. 

We can also analyze the distribution of parameter estimation $\widehat{\theta}_2$. By A19, we first have:
\begin{equation}\label{ktanalyze}
     op(T^{-1/2}) = \nabla K_{T}(\theta_2,\widehat{\theta}_1) + \nabla^2 K_{T}(\widetilde{\theta}_2,\widehat{\theta}_1)(\widehat{\theta}_2 - \theta_2).
\end{equation}
For simplifying the notation, we write:
\begin{equation}
\begin{split}
    K_{T}(\widetilde{\theta}_2,\widehat{\theta}_1) &= \left(\frac{1}{T}\sum_{t=1}^{T}\left(\frac{X_{t} - \phi(X_{t-1},\widehat{\theta}_1)}{h(X_{t-1},\widetilde{\theta}_2)}\right)^2 - 1\right)^2 \\
    &  =  \left(\frac{1}{T}\sum_{t=1}^{T}g(X_t,X_{t-1},\widetilde{\theta}_2, \widehat{\theta}_1) -1 \right)^2.\\
\end{split}
\end{equation}
We can find $ \nabla K_{T}(\theta_2,\theta_1)$ and $ \nabla^2 K_{T}(\theta_2,\theta_1)$ w.r.t $\theta_2$ have a following forms respectively:
\begin{equation}
\begin{split}
     &\nabla K_{T}(\theta_2,\theta_1) = 2\cdot\left(\frac{1}{T}\sum_{t=1}^{T}g(X_t,X_{t-1},\theta_2, \theta_1) -1 \right)\cdot\left(\frac{1}{T}\sum_{t=1}^{T}\nabla g(X_t,X_{t-1},\theta_2, \theta_1) \right) \\
      &\nabla^2 K_{T}(\theta_2,\theta_1) = 2\cdot\left(\frac{1}{T}\sum_{t=1}^{T}\nabla g(X_t,X_{t-1},\theta_2, \theta_1) \right)\cdot\left(\frac{1}{T}\sum_{t=1}^{T}\nabla g(X_t,X_{t-1},\theta_2, \theta_1) \right)^{\top}  \\
      & + 2\cdot\left(\frac{1}{T}\sum_{t=1}^{T}g(X_t,X_{t-1},\theta_2, \theta_1) -1 \right)\cdot \left(\frac{1}{T}\sum_{t=1}^{T}\nabla^2 g(X_t,X_{t-1},\theta_2, \theta_1) \right).
\end{split}
\end{equation}
Similarly to analyze $\nabla^2 L_{T}(\widetilde{\theta}_1)$, under the consistence relationship between $\widetilde{\theta}_2$ and $\theta_2$, $\widehat{\theta}_1$ and $\theta_1$, we can get:
\begin{equation}
     \nabla^2 K_{T}(\widetilde{\theta}_1,\widehat{\theta}_1) \overset{p}{\to} B_2.
\end{equation}
where $B_2 =2\cdot\mathbb{E}(\nabla g(X_{1},X_{0},\theta_2,\theta_1)) \cdot \mathbb{E}(\nabla g(X_{1},X_{0},\theta_2,\theta_1)^{\top})$. 
Looking back to \cref{ktanalyze}, it is left to analyze $\sqrt{T}\nabla K_{T}(\theta_2,\widehat{\theta}_1)$. Through Slutsky’s theorem, it is asymptotically equivalent to analyze $\sqrt{T}\left(\frac{2}{T}\sum_{t=1}^{T}[g(X_t,X_{t-1},\theta_2, \theta_1) -1]\right)\cdot B_3$; $B_3 =\mathbb{E}( \nabla g(X_{1},X_{0},\theta_2,\theta_1))$. Apply the same technique as we analyzed the distribution of $\widehat{\theta}_1$, we can get:
\begin{equation}
  \sqrt{T}\bm{a}^{\top}\nabla K_{T}(\theta_2,\widehat{\theta}_1) \overset{d}{\to} N(0,\tau_2^2),
\end{equation}
where $\tau_2^2 = 4\cdot\mathbb{E}(\bm{a}^{\top}B_3R_2B_3^{\top}\bm{a})$; $R_2 = (g(X_1,X_0,\theta_2, \theta_1) -1)^2$. Thus, we have:
\begin{equation}
    \sqrt{T}\nabla K_{T}(\theta_2,\widehat{\theta}_1) \overset{d}{\to} N(0,\Omega_2),
\end{equation}
where $\Omega_2 = 4\cdot\mathbb{E}(B_3R_2B_3^{\top})$. This further implies that:
\begin{equation}
     \sqrt{T}(\widehat{\theta}_2 - \theta_2) \overset{d}{\to} N(0,B_2^{-1}\Omega_2 B_2^{-1}).
\end{equation}
\end{proof}

\newpage
\bibliographystyle{apalike}
\bibliography{refs}
\end{document}